\documentclass[useAMS,usenatbib]{mn2e}
\usepackage[british]{babel}
\usepackage{graphicx}
\usepackage{txfonts}
\usepackage{verbatim}
\usepackage{natbib}
\usepackage{ragged2e}
\usepackage{enumerate}
\usepackage{lscape}
\usepackage{subfigure}
\usepackage{hyperref}
\usepackage{breakurl} 
\usepackage{tikz}
\usepackage{url}
\newcommand{\Lx}{\mbox{$\mathrm{L_{X}}$}}

\newcommand{\fxfk}{\mbox{$\log(\mathrm{f_{X}/f_{K_S})}$}}
\newcommand{\Rv}{\mbox{$\mathrm{R_{V}}$}}
\newcommand{\Av}{\mbox{$\mathrm{A_{V}}$}}
\newcommand{\Ak}{\mbox{$\mathrm{A_{K}}$}}
\newcommand{\Mk}{\mbox{$\mathrm{M_{K}}$}}

\newcommand{\HeII}{\mbox{${\mathrm {He\,II}}$}}
\newcommand{\ergcms}{\mbox{$\mathrm{erg\,cm^{-2}s^{-1}}$}}
\newcommand{\ergs}{\mbox{$\mathrm{erg\,s^{-1}}$}}
\newcommand{\Nh}{\mbox{$\mathrm{N_H}$}}
\newcommand{\Ks}{\mbox{$\mathrm{K_s}$}}
\newcommand{\Rsun}{\mbox{$\mathrm{R}_{\odot}$}}

\newcommand{\Lsun}{\mbox{$\mathrm{L}_{\odot}$}}
\newcommand{\Teff}{\mbox{$\mathrm{T_{eff}}$}}
\newcommand{\Lbol}{\mbox{$\mathrm{L_{bol}}$}}
\newcommand{\Pid}{\mbox{$P_{{id}}$}}
\newcommand{\ctss}{\mbox{cts\,s$^{-1}$}}
\newcommand{\sourceone}{\mbox{3XMM\,J174347.4-292309}}
\newcommand{\sourcetwo}{\mbox{3XMM\,J180920.3-201857}}
\newcommand{\sourcethree}{\mbox{3XMM\,J184541.1-025225}}
\newcommand{\sourcefour}{\mbox{3XMM\,J185210.0+001205}}
\newcommand{\sourcefive}{\mbox{3XMM\,J190144.5+045914}}
\newcommand{\one}{\mbox{Source~\#1}}
\newcommand{\two}{\mbox{Source~\#2}}
\newcommand{\three}{\mbox{Source~\#3}}
\newcommand{\four}{\mbox{Source~\#4}}
\newcommand{\five}{\mbox{Source~\#5}}

\title[]{Infrared identification of hard X-ray sources in the Galaxy\thanks{Based on service observations made with the WHT operated on the island of La Palma by the Isaac Newton Group in the Spanish Observatorio del Roque de los Muchachos of the Instituto de Astrofísica de Canarias.}}
\author[A.~Nebot G\'omez-Mor\'an et al.]{
A.~Nebot G\'omez-Mor\'an$^{1}$\thanks{E-mail: ada.nebot@astro.unistra.fr},
C.~Motch$^{1}$,
F.-X.~Pineau$^{1}$,
F.~J.~Carrera$^{2}$, 
M.~W.~Pakull$^{1}$,
F.~Riddick$^{3}$\\
$^{1}$Observatoire Astronomique de Strasbourg, Universit\'e de Strasbourg, CNRS, UMR 7550, 11 rue de l'Universit\'e, 67000 Strasbourg, France.\\
$^{2}$Instituto de Fisica de Cantabria (CSIC-UC), Avenida de los Castros, 39005 Santander, Spain.\\
$^{3}$Isaac Newton Group of Telescopes, Apartado de Correos 321, E-38700 Santa Cruz de La Palma, Spain.
}
\begin{document}
\date{}
\maketitle
\begin{abstract}
The nature of the low- to intermediate-luminosity (\Lx$\,\sim\,10^{32-34}$\,\ergs) source population revealed in hard band (2\,--\,10\,keV) X-ray surveys of the Galactic Plane is poorly understood. To overcome such problem we cross-correlated the XMM-Newton 3XMM-DR4 survey with the infrared 2MASS and GLIMPSE catalogues. We identified reliable X-ray--infrared associations for 690 sources. 
We selected 173 sources having hard X-ray spectra, typical of hard X-ray high-mass stars (kT$\,>\,5$\,keV), and 517 sources having soft X-ray spectra, typical of active coronae. About 18\,\% of the soft sources are classified in the literature: $\sim\,91\%$ as stars, with a minor fraction of WR stars. Roughly 15\,\% of the hard sources are classified in the literature: $\sim\,68\%$ as high-mass X-ray stars single or in binary systems (WR, Be and HMXBs), with a small fraction of G and B stars. 
We carried out infrared spectroscopic pilot observations at the William Herschel Telescope for five hard X-ray sources. Three of them are high-mass stars with spectral types WN7-8h, Ofpe/WN9 and Be, and \Lx$\sim\,10^{32}-10^{33}$\ergs. One source is a colliding-wind binary, while another source is a colliding-wind binary or a supergiant fast X-ray transient in quiescence. The Be star is a likely $\gamma$-Cas system. The nature of the other two X-ray sources is uncertain. 
The distribution of hard X-ray sources in the parameter space made of X-ray hardness ratio, infrared colours and X-ray-to-infrared flux ratio suggests that many of the unidentified sources are new $\gamma$-Cas analogues, WRs and low \Lx\ HMXBs. However, the nature of the X-ray population with \Ks $\geq$ 11 and average X-ray-to-infrared flux ratio remains unconstrained.
\end{abstract}

\begin{keywords}
Stars: emission-line, Be; Wolf-Rayet - 
X-rays: stars, binaries. 
\end{keywords}
\section{Introduction}
The landscape of the Galaxy is dominated by relatively nearby stellar coronae at low X-ray luminosities (\Lx$\,<\,10^{31}$\,\ergs) and soft energies ($<\,$2 keV) \citep[e.g.][]{motchetal97-1,nebotgomezmoranetal13-1}. In contrast, at high X-ray luminosities (\Lx$\,>\,10^{35}$\,\ergs), the make-up of the Milky Way is dominated by high-mass X-ray binaries (HMXB) and low-mass X-ray binaries (LMXB) \citep{grimmetal02-1,gilfanov04-1}. On the other hand, the nature of the low- to intermediate-luminosity source population revealed in hard band (2\,--\,10\,keV) X-ray surveys of the Galactic Plane, such as those carried out by ASCA \citep{sugizakietal01-1} and more recently by Chandra \citep[e.g.][]{ebisawaetal05-1} and XMM-Newton \citep[e.g.][]{handsetal04-1} is only partly understood. Importantly, in this \Lx\ range we expect to find systems predicted by evolutionary scenarios of low and high mass X-ray binaries. For instance, we have not yet detected the long-lived wind-accreting low X-ray luminosity stages preceding or following the bright phase during which they become conspicuous. The common envelope spiral-in creation channel for low-mass X-ray binaries predicts the existence of pre-LMXBs which could radiate as much as $\sim\,10^{32}$\,\ergs\ in hard X-rays through accretion of stellar wind onto the neutron star or its magnetosphere \citep[see e.g.][]{tauris+vanheuvel06-1}. Likewise, evolution theories of high mass X-ray binaries foresee that about $10^{6}$ wind accreting binaries containing a main sequence star and a neutron star or a black hole populate the Galaxy \citep{pfahletal02-1}. Moreover, Be + white dwarf systems are expected to be ten times more frequent than Be + neutron stars binaries \citep{raguzova01-1}, whereas none is known so far in our Galaxy. With mass accretion rates potentially similar to cataclysmic variables (CVs), many of these systems should emit copious amounts of hard X-rays. A census of the contribution of the different populations to the Galactic hard X-ray emission is hence needed if we want to test the validity of these evolutionary models.

Active binaries such as RS\,CVn systems, and CVs are expected to contribute significantly to the low luminosity hard X-ray source population \citep{sazonovetal06,revnivtsevetal09-1}. Massive objects such as magnetic OB stars \citep{gagneetal11-1}, colliding-wind binary O or Wolf Rayet stars \citep{skinneretal10-1,mauerhanetal10-1}, $\gamma$-Cas like objects, being so far the best Be + WD candidates \citep{motchetal07-1}, and X-ray transient binaries in quiescent state have also been identified in this luminosity range. However, despite significant efforts \citep{motchetal10-1,andersonetal11-1}, our knowledge of the makeup of the Galactic low- to medium-luminosity hard X-ray source population associated to massive stars is restricted due to the lack of fully identified X-ray source samples. First, the space density of each of these sub-classes of X-ray sources is basically unconstrained and, second, it remains to be seen whether any new types of Galactic X-ray-emitting massive stars, such as those predicted by evolutionary theories (e.~g. Be + white dwarf systems), are present within current hard X-ray Galactic surveys. Owing to sensitivity limits and localisation accuracies such studies can only be carried out in our Galaxy.

In this paper we investigate the properties of this important class of massive X-ray sources of low- to intermediate-luminosity by observing candidates selected from the most recent X-ray and infrared surveys. Our target selection, based on the cross-correlation of the 3XMM-DR4 catalogue with the 2MASS and GLIMPSE catalogues, is described in Section~\ref{sec:target}. As part of a pilot study we obtained infrared spectroscopic observations of a small sample of hard X-ray sources (see Section~\ref{sec:telescopes}), and produced images of the target environment (see Section~\ref{sec:images}), which helped to classify the sources (see Section~\ref{sec:SpT}). We discuss our results in Section~\ref{sec:dis} and conclude in Section~\ref{sec:summ}. 

\section{Target selection}
\label{sec:target}
In this section we briefly describe the cross-matched catalogues and the cross-matching procedure. We explain our selection criterion and list the X-ray sources chosen for further study at the telescope.
\subsection{Cross-matched catalogues}
\subsubsection{3XMM-DR4}
ESA's X-ray Multi-Mirror observatory XMM-Newton was launched in December 1999, and started operations at the beginning of 2000 \citep{jansenetal01-1}. It has three telescopes, each equipped with one X-ray CCD camera, comprising the European Photon Imaging Camera (EPIC). Two of the cameras are MOS CCD detectors \citep[referred to as MOS-1 and MOS-2 cameras,][]{turneretal01-1} and the third is a pn CCD detector \citep[referred to as EPIC-pn camera,][]{struderetal01-1}. While the EPIC-pn camera receives all the incident light, a grating disperses about 40\%\ of the light going to the EPIC-MOS cameras, diverging it to two Reflection Grating Spectrometers \citep[RGS,][]{denherderetal01-1}. The field of view (FOV) of the telescope is about 30\arcmin\ diameter and the EPIC cameras are sensitive in the energy range 0.2--12 keV with a spectral resolution E/$\Delta$E of about 20--50 and a spatial resolution of 5\arcsec\ (FWHM). 
The XMM-Newton Survey Science Centre (SSC) compiles catalogues of serendipitous sources detected by the EPIC cameras on board the XMM-Newton satellite \citep{watsonetal09-1}. The 3XMM-DR4 catalogue was released on the 15th September 2013\footnote{\url{http://xmmssc-www.star.le.ac.uk/Catalogue/3XMM-DR4/UserGuide_xmmcat.html}} and contains 531\,261 detections for 372\,728 unique X-ray sources. Taking into account overlaps the sky coverage is 794 deg$^2$, about 1.4\% of the sky. The limiting X-ray flux in the 0.2-12 keV band is $\sim2\times10^{-15}$\,\ergcms\ and source positions have a typical accuracy better than 3\arcsec\ (90\%\ confidence radius). Among other parameters, the catalogue contains information on the count rate, some flags describing the quality of the observation, the extension of the source, and the hardness ratios (HR), X-ray colours defined as: 
\begin{equation}
\mathrm{HR_{i} = \frac{C_{i+1}-C_{i}}{C_{i+1}+C_i}},
\end{equation}
where $C_i$ stands for the count rate in the energy band i, where i\,=\,0.2--0.5, 0.5--1.0, 1.0--2.0, 2.0--4.5, 4.5--12.0\,keV. 

\subsubsection{GLIMPSE}
The Galactic Legacy Infrared Mid-Plane Survey Extraordinaire \citep[GLIMPSE][]{benjaminetal03-1} is a Spitzer Space Telescope Mission program that aims to map the inner regions of the Galaxy in the infrared. GLIMPSE source catalogue (I, II + 3D) was published in June 2009 \citep{glimpse09-1} through VizieR \citep{ochsenbeinetal00-1}. It provides images and magnitudes for over 104 million sources in four bands: 3.6, 4.5, 5.8, and 8.0\,$\mu$m (I1, I2, I3 and I4). The catalogue covers Galactic longitudes $\pm65^\circ$ and latitudes $\pm1^\circ$ (up to $\pm3^\circ$ in the inner regions of the Galaxy). Spatial resolution is $<\,2$\arcsec, and magnitude limits are about 15.5, 15.0, 13.0, and 13.0 in each band respectively \citep{churchwelletal09-1}. GLIMPSE covers about 1\% of the sky with a typical position accuracy of 0.3\arcsec. 

\subsubsection{2MASS PSC}
The Two Micron All Sky Survey \citep[2MASS,][]{skrutskieetal06-1} used two 1.3\,m\ telescopes located in the northern and southern hemispheres equipped with near infrared cameras to obtain simultaneous photometry in three bands: J (1.25\,$\mu$m), H (1.65\,$\mu$m) and K (2.17\,$\mu$m), for the entire sky. The 2MASS point source catalogue was released in 2003 and contains about 471 million sources. The angular resolution is 2\arcsec\ and the survey completeness limits are J\,=\,15.8, H\,=\,15.1 and \Ks\,=\,14.3 mag. The mean position accuracy is $\sim80$\,mas rms \citep{cutrietal03-1}.  

\begin{table}
\small\addtolength{\tabcolsep}{-2pt}
\begin{center}
\caption{Completeness and spurious fraction of X-ray-infrared associations.}\label{t:table1}
\begin{tabular}{ccccccc}
\hline 
             & \multicolumn{5}{c}{3XMM-2MASS \Pid$^\dag$} \\
\hline
             & 50\% & 60\% & 70\% & 80\% & 90\% \\
Completeness & 89\% & 85\% & 80\% & 72\% & 58\% \\
Spurious fr. & 12\% & 10\% &  8\% &  6\% &  3\% \\
\hline
             & \multicolumn{5}{c}{3XMM-GLIMPSE \Pid} \\
\hline
             & 50\% & 60\% & 70\% & 80\% & 90\% \\
Completeness & 68\% & 59\% & 49\% & 35\% & 19\% \\
Spurious fr. & 24\% & 19\% & 15\% & 10\% &  5\% \\
\hline
\end{tabular}
\end{center}
\textbf{Notes.}$^\dag$ Limited to the GLIMPSE footprint.\\
\end{table}

\subsection{Cross-match procedure}
\label{sec:cross-match}
We cross-matched the 3XMM-DR4 catalogue with the 2MASS and GLIMPSE catalogues using the method developed by \cite{pineauetal11-1}. In brief this method looks for all the catalogue sources around the X-ray source position within a distance lower than 3.439$\sigma_{XC}$, corresponding to a completeness of 99.7\%. Here $\sigma_{XC}$ is the combined positional error of the two catalogues that are being cross-matched added in quadrature. For all the matches found within this distance, i.~e for all infrared candidates associated to a X-ray source, the method computes the likelihood ratio (LR) between the probability that the two sources (X-ray and infrared sources) are associated (\Pid) and thus have the same position, and the probability of finding a spurious association at the same angular distance and with the same magnitude or brighter than the candidate. For that purpose, local densities are calculated for sources brighter than the candidate. The LR depends on the probability of an X-ray source to have a counterpart in the considered catalogue, which depends on the object type considered, its properties and the relative fractions of the different populations. To estimate \Pid, it is common to build LR histograms and determine the contribution of spurious associations through time-consuming Monte Carlo simulations where positions and errors are randomised. In the method developed by \cite{pineauetal11-1}, the number of spurious associations in each LR bin is estimated, assuming a mean local surface density of X-ray sources, as the sum over all possible associations of the ratio between the surface element (convolution ellipse defined by the positions and their errors) and the field of view area. Once the rate of spurious associations is calculated \Pid\ can be estimated for each LR bin. 

\begin{figure}
\begin{center}
\includegraphics[width=\linewidth]{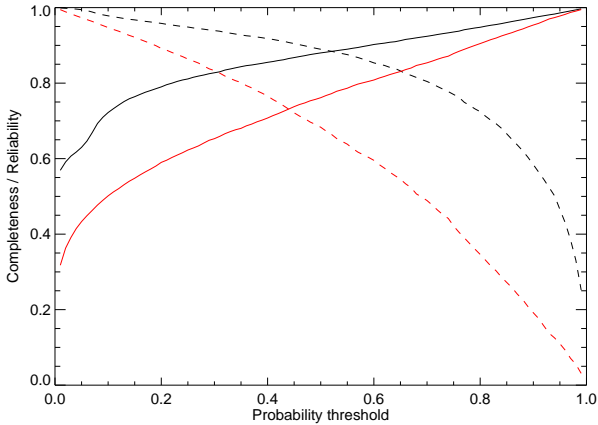}
\caption{Completeness (dashed line) and reliability (solid line) of the 3XMM-2MASS and 3XMM-GLIMPSE associations as a function of the threshold in the probability is shown in black and red respectively. }\label{g:compl_relia}
\end{center}
\end{figure}
\begin{table*}
\addtolength{\tabcolsep}{-5.0pt}
\begin{center}
\caption{Infrared counterparts of X-ray sources.}\label{t:table2}
\begin{tabular}{cccccccc}
\hline
Col.& Name             &  Contents                             &                         &                         &                           &                           &                          \\%
\hline                                                                                                                                                                                                  
1   &	 3XMM	       &  3XMM IAUNAME                        &  J174347.4-292309       & J180920.3-201857        & J184541.1-025225          &    J185210.0+001205       &  J190144.5+045914        \\%
2   &    RA	       &  SC\_RA                              &  265.94780              & 272.33492               & 281.42126                 &    283.04190              &  285.43566               \\%
3   &    DEC	       &  SC\_DEC                             &  -29.38609              & -20.31585               &  -2.87375                 &      0.20143              &    4.98748               \\%
4   &    POSERR	       &  SC\_POSERR                          &  1.21                   & 0.35                    & 0.44                      &    0.81                   &  0.58                    \\%
5   &    CTS1	       &  Count rate in 0.2-0.5 keV           &  0.00095                & 0.00037                 & 0.00017                   &    0.00093                &  0.00000                 \\%
6   &    CTS2	       &  Count rate in 0.5-1.0 keV           &  0.00000                & 0.00014                 & 0.00025                   &    0.00001                &  0.00014                 \\%
7   &    CTS3	       &  Count rate in 1.0-2.0 keV           &  0.00273                & 0.00065                 & 0.00202                   &    0.00138                &  0.00427                 \\%
8   &    CTS4	       &  Count rate in 2.0-4.5 keV           &  0.00288                & 0.00516                 & 0.01109                   &    0.00445                &  0.01528                 \\%
9   &    CTS5	       &  Count rate in 4.5-12.0 keV          &  0.01588                & 0.00583                 & 0.03401                   &    0.00247                &  0.01325                 \\%
10  &    CTS8	       &  Count rate in 0.2-12.0 keV          &  0.02245                & 0.01215                 & 0.04753                   &    0.00930                &  0.03295                 \\%
11  &    CTS9	       &  Count rate in 0.5-4.5 keV           &  0.00401                & 0.00536                 & 0.01321                   &    0.00599                &  0.01944                 \\%
12  &    HR1	       &  Hardness ratio 1                    &  -0.690                 & -0.572                  & -0.695                    &    -0.833                 &  1.000                   \\%
13  &    HR2	       &  Hardness ratio 2                    &  0.997                  & 0.963                   & 0.947                     &    0.948                  &  0.971                   \\%
14  &    HR3	       &  Hardness ratio 3                    &  0.557                  & 0.831                   & 0.902                     &    0.751                  &  0.581                   \\%
15  &    HR4	       &  Hardness ratio 4                    &  0.653                  & 0.146                   & -0.014                    &    -0.358                 &  -0.233                  \\%
16  &    HR1\_ERR      &  Error in hardness ratio 1           &  0.545                  & 0.215                   & 0.283                     &    0.265                  &  1.346                   \\%
17  &    HR2\_ERR      &  Error in hardness ratio 2           &  0.158                  & 0.058                   & 0.085                     &    0.096                  &  0.047                   \\%
18  &    HR3\_ERR      &  Error in hardness ratio 3           &  0.163                  & 0.033                   & 0.021                     &    0.062                  &  0.042                   \\%
19  &    HR4\_ERR      &  Error in hardness ratio 4           &  0.099                  & 0.038                   & 0.034                     &    0.085                  &  0.044                   \\%
20  &    2MASS	       &  2MASS IAUNAME                       &  17434750-2923098       & 18092038-2018572        & 18454111-0252257          &    18521006+0012073       &  19014457+0459147        \\%
21  &    d\_2MASS      &  Distance to 2MASS assoc.            &  0.4                    & 0.18                    & 0.3                       &    2.24                   &  0.290                   \\%
22  &    PROB\_2MASS   &  Probability of 2MASS assoc.         &  0.989                  & 0.998                   & 0.994                     &    0.670                  &  0.998                   \\%
23  &    J	       &  2MASS J magnitude                   &  11.145                 & 11.923                  & 11.562                    &    11.333                 &  11.148                  \\%
24  &    H	       &  2MASS H magnitude                   &  7.897                  & 9.225                   & 8.700                     &    9.487                  &  9.763                   \\%
25  &    \Ks           &  2MASS \Ks\ magnitude                &  6.159                  & 7.719                   & 7.014                     &    7.852                  &  8.841                   \\%
26  &    eJ	       &  2MASS J magnitude error             &  0.020                  & 0.027                   & 0.024                     &    0.024                  &  0.022                   \\%
27  &    eH	       &  2MASS H magnitude error             &  0.042                  & 0.023                   & 0.038                     &    0.028                  &  0.021                   \\%
28  &    eK	       &  2MASS \Ks\ magnitude error          &  0.017                  & 0.023                   & 0.027                     &    0.027                  &  0.021                   \\%
29  &    GLIMPSE       &  GLIMPSE IAUNAME  	              &  G359.4078+00.1051      & G010.1568-00.3361       & G029.7187-00.0316         &    G033.1949-00.0701      &  G038.5439-00.0120       \\%
30  &    d\_GLIMPSE    &  Distance to GLIMPSE assoc.          &  0.320                  & 0.34                    & 0.12                      &    2.35                   &  0.110                   \\%
31  &    PROB\_GLIMPSE &  Probability of GLIMPSE assoc.       &  0.998                  & 0.999                   & 0.999                     &    0.966                  &  0.995                   \\%
32  &    I1	       &  GLIMPSE 3.6$\mu$m mag.              &  4.891                  & 6.893                   &                           &    6.537                  &  7.873                   \\%
33  &    I2	       &  GLIMPSE 4.5$\mu$m mag.              &  4.798                  & 6.227                   & 5.346                     &    5.462                  &  7.589                   \\%
34  &    I3	       &  GLIMPSE 5.8$\mu$m mag.              &  4.152                  & 5.516                   & 4.578                     &    4.781                  &  7.290                   \\%
35  &    I4	       &  GLIMPSE 8.0$\mu$m mag.              &  4.137                  & 5.322                   & 4.284                     &    4.469                  &  6.981                   \\%
36  &    eI1	       &  GLIMPSE 3.6$\mu$m mag. error        &  0.133                  & 0.064                   &                           &    0.128                  &  0.046                   \\%
37  &    eI2	       &  GLIMPSE 4.5$\mu$m mag. error        &  0.046                  & 0.086                   & 0.067                     &    0.066                  &  0.043                   \\%
38  &    eI3	       &  GLIMPSE 5.8$\mu$m mag. error        &  0.015                  & 0.024                   & 0.024                     &    0.030                  &  0.037                   \\%
39  &    eI4	       &  GLIMPSE 8.0$\mu$m mag. error        &  0.019                  & 0.024                   & 0.027                     &    0.021                  &  0.025                   \\%
40  &    $fx^\dag$      &  X-ray flux in the 2-12 keV          &  2.12e-13               & 1.15e-13                & 4.49e-13                  &    8.78e-14               &  3.11e-13                \\%
    &                  &  (ecf\,$\sim\,9.44\,\times\,10^{-12}$)&    	                &                         &                           &                           &                          \\%
41  &    $fk$	       &  Infrared \Ks\ flux   	              &  3.18e-09               & 7.58e-10                & 1.45e-09                  &    6.69e-10               &  2.69e-10                \\%
42  &    $\log(fx/fk)$  &  X-ray to infrared flux ratio       &  -4.25                  & -3.86                   & -3.53                     &    -4.01                  &  -3.00                   \\%
43  &    X-Type	       &  S for soft or H for hard            &  H                      & H                       & H                         &    H                      &  H                       \\%
44  &    NAME          &  Simbad name   	              &                         &                         &                           &                           &                          \\%
45  &    Class	       &  Object class  	              &  Star(?)                & WR*                     & WR*                       &    Star(?)                &  Be*                     \\%
46  &    SpT	       &  Spectral type  	              &  M2I                    & WN7-8h                  & Ofpe/WN9                  &    M2I                    &  09e-B3e               \\%
47  &    References    &  Spectral type reference             &  This paper             & This paper              & This paper                &    This paper             &  This paper              \\%
\hline\\
\end{tabular}
\end{center}
\begin{justify} 
  \textbf{Notes.}The meaning of each line is given under the column \emph{Contents}. $^\dag$ We estimated the X-ray flux from the EPIC-pn count rate in the 2\,--\,12\,keV band using a certain count rate to energy conversion factor (\emph{ecf}). Among others the ecf depends on the combination of camera and filter used, the X-ray emission mechanism, and the Galactic absorption. The ecf was computed assuming a thermal Bremsstrahlung model with kT\,=\,5\,keV and an absorption of \Nh\,$=\,10^{22}$ atoms\,/\,cm$^2$ (\emph{ecf}$\,\sim\,9.44\,\times\,10^{-12}$\,count\,/\,s\,\,/\,\ergcms). The choice of the X-ray model and \Nh\ value was done so as to be representative of hard X-ray emitting high-mass stars absorbed by the mean Galactic absorption of the sample, calculated from the mean (H-\Ks) and assuming sources have no intrinsic absorption and a colour (H-\Ks)\,=\,0 independent of their spectral type. A different \Nh\ value yields a different ecf, X-ray fluxes are thus to be taken with caution. Class and spectral type come from the literature with exception of the five sources studied in this paper. When available the main Simbad name is given. We present here an excerpt of the table which will be fully available in the electronic version.
\end{justify} 
\end{table*}

For sources above a given threshold in the identification probability the method provides us with the completeness and the fraction of possible spurious associations in the cross-matched survey (see Fig.~\ref{g:compl_relia}). In Table~\ref{t:table1} we list the survey completeness and the fraction of spurious catalogue identifications as a function of \Pid. 

\begin{figure*}
\begin{center}
\includegraphics[width=0.49\linewidth]{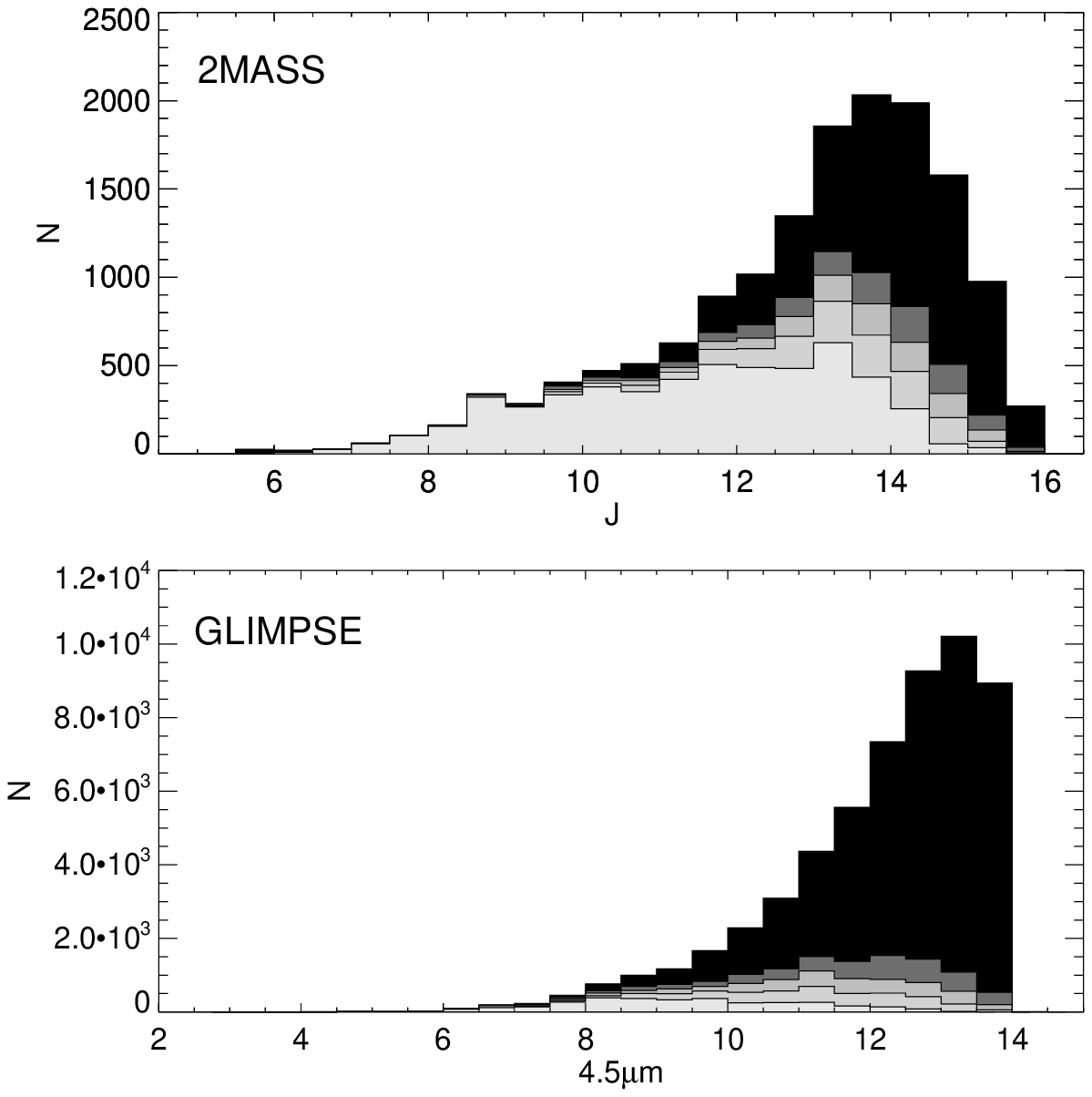}
\includegraphics[width=0.49\linewidth]{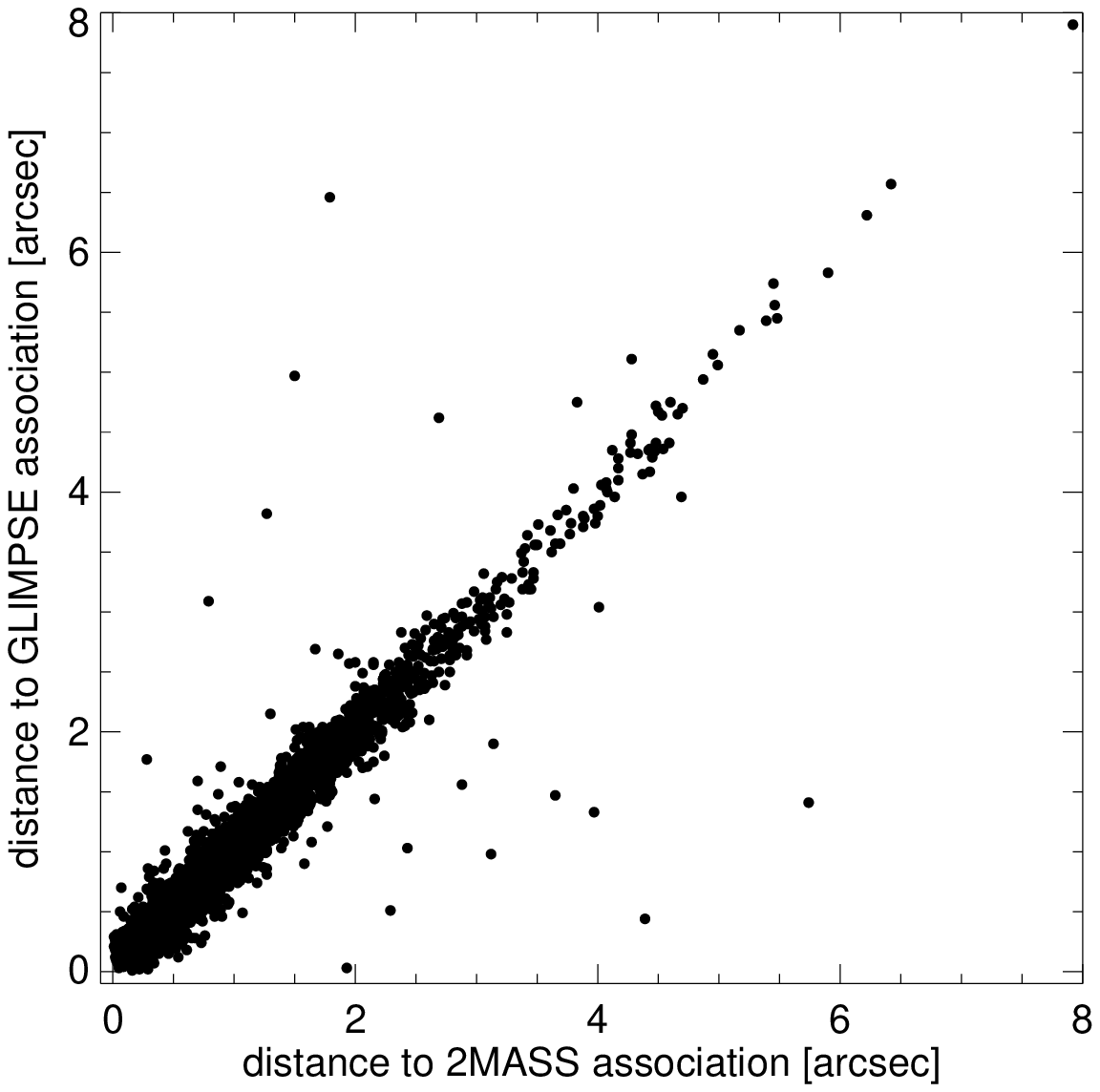}\hfill
\caption{Left panel: number of possible 3XMM-2MASS (top) and 3XMM-GLIMPSE (bottom) associations per magnitude bin for \Pid$\,>\,0.2,0.4,0.6,0.8,$ and $1$ (black, dark grey, grey, medium grey and light grey respectively). Right panel: distance between the 3XMM-GLIMPSE possible associations against the 3XMM-2MASS associations for pairs with a common 3XMM source. \label{g:histograms_Pid}}
\end{center}
\end{figure*}
In practice, we excluded sources uncorrected for satellite attitude errors (task EPOSCOR) and with positional errors (SC\_POSERR) in the 3XMM-DR4 catalogue larger than 5\arcsec. 
For the GLIMPSE catalogue we took into account all sources from the v2.0 Archive (v2.0\_GLMIIA), i.e. sources detected twice in at least one band with S/N\,$>\,5$ and we limited our sample to sources brighter than 14 magnitudes in the 4.5\,$\mu$m band, corresponding to a catalogue completeness of 90\% \citep{kobulnickyetal13-1}. 
For the 2MASS-PSC catalogue we only considered sources with the best photometric quality, Qflag = 'AAA', which corresponds to a S/N\,$>\,10$ and brighter than 15.8 magnitudes in the J band, level at which the catalogue is 99\% complete\footnote{\url{http://www.ipac.caltech.edu/2mass/releases/allsky/doc/sec2_2.html}}. We only cross-matched the 3XMM-DR4 and the 2MASS catalogues in areas also covered by GLIMPSE. 

To speed up the process we made use of HEALPix Multi Order Coverage maps (MOC)\footnote{\url{http://www.ivoa.net/documents/MOC/index.html}} which provided us with the coverage of a survey within a given minimum pixel resolution. Before performing the cross-match we pre-selected XMM-Newton sources within the footprint of the GLIMPSE survey. We accessed corresponding MOCs from the interactive software sky atlas \emph{Aladin} \citep{bonnareletal00-1}, and selected sources within the MOC making use of the function \emph{inMoc} in TOPCAT\footnote{\url{http://www.star.bris.ac.uk/~mbt/topcat/}} \citep{taylor11-1}. The overlap between 3XMM-DR4 and GLIMPSE footprints covers 0.1\% of the sky, and contains 31\,115 unique X-ray sources. 

We found 15\,018 possible 3XMM-2MASS pairs (for 9\,437 unique 3XMM sources) and 56\,751 3XMM-GLIMPSE possible pairs (for 16\,668 unique 3XMM sources). In the left panel of Fig.~\ref{g:histograms_Pid} we show magnitude histograms of all the possible pairs above a certain associated probability. The higher the cut in \Pid\ the brighter the sub-sample is, implying that we can only find reliable X-ray--infrared associations, i.~e. a small fraction of spurious associations, for relatively bright infrared sources. 

\subsubsection{3XMM, 2MASS, GLIMPSE associations}
Once all the possible 3XMM-2MASS and 3XMM-GLIMPSE associations were obtained, the next step was to identify each 3XMM-2MASS-GLIMPSE unique association. For that purpose, we proceeded as follows. 

We limited the sample to X-ray good quality point-like sources above a $4\sigma$ detection. For assuring a $4\sigma$ detection we selected sources with parameter sc\_det\_ml\,$>$\,10. We defined sources as good X-ray detections if sum\_flag\,$\le$\,2, where sum\_flag contains the summary information on 12 flags set automatically and manually which indicate the quality of the X-ray detection. We restricted our sample to point-like X-ray sources, for which we imposed SC\_EXTENT\,=\,0, i.e. no additional source extent model was convolved with the point spread function fit to the source count distribution. 

We restricted our study to associations with 3XMM-2MASS $\Pid\,>\,0.6$, value at which the expected fraction of spurious associations is about 10\% (see Table~\ref{t:table1}). For each of these unique 3XMM-2MASS associations, we looked for all possible 3XMM-GLIMPSE counterparts, based on their unique 3XMM-IAUNAME identifier, and we considered the right match the one with the highest 3XMM-GLIMPSE \Pid. In this way, we found 2\,988 possible 3XMM-2MASS-GLIMPSE associations. Being aware that this could eventually lead to wrong associations we created finding charts and visually inspected them to ensure that the chosen GLIMPSE counterpart is the correct one. In the right panel of Fig.~\ref{g:histograms_Pid} we plot the distance between 3XMM-2MASS and the 3XMM-GLIMPSE pairs. There are a few cases where the distance to the 2MASS association is clearly larger than that of the GLIMPSE association and viceversa. Visual inspection in these cases revealed that the 2MASS and GLIMPSE sources are possibly not associated. We prefer to be cautious about these cases and flag them as dubious for any further study. We also checked whether our associations between 2MASS and GLIMPSE sources were in agreement with those given in the GLIMPSE catalogue. In ten cases, we had different associations and in another 29 cases no association was given in the GLIMPSE catalogue. All these 39 cases corresponded to associations already flagged as being dubious based on visual inspection of their finding charts. We calculated the distance between 2MASS and GLIMPSE sources and found that most of the 3XMM-2MASS-GLIMPSE associations with high $\Pid$ have a distance smaller than 1\arcsec. 

\subsection{Selection of hard X-ray high-mass star candidates}
\label{sec:cands}
To distinguish sources that are likely active coronae with soft X-ray spectra from high-mass stars with hard X-ray spectra we compared the observed X-ray properties of our sample with modelled data. Following the same approach as in \cite{nebotgomezmoranetal13-1} we modelled X-ray spectra for different types of objects: 
\begin{itemize}
\item Model 1: a young population (70\,Myr) of active coronae, for which we assumed two temperature thermal emission components, one componenent (kT$_1\,=\,0.5$\,keV) representing the hot plasma and a second hotter component (kT$_2\,=\,0.8$\,keV) \citep{guedeletal97-1}. 
\item Model 2: RS\,CVn binaries, for which we took as a model values of WW\,Dra \citep{dempseyetal93-1}, with (kT$_1$, kT$_2$)\,=\,(0.2, 2.1) and emisivities EM$_2$/EM$_1$\,=\,0.67. 
\item Model 3: a population of hard X-ray massive stars. We chose one single temperature plasma kT\,=\,5\,keV as a representative value of a population of hard X-ray massive stars containing WR stars \citep[see e.~g. WR\,142,][]{oskinovaetal09-1,sokaletal10-1} and $\gamma$-Cas analogues \citep{lopesdeoliveira07-1}.
\end{itemize}
We modelled the emission of these three different populations as a function of Galactic absorption and we computed the corresponding hardness ratios. For more details we refer the reader to \cite{nebotgomezmoranetal13-1}. 

\begin{figure*}
\begin{center}
\includegraphics[width=0.49\linewidth]{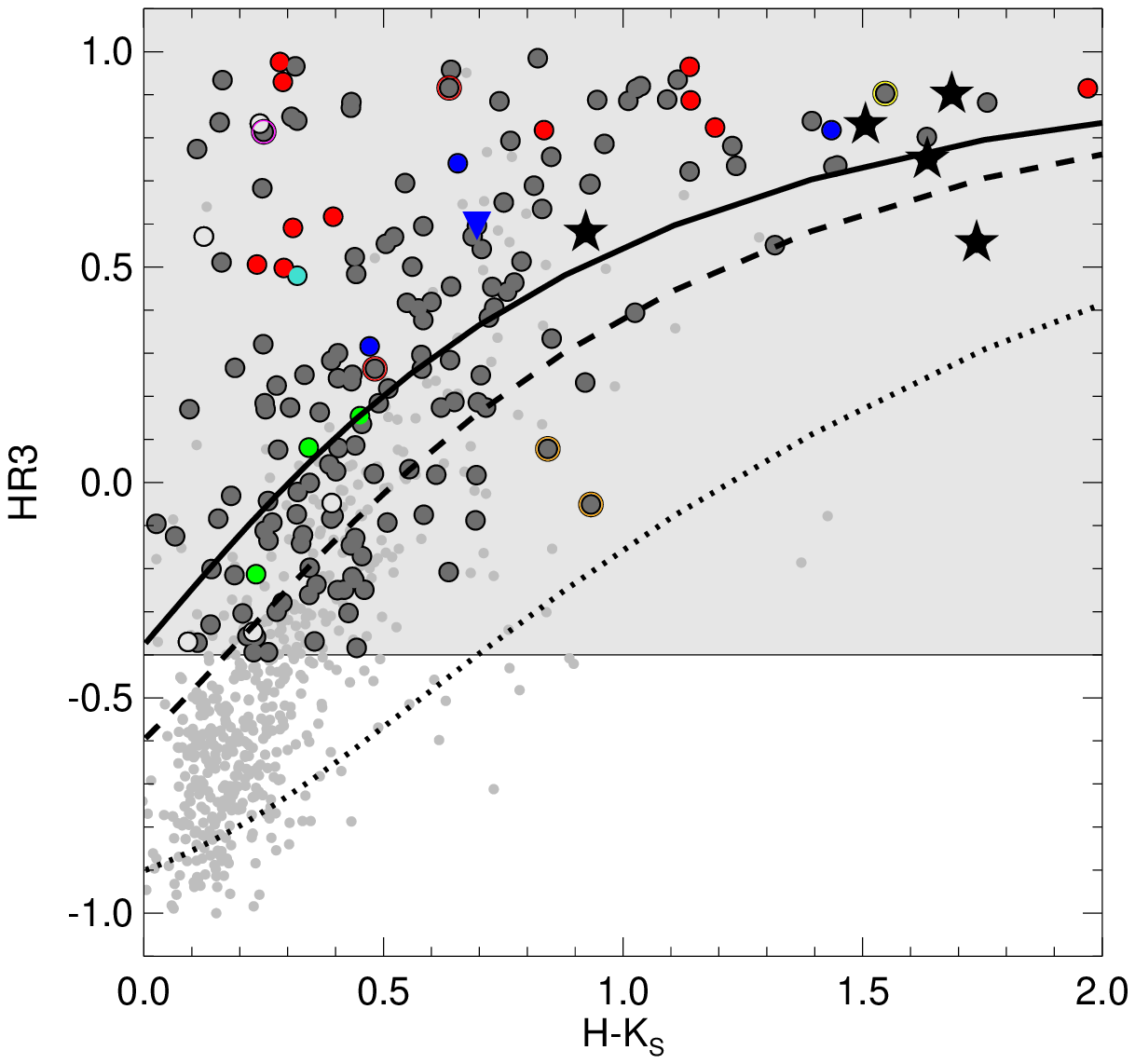}
\includegraphics[width=0.49\linewidth]{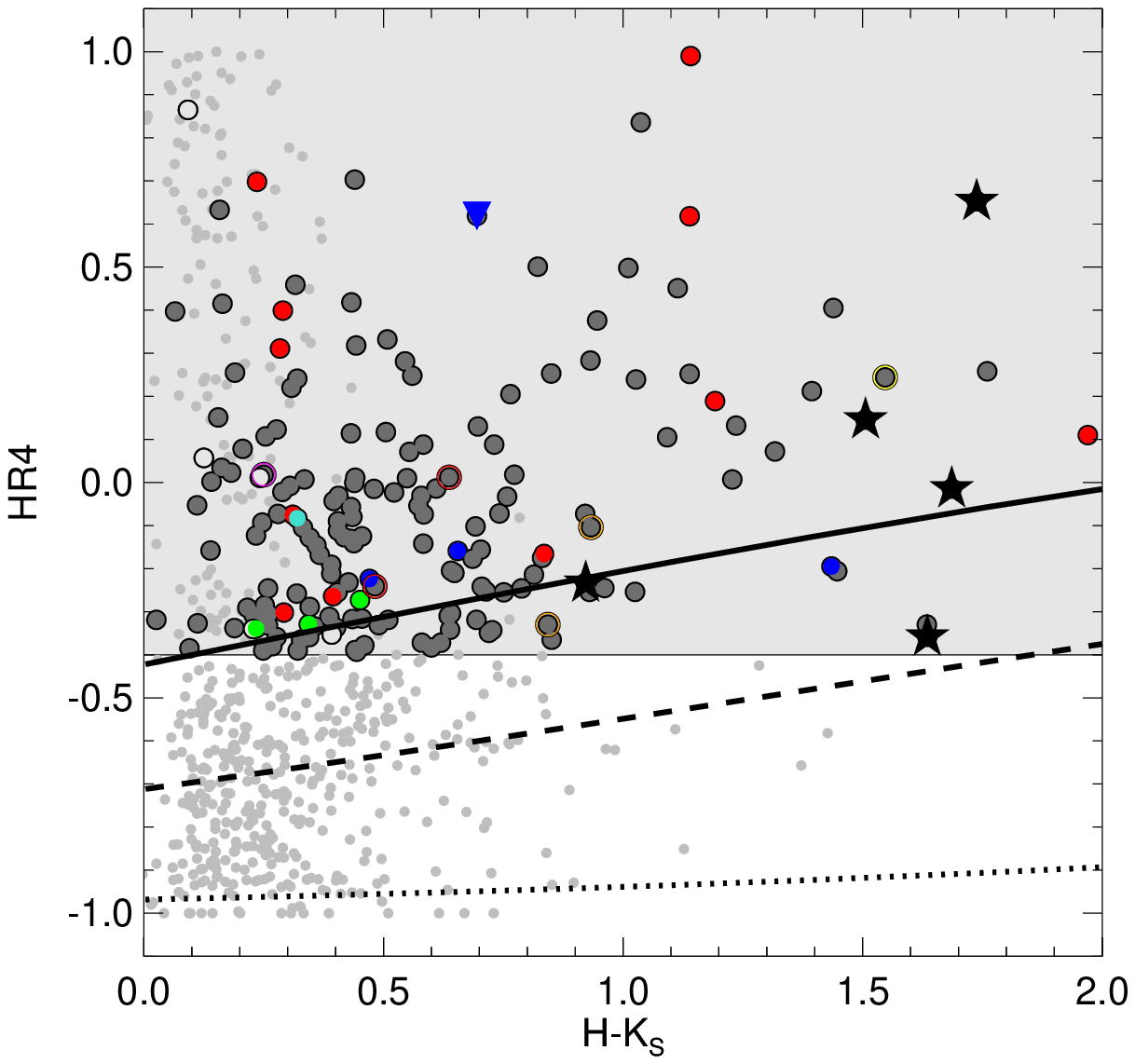}\hfill
\caption{X-ray versus infrared colour-colour diagrams for soft (light grey) and hard sources (dark grey) compared to modelled values for active coronae (dotted), RS\,CVn binaries (dashed) and hard X-ray massive stars (solid). Note that we classified as hard sources those having both hardness ratios HR3 and HR4 greater than -0.4. Plotted with filled circles are confirmed stars (2 B and 3 G stars) (light grey), HMXBs (red), Be stars (green), and Wolf-Rayet stars (blue). The only AGN from the sample is shown as a filled blue upside down triangle and a Nova as a turquoise filled circle. Unidentified hard X-ray sources are shown in dark grey, surrounded by a coloured ring when a candidate classification of the source is available: candidate HMXBs (red), candidate YSO (yellow), candidate AGB stars (orange), candidate LMXB (magenta). Targets observed at the telescope are shown with a star symbol. See text for more details.}\label{g:hr3hr4}
\end{center}
\end{figure*}

Among the 3XMM-2MASS-GLIMPSE associations, a large fraction have rather large hardness ratio errors, making it difficult to differenciate between soft and hard sources. To avoid such problem we only considered sources in our sample with errors smaller than 0.2 in HR3 and HR4, a value close to the mean error in the 3XMM-DR4 catalogue. We finally restricted the sample to sources with errors in the HR2 smaller than 1 and that have been detected in the EPIC-pn camera, so as to compute fluxes avoiding any possible problems related to the merging of the EPIC-MOS and EPIC-pn individual detections. We prefered the EPIC-pn camera due to its higher sensitivity compared to the two EPIC-MOS cameras. 
After applying these cuts, we were left with 690 3XMM-2MASS-GLIMPSE unique associations\footnote{Associations 3XMM\,J174505.4-285117--2MASS\,17450537-2851166--G000.0088+00.1411, 3XMM\,J175703.9-295921--2MASS\,17570406-2959219--G000.3666-02.6797 and 3XMM\,J174912.4-272537--2MASS\,17491250-2725357--G001.7001+00.1051 are dubious, the distance between the 2MASS and the GLIMPSE possible counterparts is larger than 1\arcsec.}. 

In Figure~\ref{g:hr3hr4} we show the location of all the 690 sources in two hardness ratio versus (H-\Ks) diagrams. Since the colour (H-\Ks) remains within 0 to +0.1 from A to K spectral types, independent on the luminosity class \citep{coveyetal07-1}, variations in colour mainly reflect variations in Galactic absorption or intrinsic emission by circumstellar matter. Under the assumption of no intrinsic colour, i.~e. (H-\Ks)\,=\,0, we calculated the expected hardness ratios as a function of (H-\Ks), using the formula \mbox{$\Nh\,=\,3.5\,\times\,10^{22}\,\times$\,E(H-\Ks)}, for the different kind of objects for which we modelled the X-ray spectra. We compared the observed values to modelled values for active corona, RS\,CVn binaries and hard X-ray massive stars. Among these three type of objects, only hard X-ray massive stars are expected to have HR4 above -0.4 at any Galactic absorption (see right panel in Fig.~\ref{g:hr3hr4}). 

We selected sources with hard spectra, defined by HR3 and HR4, indicating thermal temperatures greater than 5\,keV for \Nh\ varying from $1\times10^{20}$ to $3\times10^{22}$\,cm$^{-2}$. The minimum temperature limit (HR limit) applied to exclude all normal stellar coronal emitters is the following: 
\begin{equation}
\begin{array}{l}
  \mathrm{HR3 > -0.4}\,  \&\,  \mathrm{HR4 > -0.4}  \label{eq:hr}
\end{array}
\label{eq:crit}
\end{equation}
Selecting sources with HR3 and HR4 greater than $-0.4$ implies that sources are detected in hard X-rays due to either intrinsic hard X-ray emission (if (H-\Ks)$\sim$\,0--0.1) or a combination of intrinsic X-ray emission and high extinction (if (H-\Ks)$>$\,0.1). Sources with high values of HR3, HR4 and (H-\Ks) will probably have a high extinction. 

A total of 173 X-ray sources fulfill the selection criteria~(\ref{eq:hr}) and we thus consider them hard X-ray sources. The remaining sources are considered soft X-ray sources in what follows. In Fig.~\ref{g:hr3hr4} we show the location of soft and hard X-ray sources in the X-ray versus infrared coulour-colour diagrams. 

We investigated the nature of these 690 sources. We looked for possibles entries in SIMBAD making use of XCatDB\footnote{\url{http://xcatdb.unistra.fr/3xmm/}}, and we complemented our search of known sources making use of VizieR. In Table~\ref{t:table2} we list all the 690 associations. The content of the table is as follows. The XMM-Newton IAUNAME is followed by the X-ray positions, errors, count rates and hardness ratios. For each XMM-Newton source the best 2MASS and GLIMPSE associations are given. 2MASS and GLIMPSE IAUNAMES, magnitudes, the distance between the X-ray and the infrared positions and the probability of identification are listed. The X-ray (2-12\,keV) and the \Ks\ fluxes and the X-ray-to-infrared flux ratio are also given. The main SIMBAD name, the type of source and the spectral class are given when known. We finally give the spectral type reference used for the classification of each source. We present here an excerpt of the table which will be fully available in the electronic version.  

We estimated the X-ray flux from the EPIC-pn count rate in the 2\,--\,12\,keV band and using a count rate to energy conversion factor (\emph{ecf}). Among others the ecf depends on the combination of camera and filter used, the X-ray emission mechanism, and the Galactic absorption \citep[see][]{mateosetal09-1}. The ecf was computed assuming a thermal Bremsstrahlung model with kT\,=\,5\,keV and an absorption of \Nh\,$=\,10^{22}$ atoms\,/\,cm$^2$ (\emph{ecf}$\,\sim\,9.44\,\times\,10^{-12}$\,count\,/\,s\,\,/\,\ergcms). The choice of the X-ray model and \Nh\ value was done so as to be representative of hard X-ray emitting high-mass stars absorbed by the mean Galactic absorption of our sample, calculated from the mean (H-\Ks) and assuming sources have no intrinsic absorption and a colour (H-\Ks)\,=\,0 independent of their spectral type. A different \Nh\ value and/or a different assumed model yields a different ecf, and thus X-ray fluxes presented in Table~\ref{t:table2} are to be taken with caution. 

\begin{figure*}
\begin{center}
\includegraphics[width=0.195\linewidth]{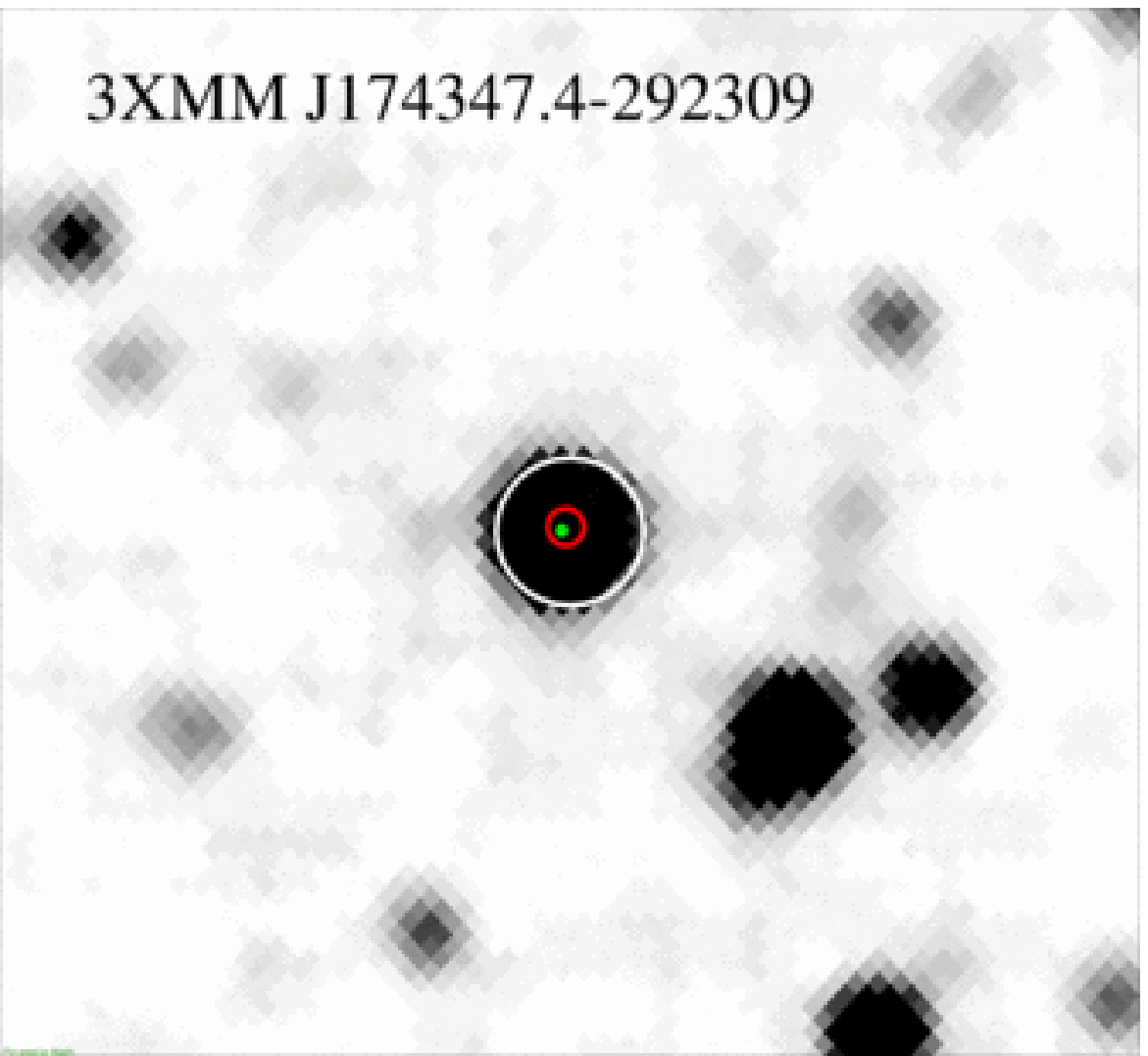}
\includegraphics[width=0.195\linewidth]{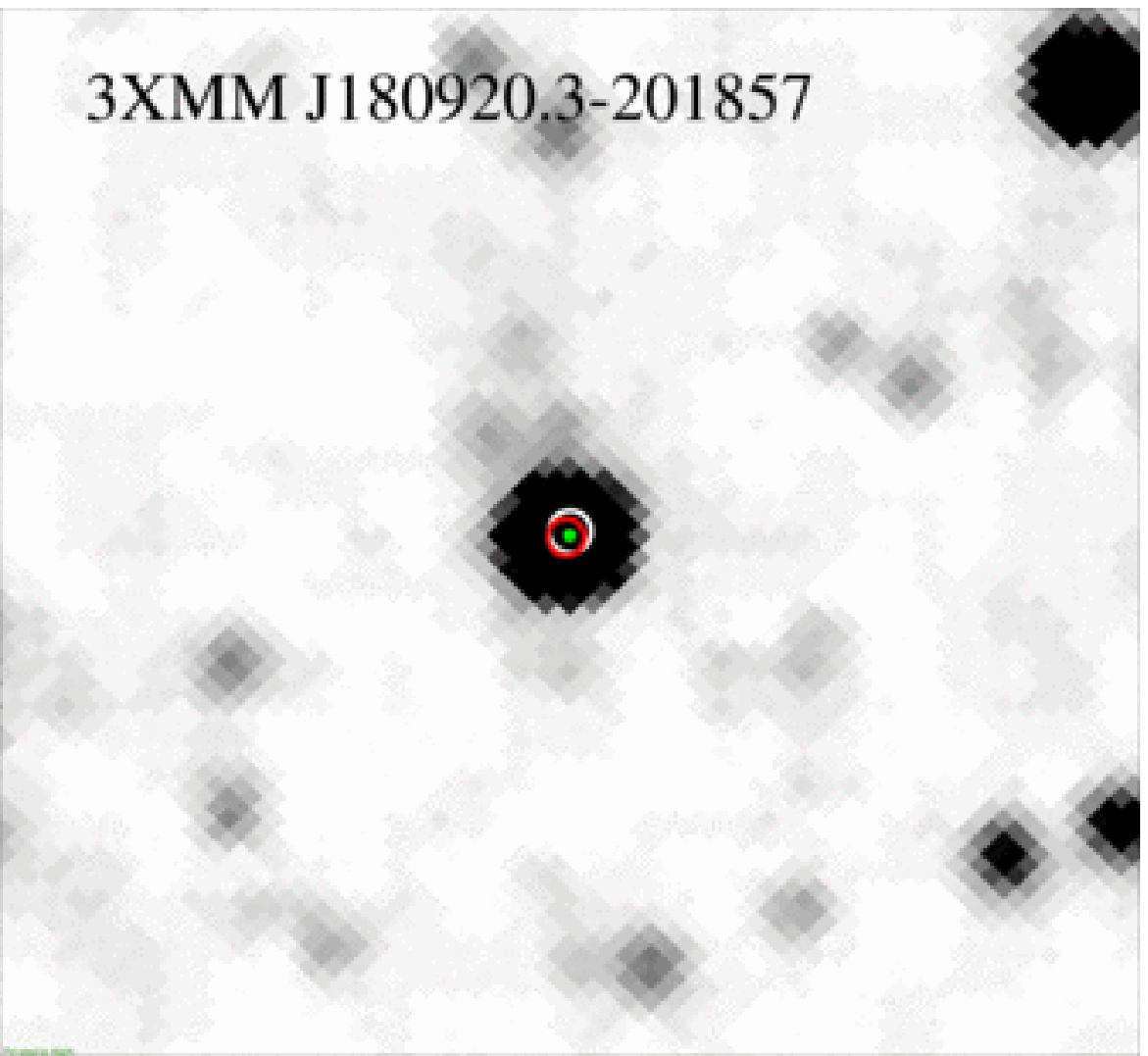}
\includegraphics[width=0.195\linewidth]{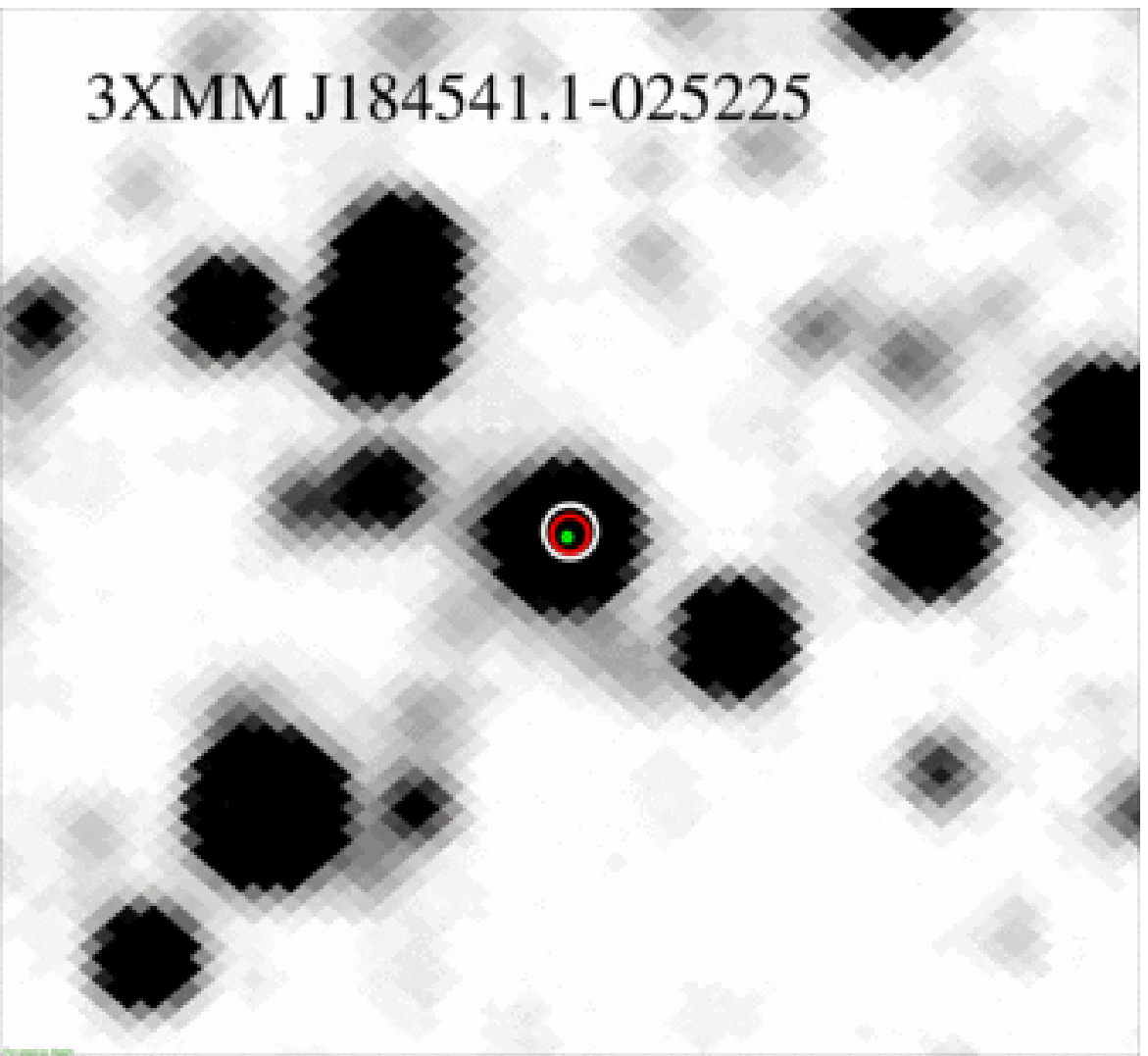}
\includegraphics[width=0.195\linewidth]{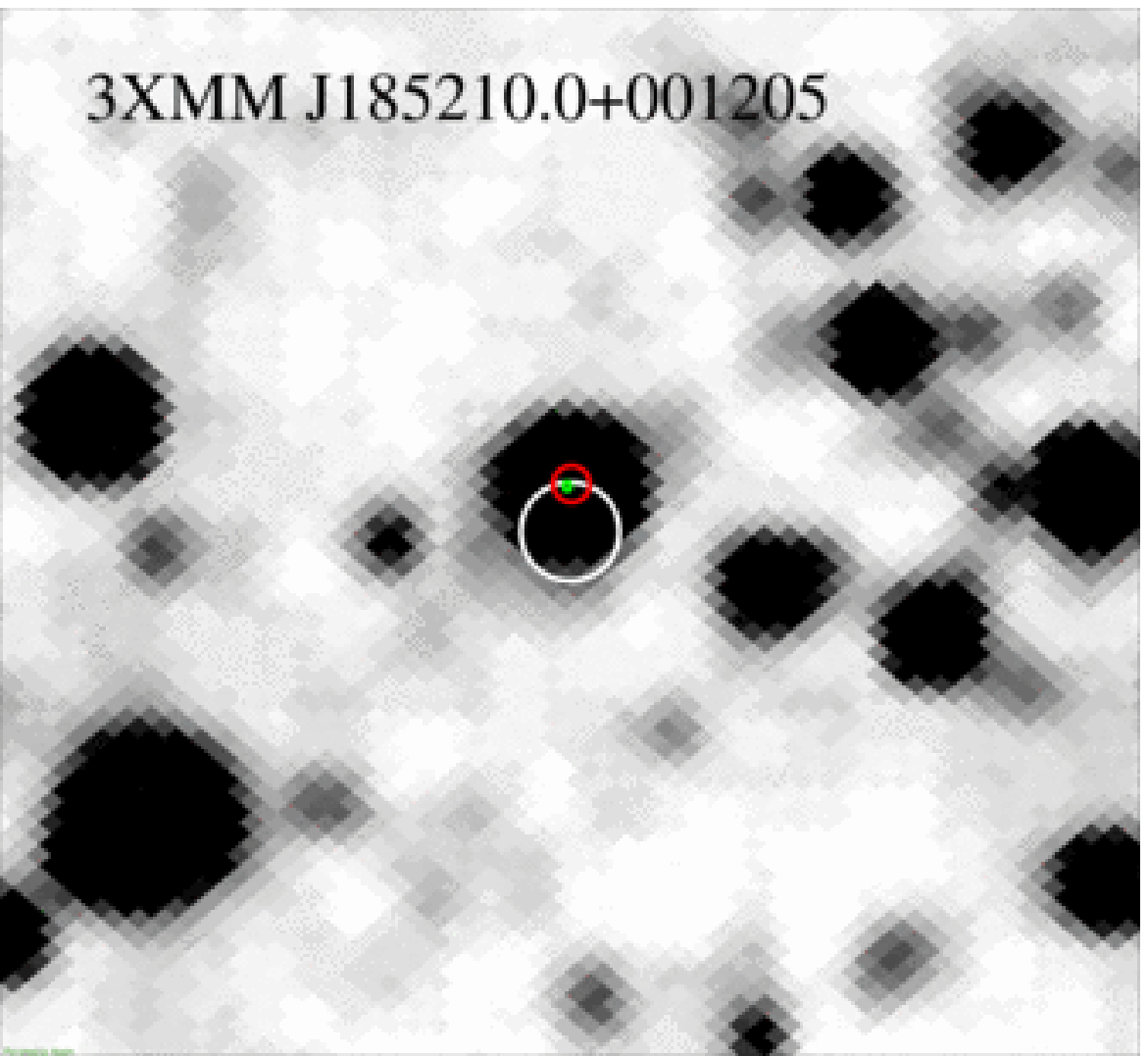}
\includegraphics[width=0.195\linewidth]{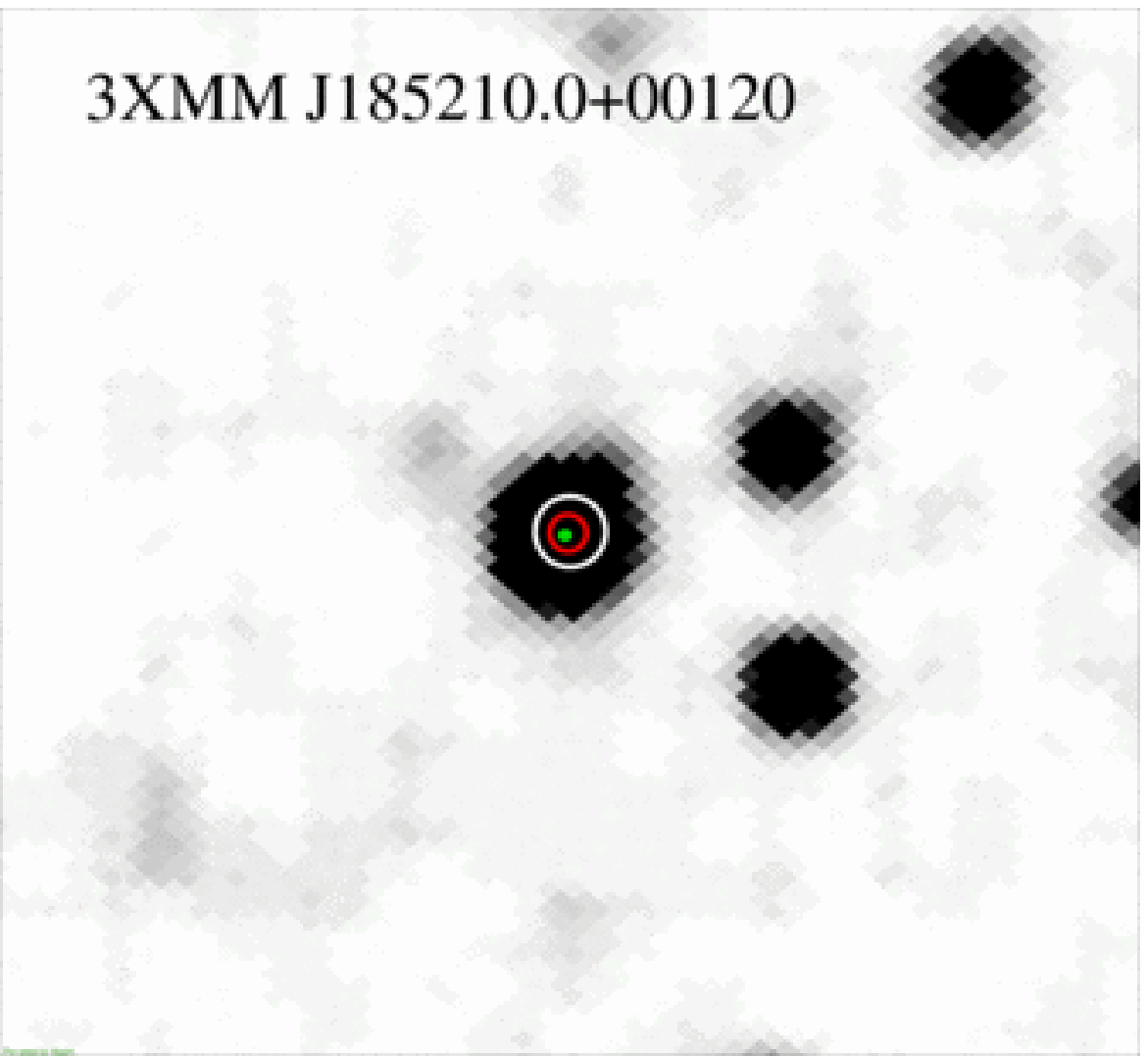}\hfill
\caption{Two arc-minute square 2MASS J-band image centred around the five observed targets at the WHT. Positions and 3$\sigma$ errors are shown with circles, in white for 3XMM, in green for 2MASS and in red for GLIMPSE detections. Following the standard convention, North is up and East is left. \label{g:FC}}
\end{center}
\end{figure*}

As part of a pilot study we were awarded with one night at the William Herschel Telescope (see Section~\ref{sec:telescopes}). We selected targets visible for at least one hour (DEC\,$>$\,--30$^\circ$, 15\,h\,$<$\,RA\,$<$\,20\,h). We prioritised infrared bright sources and went down the list until the end of the observations. We could acquire infrared spectroscopy for the following five sources: \sourceone, \sourcetwo, \sourcethree, \sourcefour\ and \sourcefive. From now on we will refer to these sources as \one, \two, \three, \four\ and \five\ respectively). Finding charts used to identify the infrared target at the telescope are shown in Fig.~\ref{g:FC}. Observations and data reduction are described in the next section.

\section{Infrared spectroscopy}
\label{sec:telescopes}
\begin{figure*}
\includegraphics[width=0.245\linewidth,angle=0]{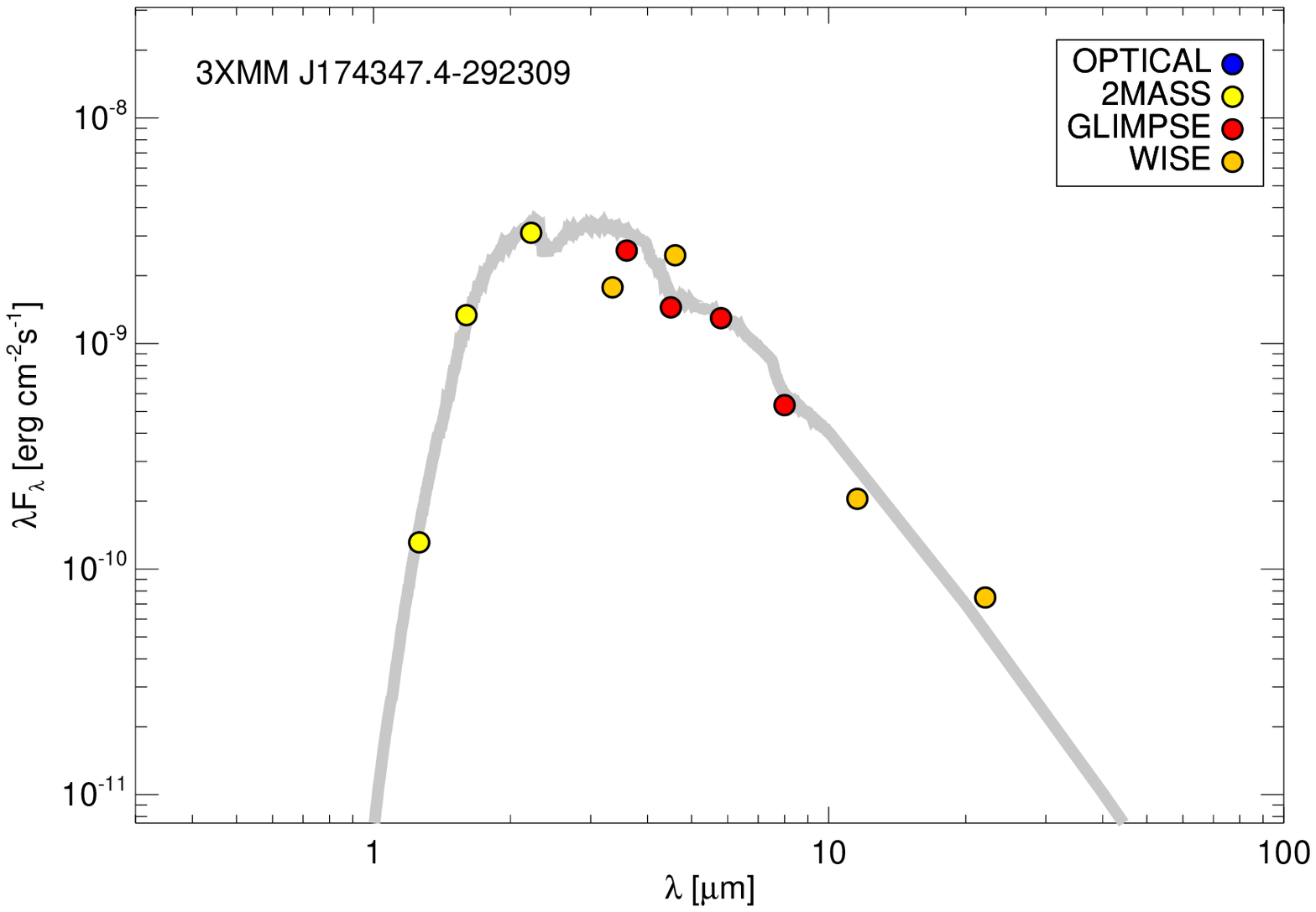}
\includegraphics[width=0.245\linewidth,angle=0]{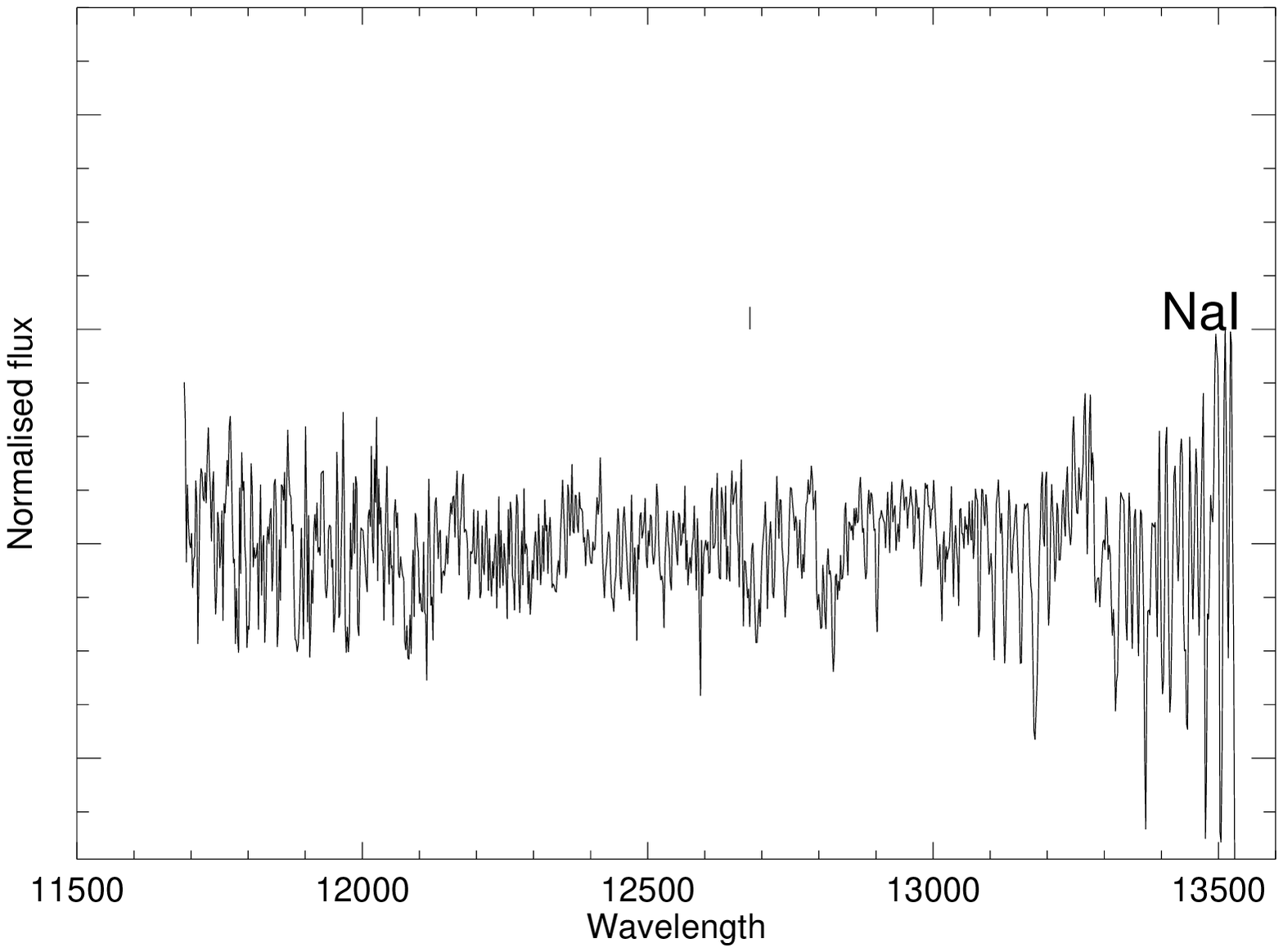}
\includegraphics[width=0.245\linewidth,angle=0]{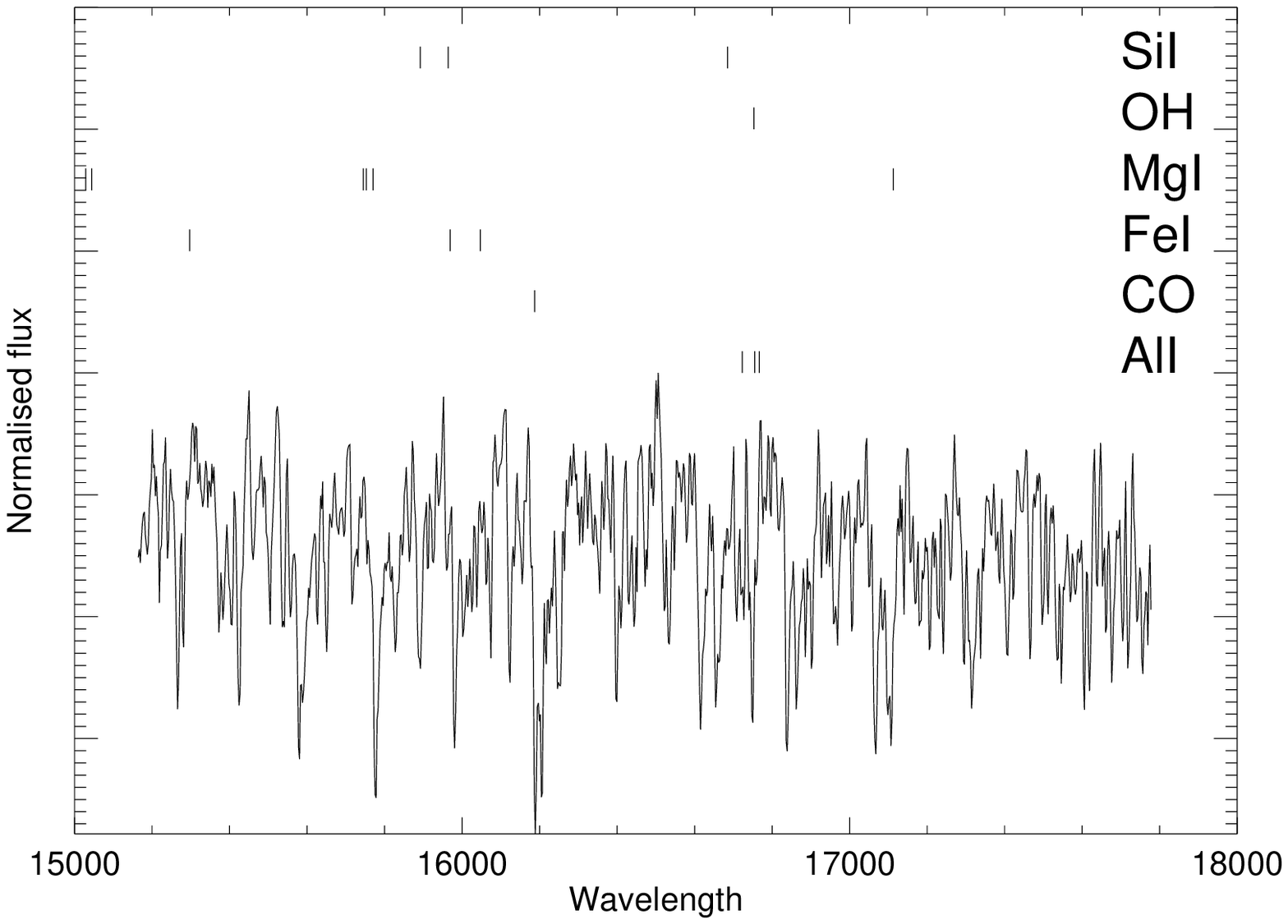}
\includegraphics[width=0.245\linewidth,angle=0]{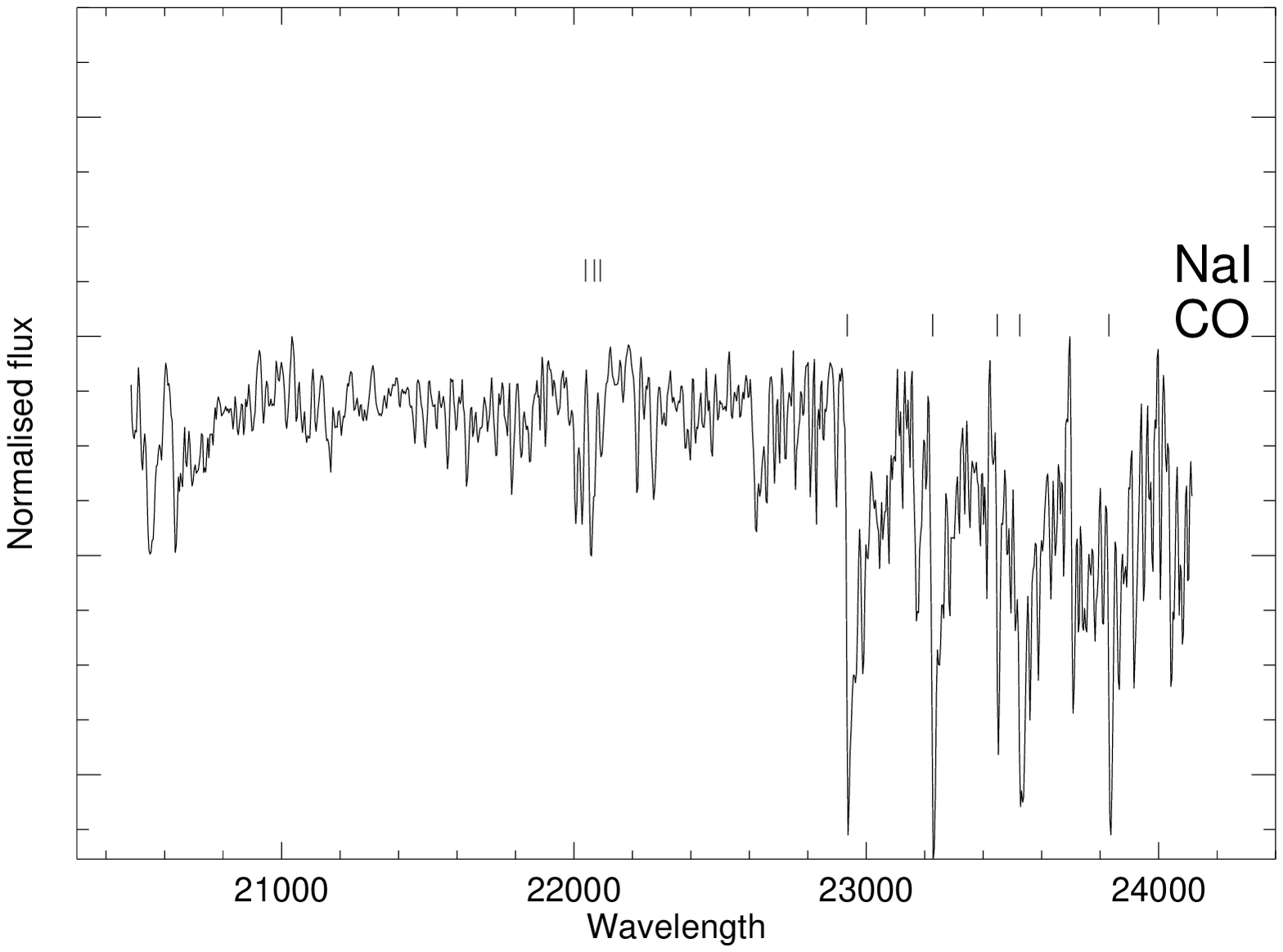}\vfill
\includegraphics[width=0.245\linewidth,angle=0]{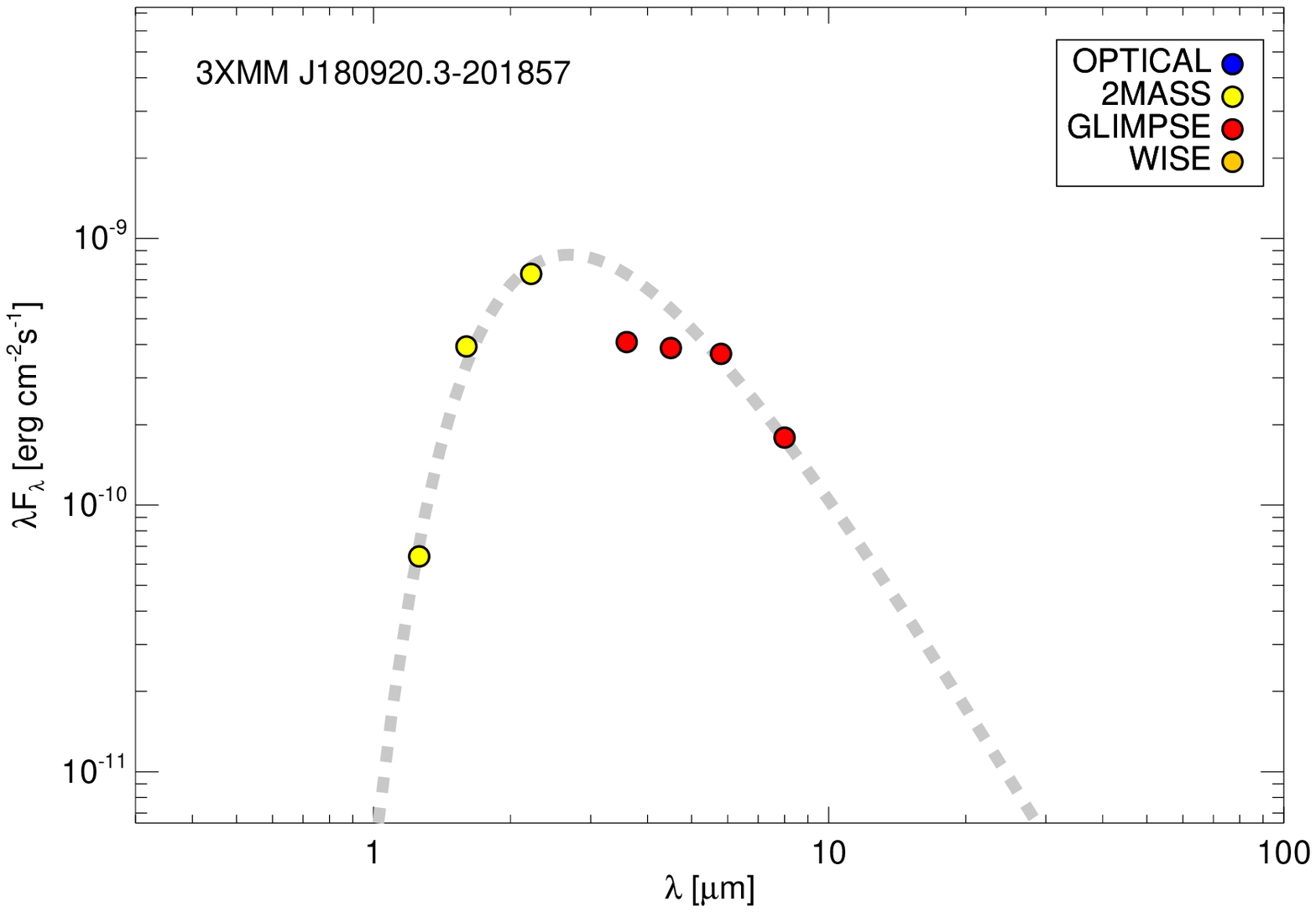}
\includegraphics[width=0.245\linewidth,angle=0]{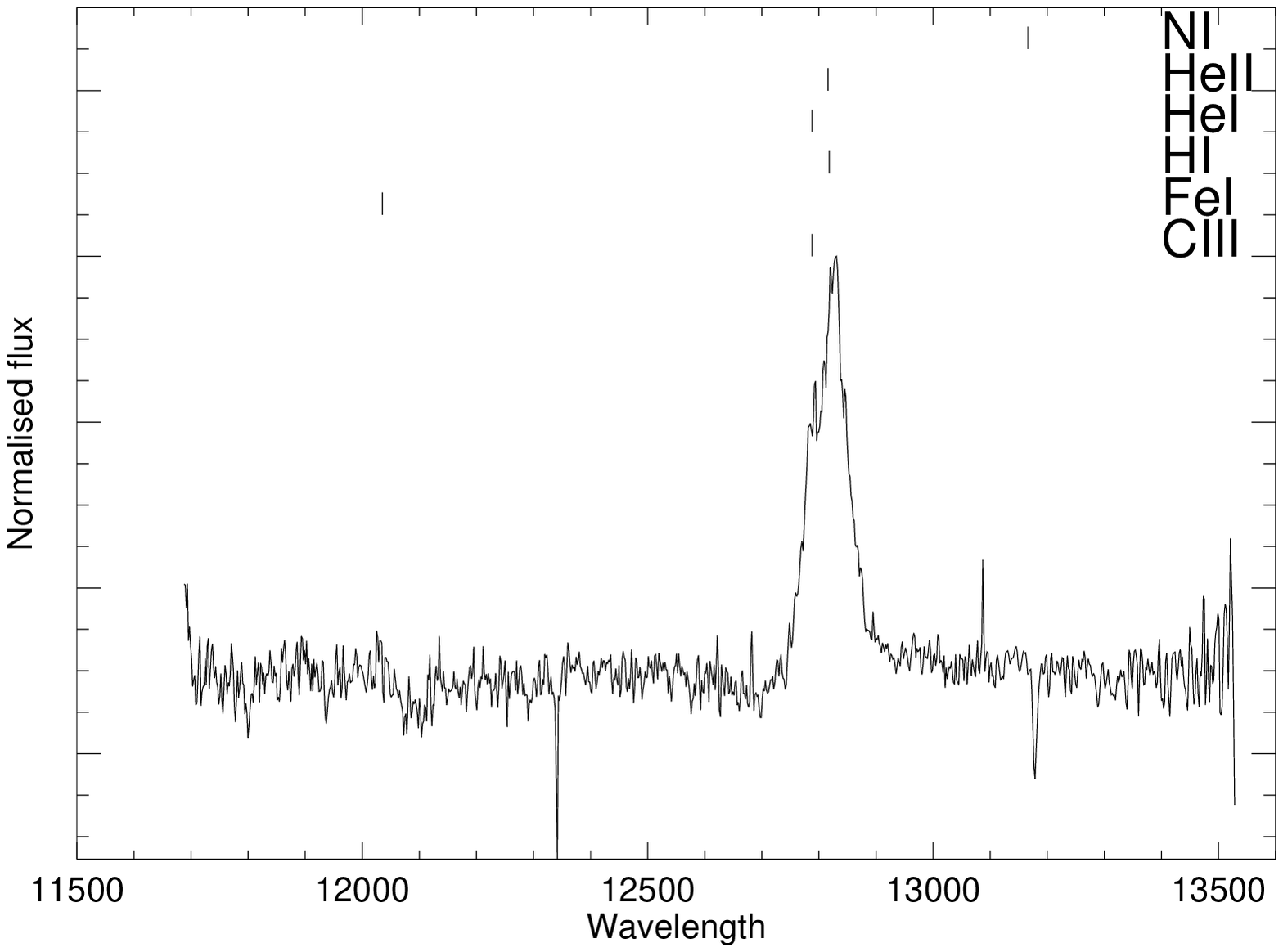}
\includegraphics[width=0.245\linewidth,angle=0]{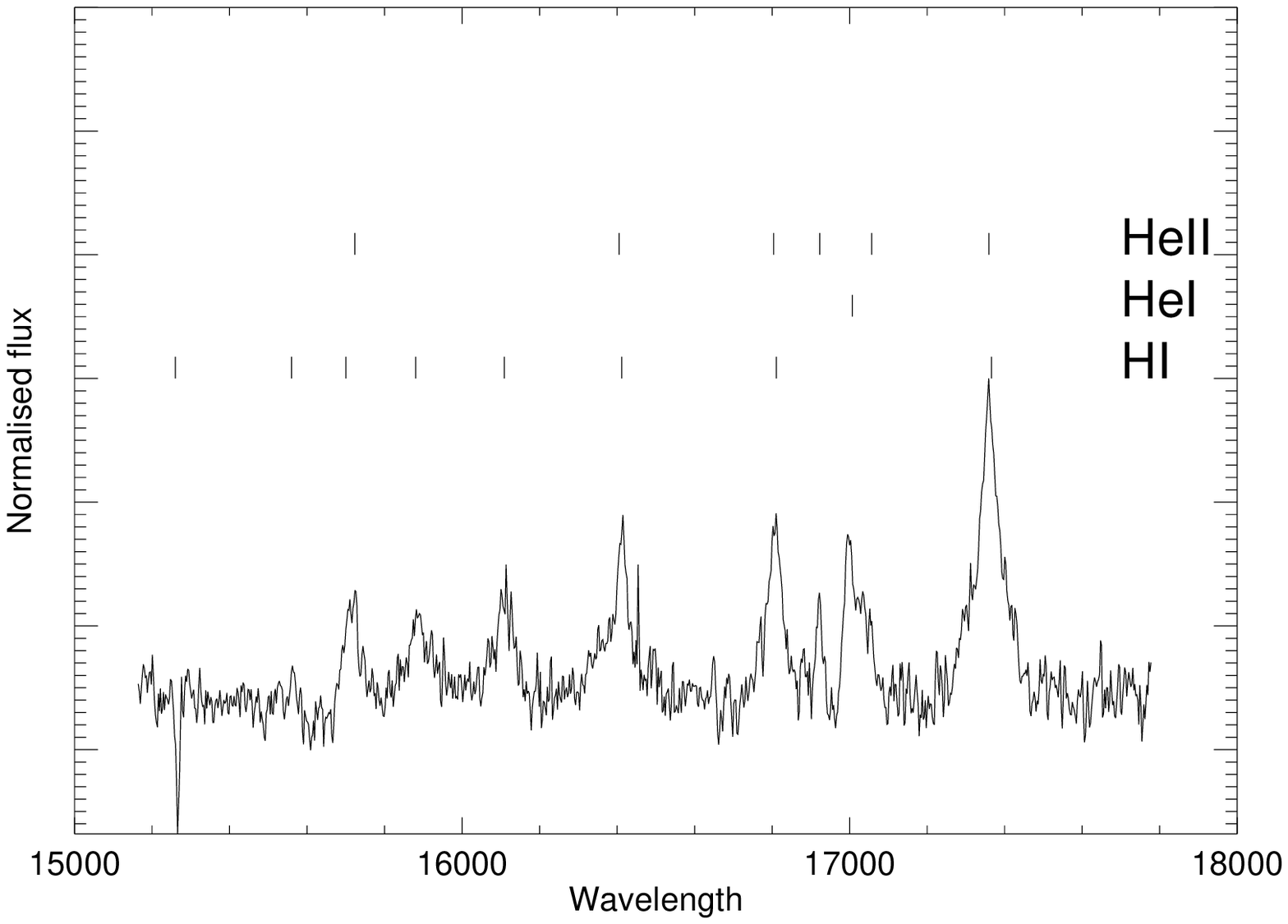}
\includegraphics[width=0.245\linewidth,angle=0]{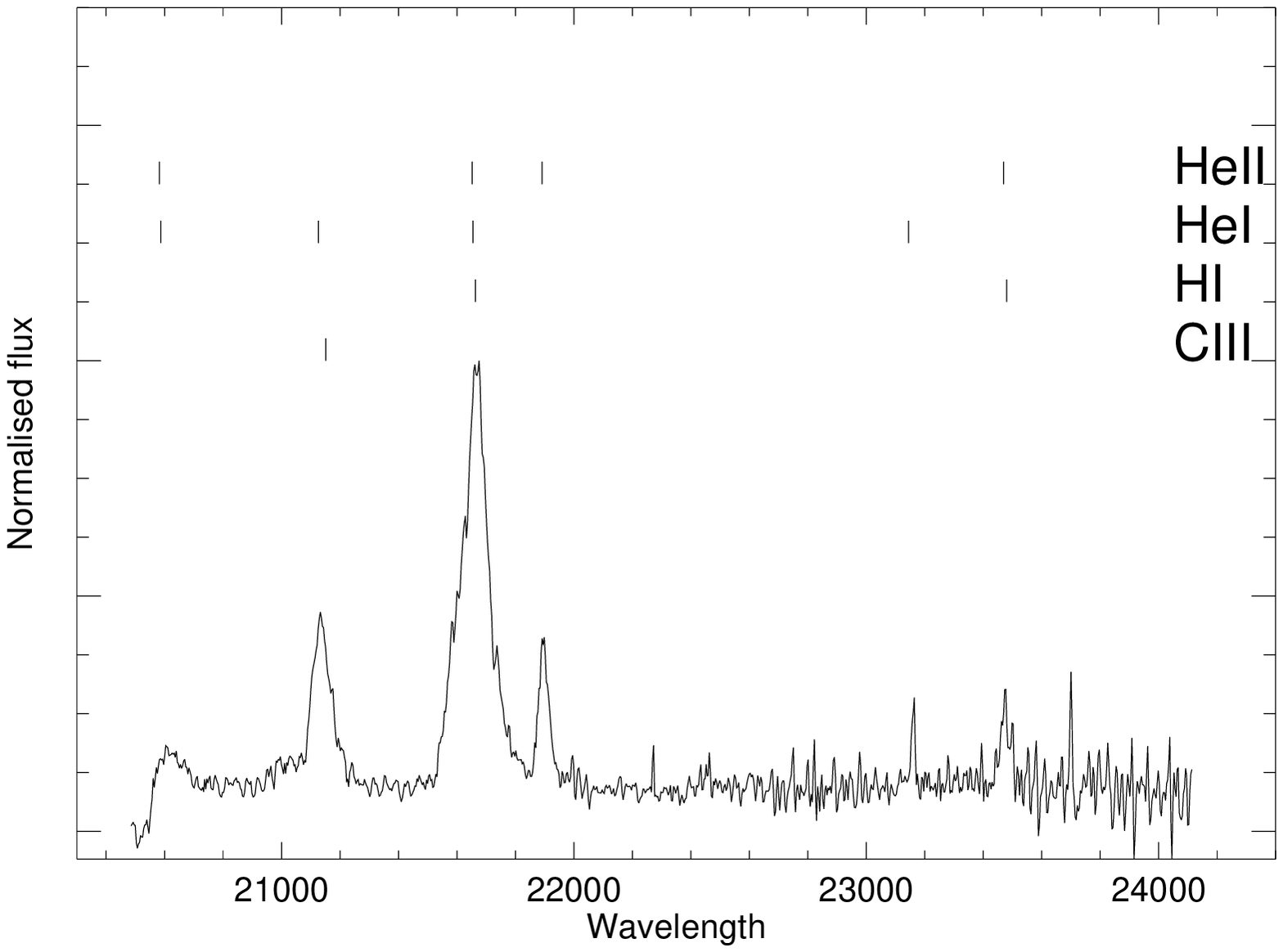}\vfill
\includegraphics[width=0.245\linewidth,angle=0]{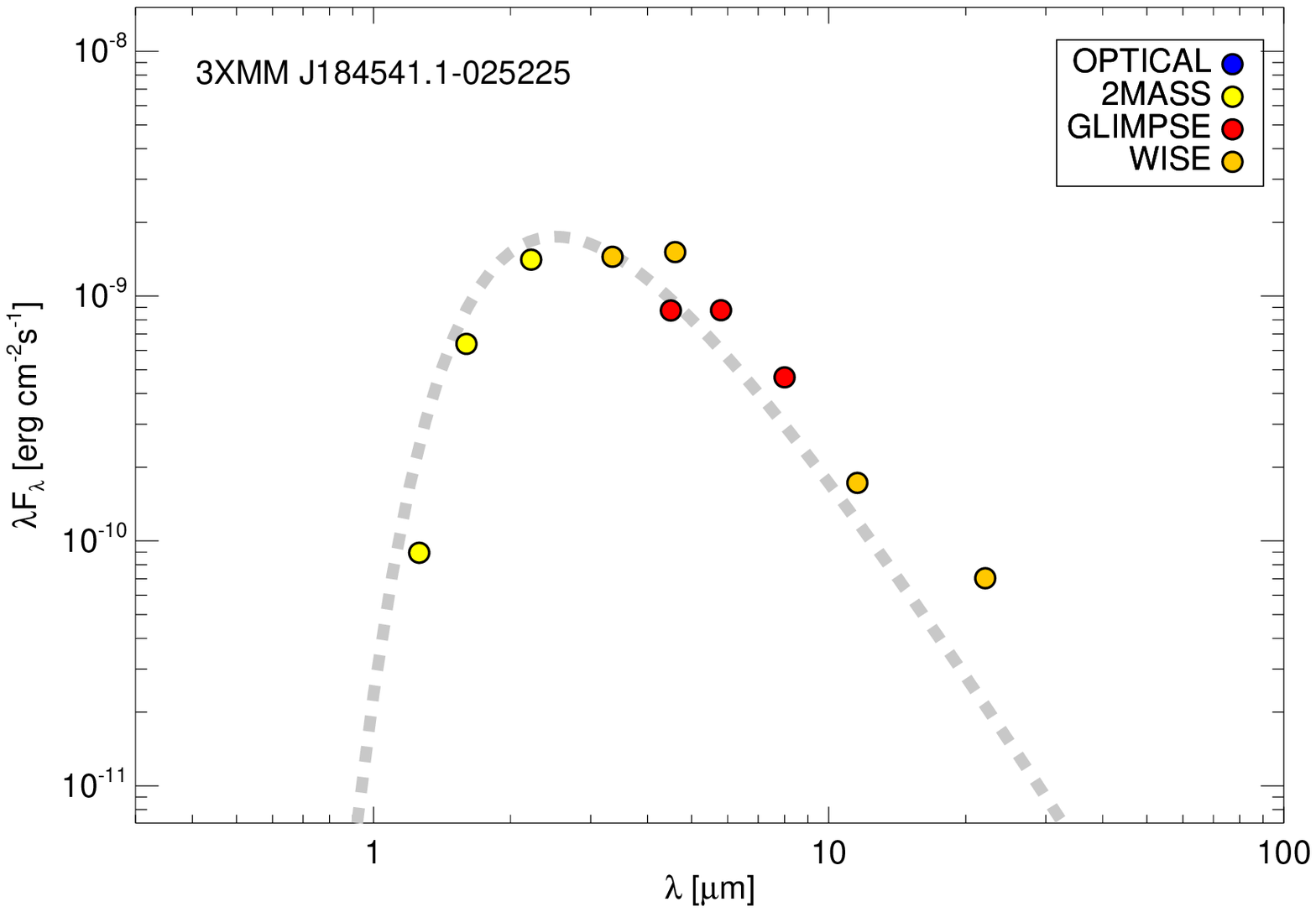}
\includegraphics[width=0.245\linewidth,angle=0]{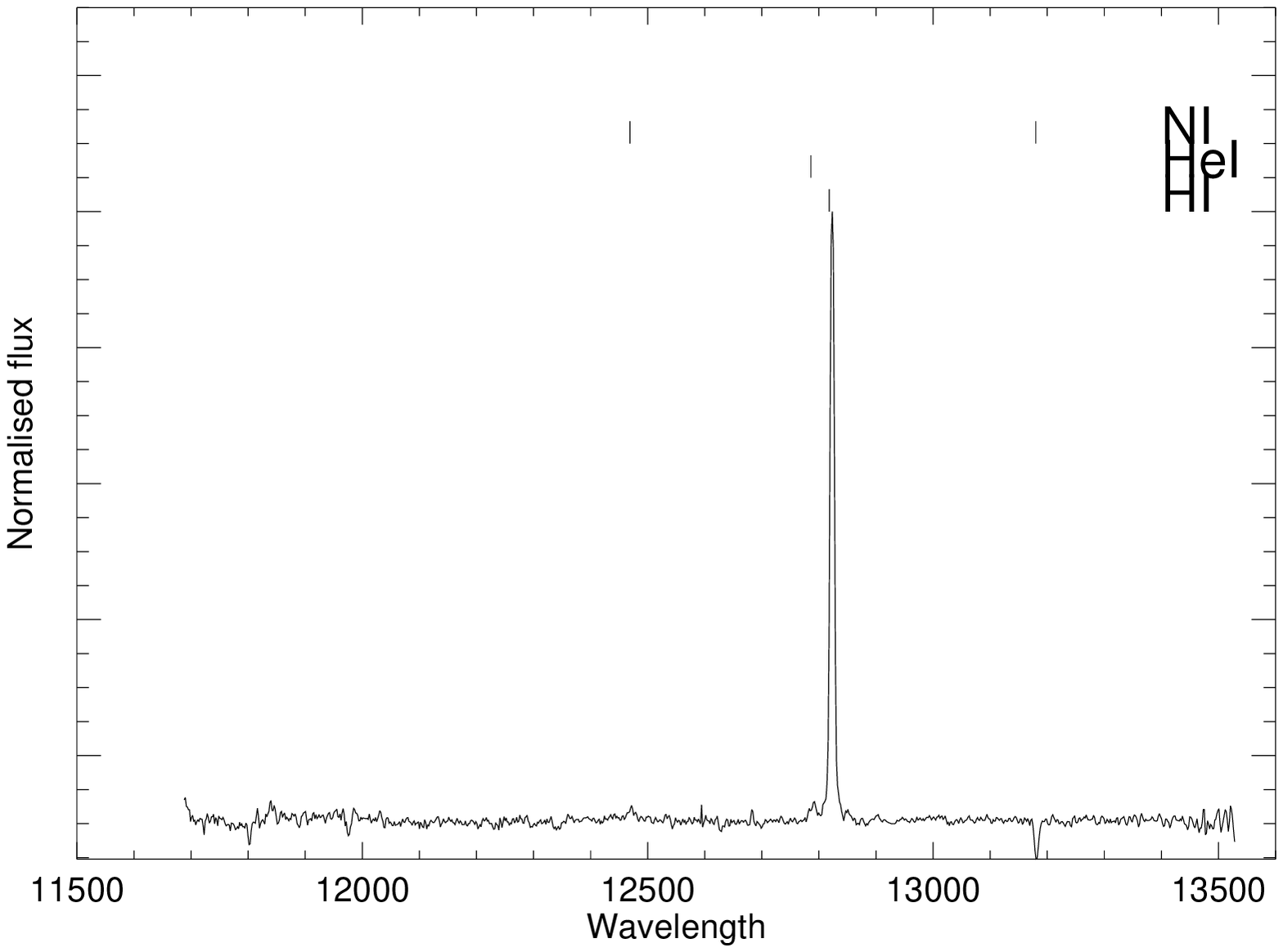}
\includegraphics[width=0.245\linewidth,angle=0]{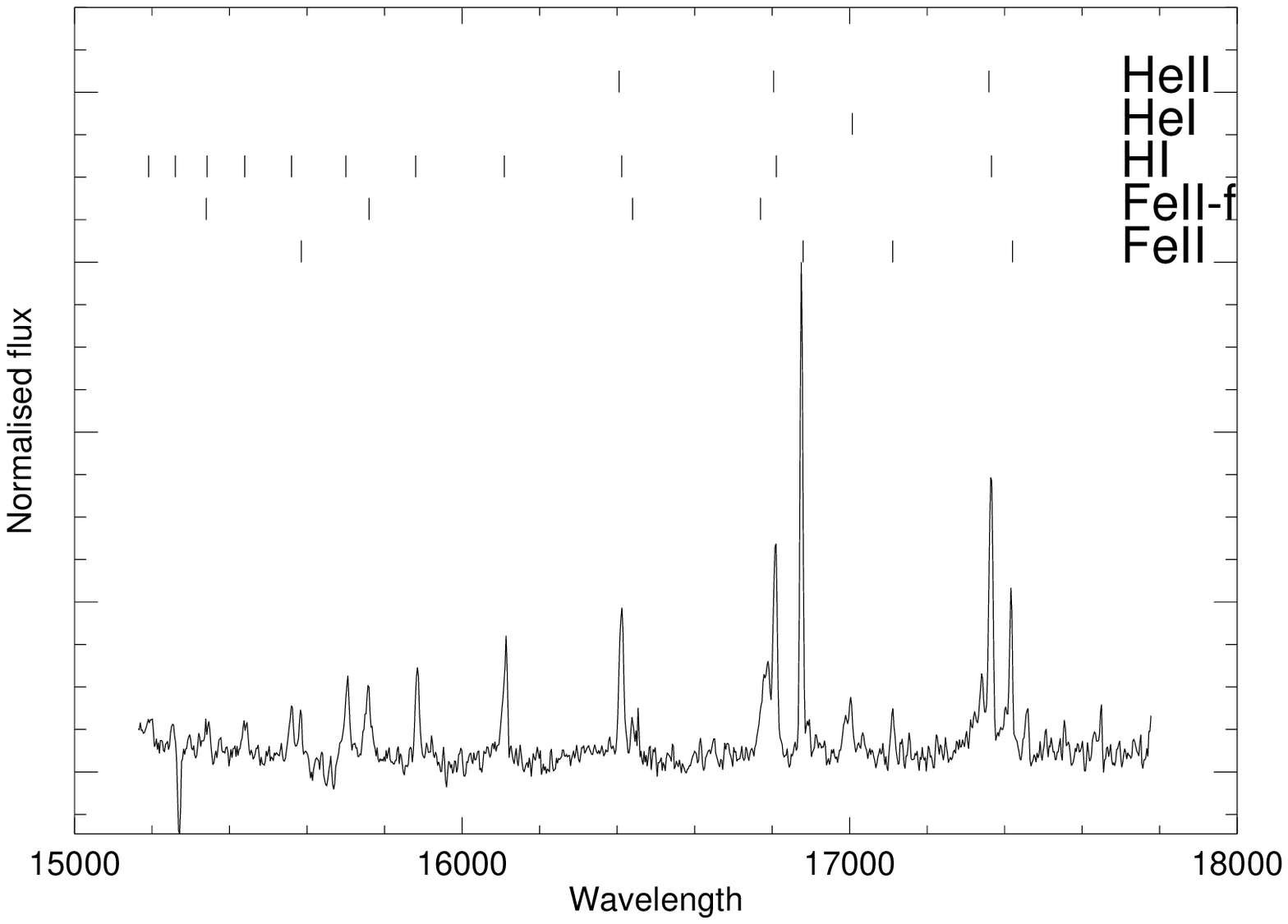}
\includegraphics[width=0.245\linewidth,angle=0]{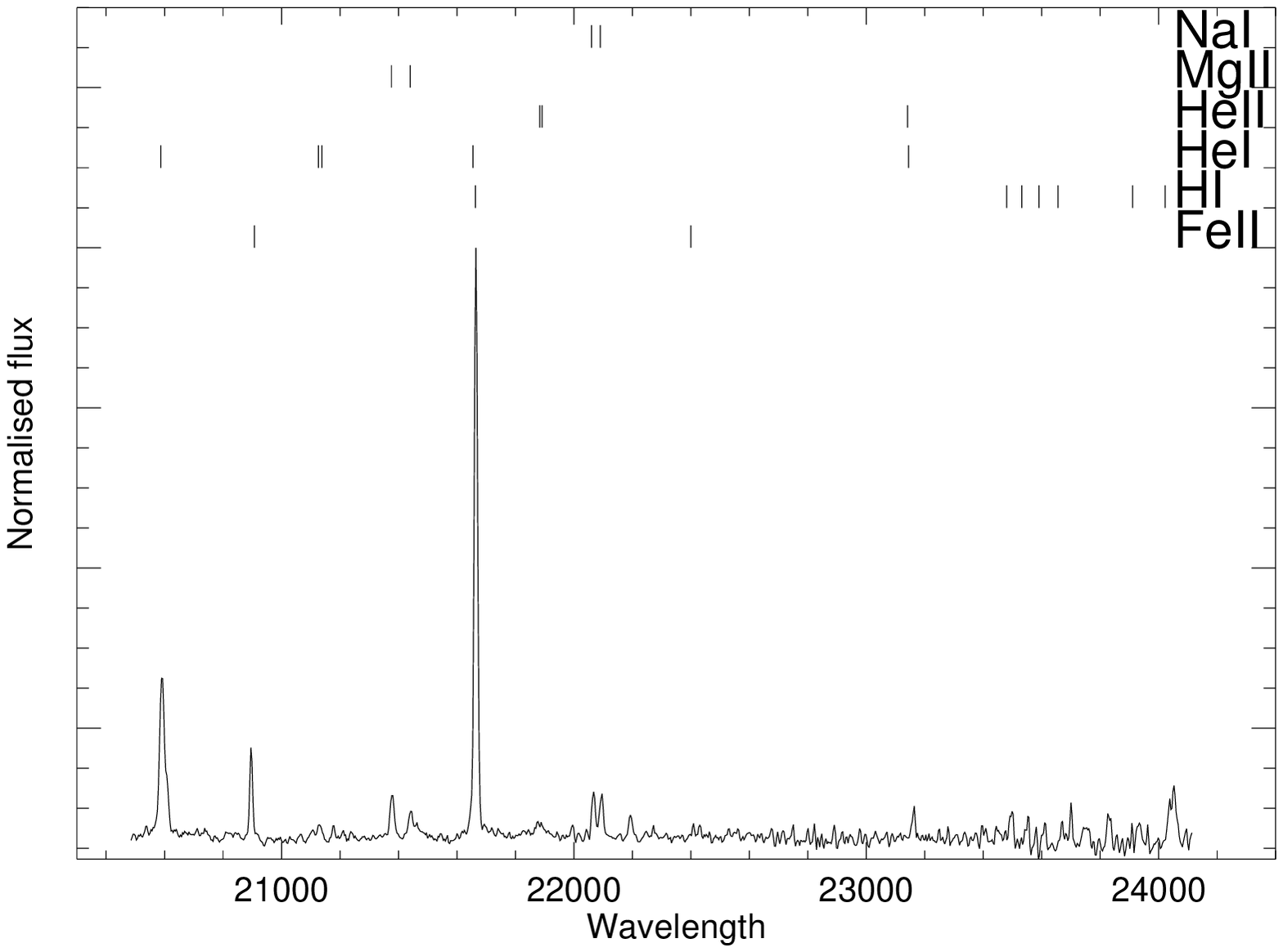}\vfill
\includegraphics[width=0.245\linewidth,angle=0]{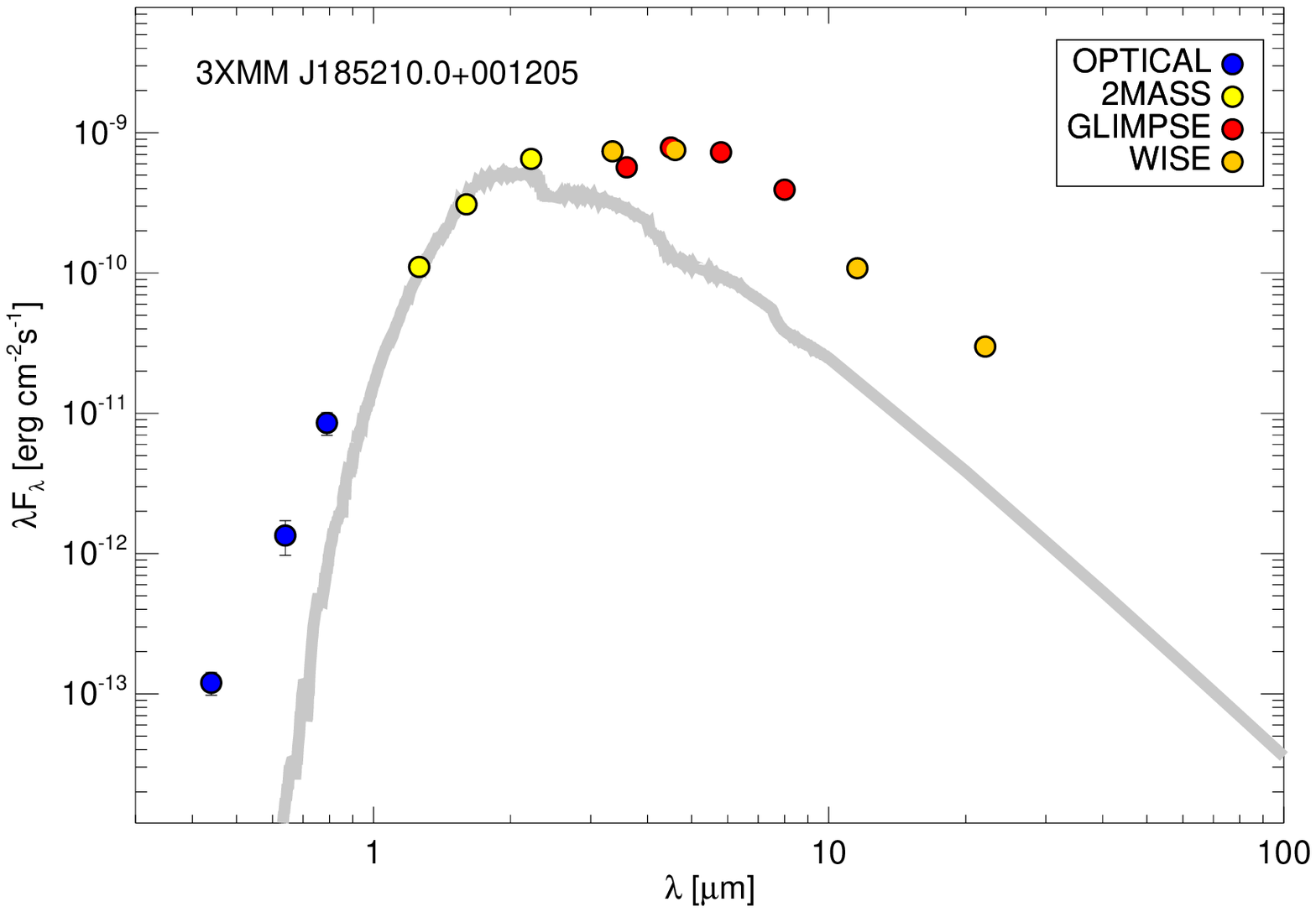}
\includegraphics[width=0.245\linewidth,angle=0]{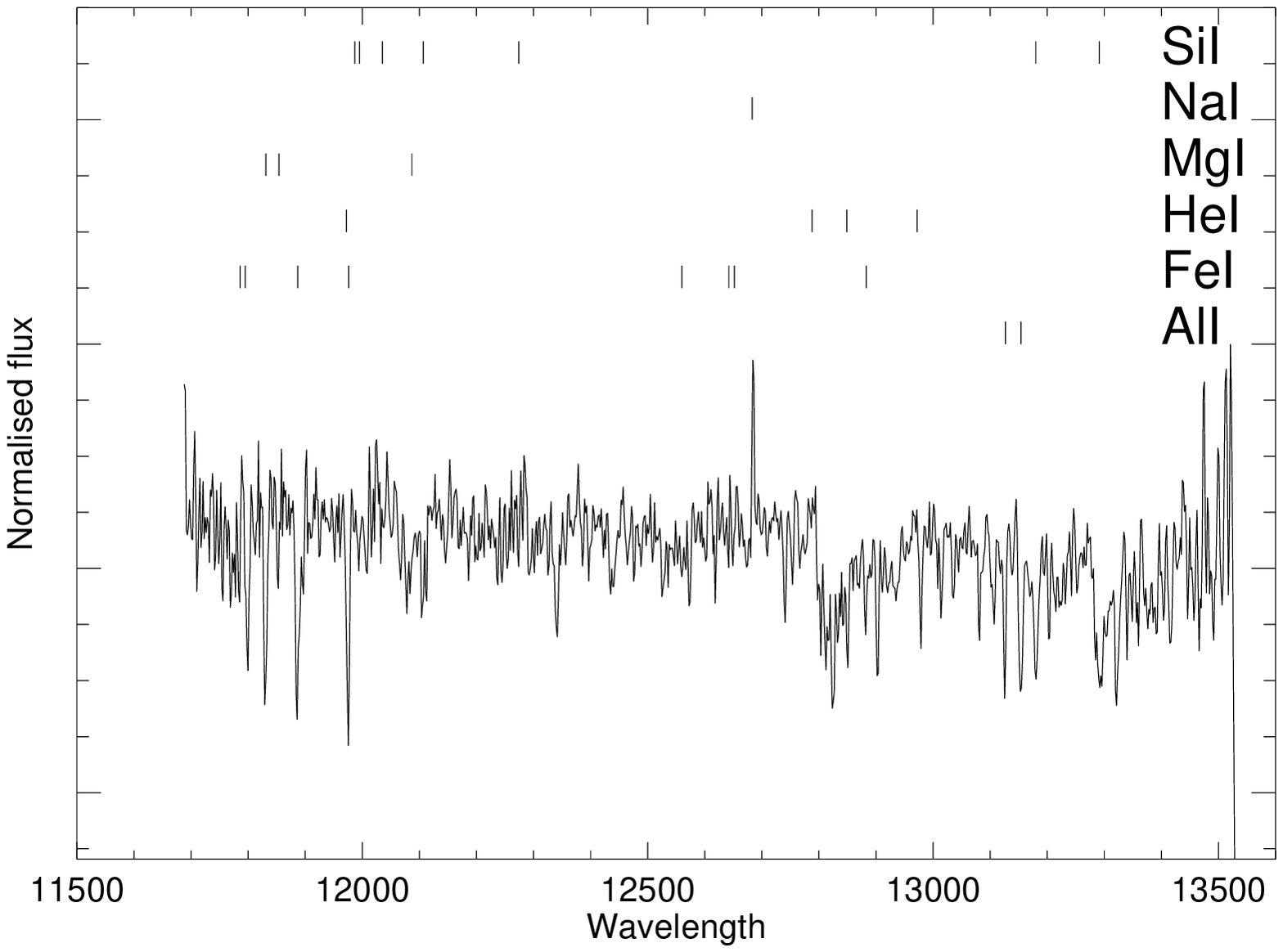}
\includegraphics[width=0.245\linewidth,angle=0]{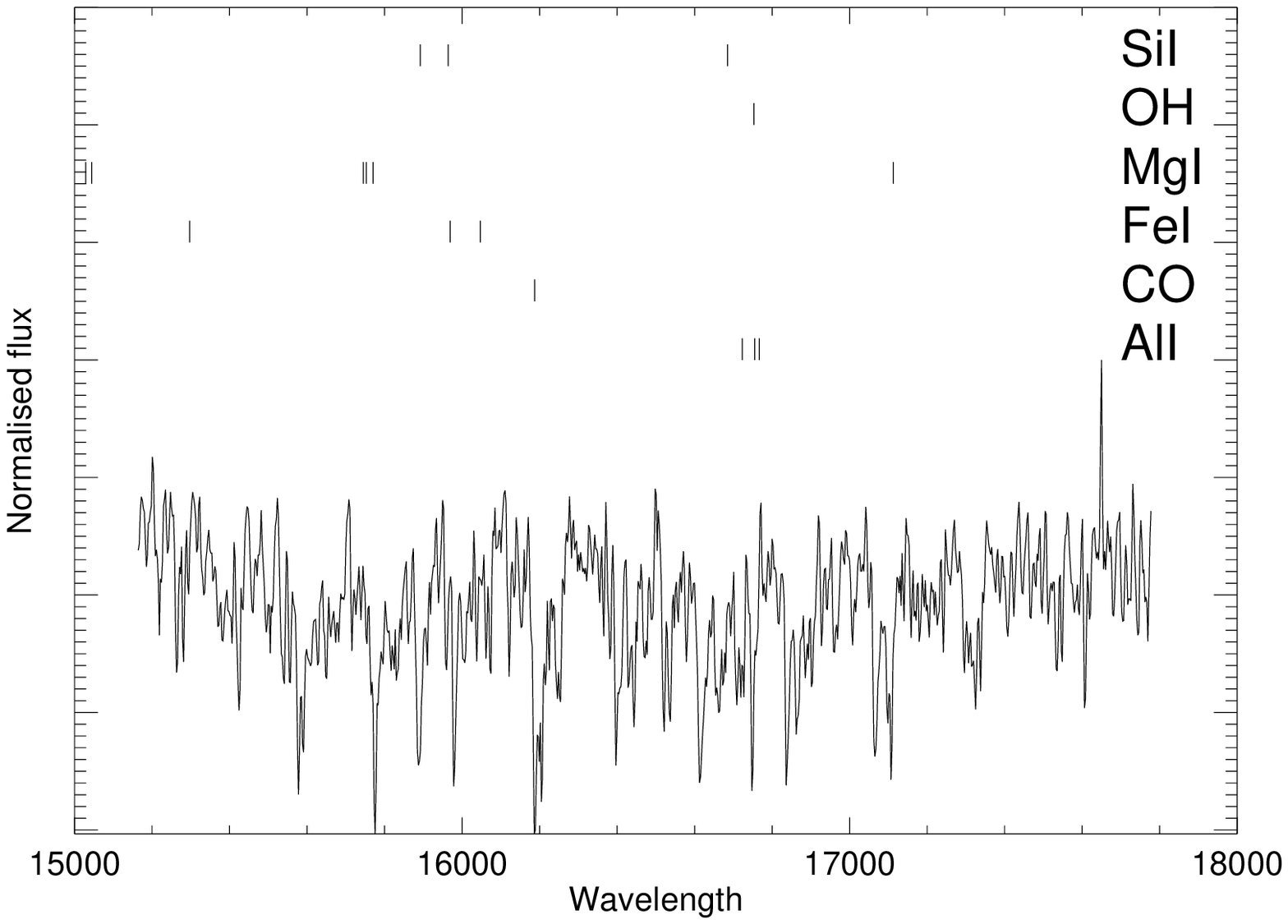}
\includegraphics[width=0.245\linewidth,angle=0]{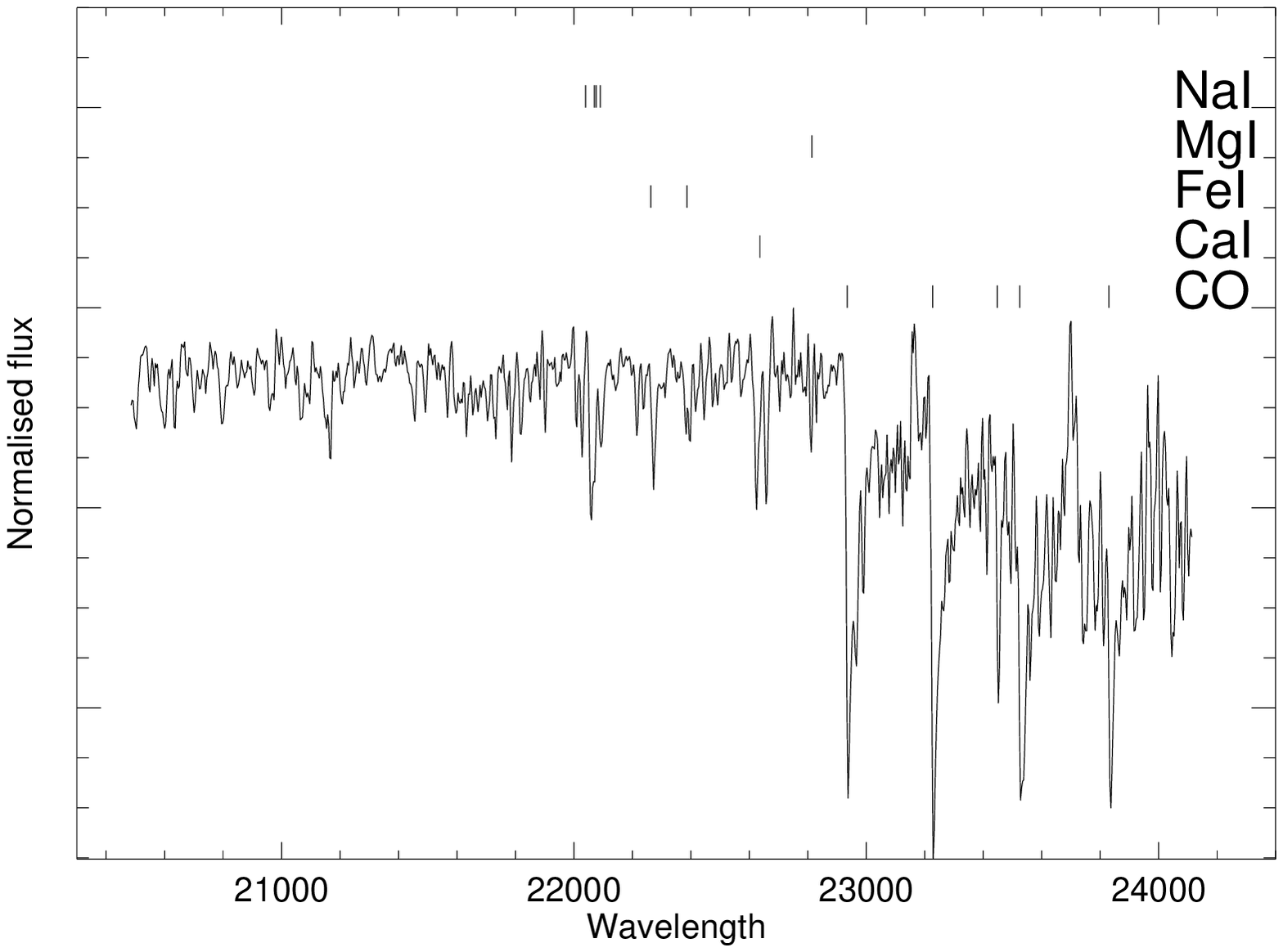}\vfill
\includegraphics[width=0.245\linewidth,angle=0]{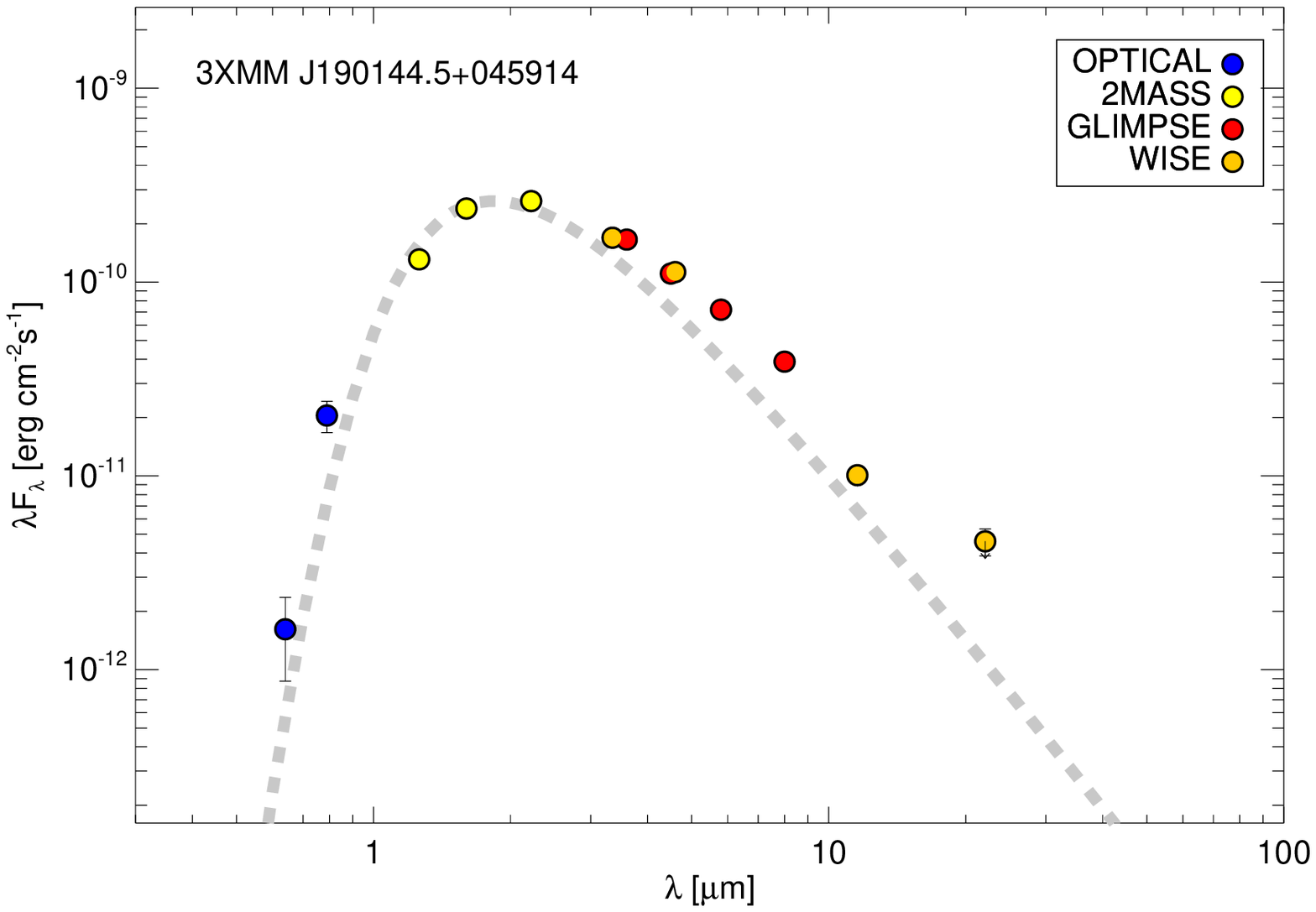}
\includegraphics[width=0.245\linewidth,angle=0]{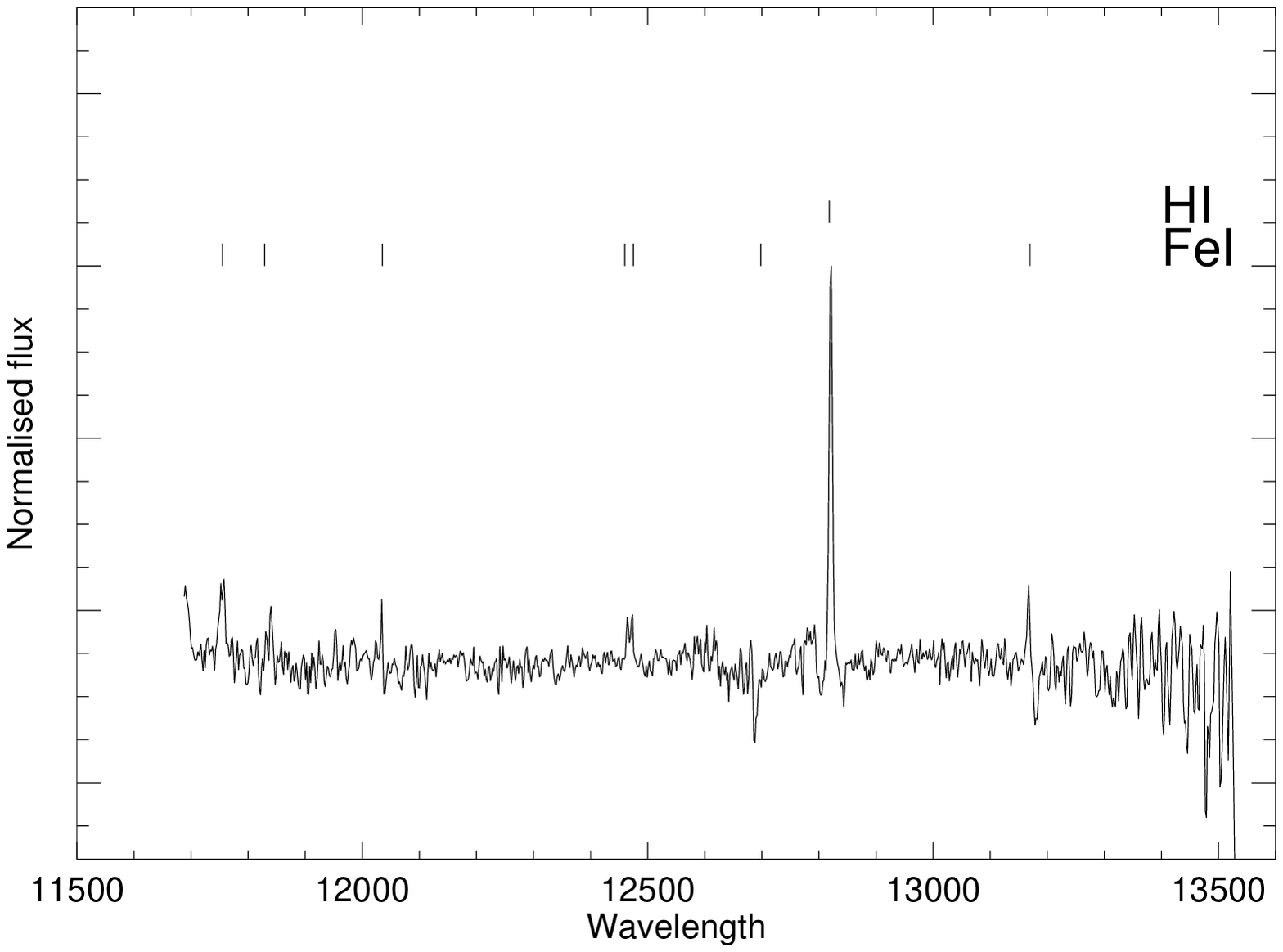}
\includegraphics[width=0.245\linewidth,angle=0]{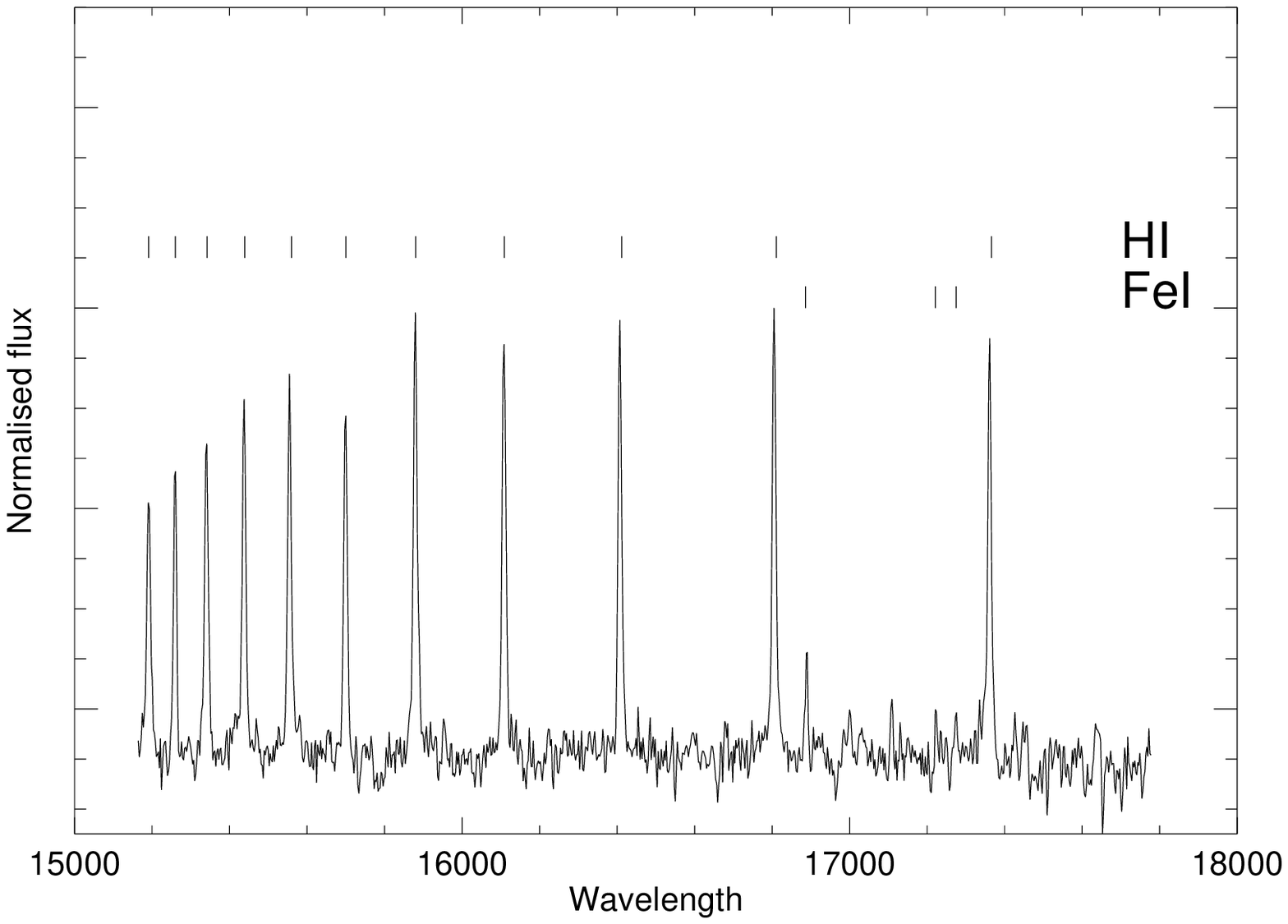}
\includegraphics[width=0.245\linewidth,angle=0]{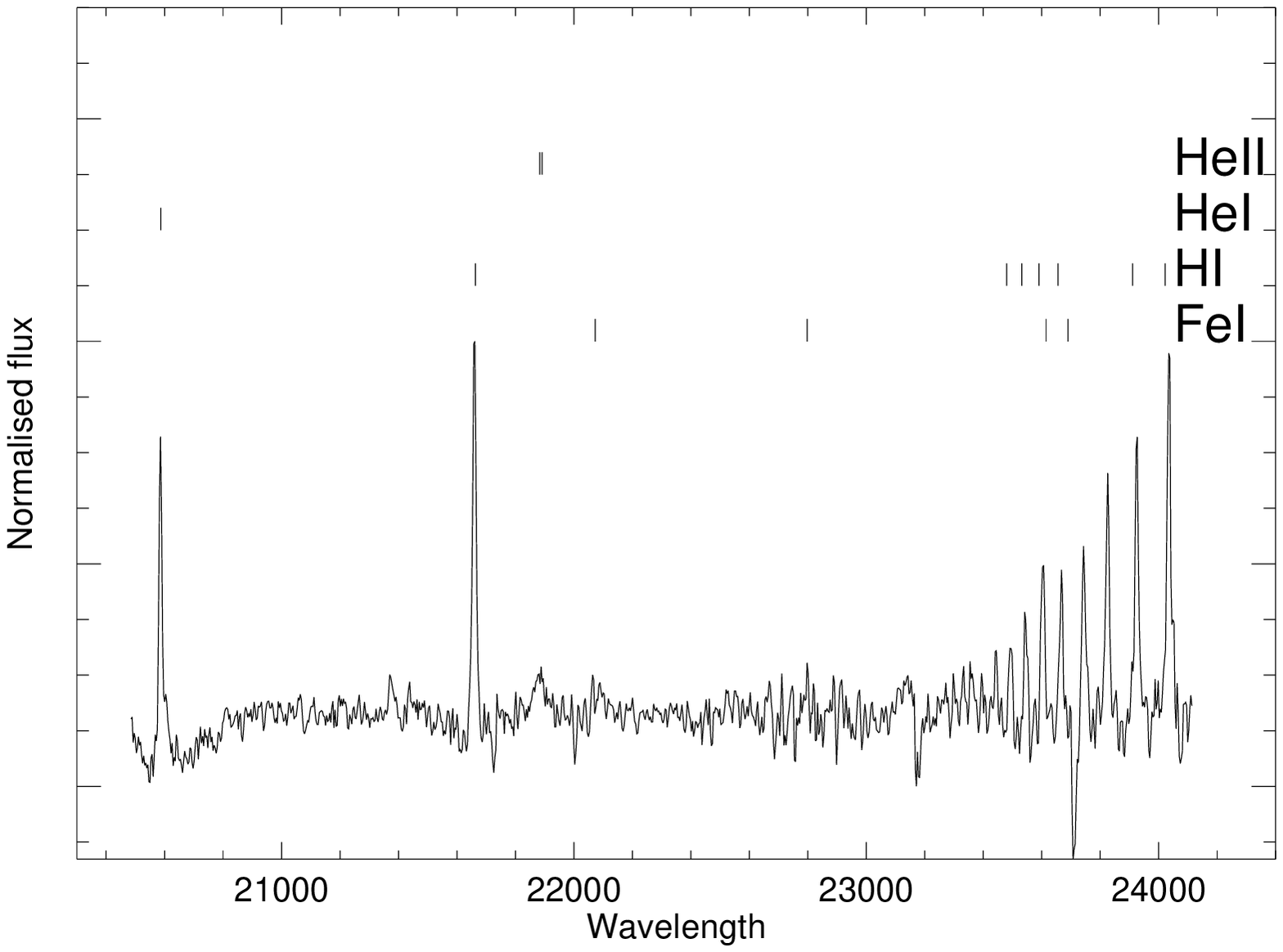}\vfill
\caption{Observed SED (coloured dots) compared to the best fitting absorbed model or Rayleigh Jeans distribution (solid or dashed grey line), followed by the WHT infrared spectra in the J, H and K bands. From top to bottom: \one, \two, \three, \four, and \five.} \label{g:spectra}
\end{figure*}

\emph{WHT/LIRIS.} Infrared spectroscopic observations were carried out at the William Herschel Telescope (WHT) equipped with the near-IR imager/spectrograph LIRIS (Long-slit Intermediate Resolution Infrared Spectrograph) on the 3rd of May 2012 (UT) for five of our sources. We obtained spectra with gratings J, H and K with a slit width of 0.75\arcsec\ (see Table~\ref{t:table3}). We positioned our targets at two nodding points around which we applied small offsets in order to subtract the sky and telescope/mirror emissions. The total exposure times were adjusted to obtain a signal to noise of about 100. We used standard MIDAS procedures to flat-field correct and extract our spectra, using the optimal extraction method described by \cite{horne86-1}. We took Xenon-Argon arc lamp exposures to calibrate in wavelength, and obtained spectra of the standard stars HIP89218 and HIP93118 at the beginning, middle and end of the observations to clean our spectra from telluric lines. Flux calibration was not possible due to a technical problem of the CCD. Spectra are shown in Fig.~\ref{g:spectra}. 

We identified spectral lines in the infrared spectra using a diversity of references \citep{morrisetal96-1,figeretal97-1,hansonetal98-1,foersteretal00-1,clark+steele00-1,steele+clark01-1} and with help of the atomic databases NIST\footnote{\url{http://www.nist.gov/pml/data/asd.cfm}} and \url{http://www.pa.uky.edu/~peter/atomic/}. 
\begin{table}
\begin{center}
\caption{Observational technical details.}\label{t:table3}
\small\addtolength{\tabcolsep}{-2pt}
\begin{tabular}{ccccccc}
\hline
Tel. & Inst. & Grism & Slit      & Spectral    & Spectral   & Date \\
     &       &       & [\arcsec] & range [\AA] & res. [\AA] & [dd/mm/yyyy]     \\
\hline
\noalign{\smallskip}
WHT      & LIRIS & J     & 0.75      & 11690--13530 &    3.6     & 03/05/2012 \\
WHT      & LIRIS & H     & 0.75      & 15160--17790 &    5.2     & 03/05/2012 \\
WHT      & LIRIS & K     & 0.75      & 20490--24110 &    7.2     & 03/05/2012 \\
\hline
\end{tabular}
\end{center}
\end{table}

\section{MAGPIS mid-IR and radio images}
\label{sec:images}
\begin{figure}
\includegraphics[width=\linewidth,angle=0]{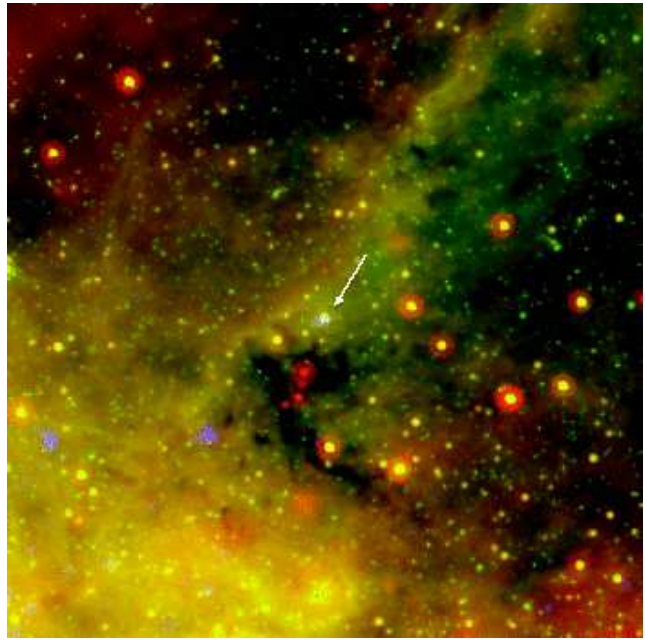}
\caption{Ten arc-minute square RGB image centred around \one\ (red = 24\,$\mu$m, green = 8\,$\mu$m GLIMPSE, blue = EPIC-pn XMM-Newton). North is up and East is left. }\label{g:figure1}
\end{figure}

We looked for possible mid-IR and radio emission of our sources in the Multi-Array Galactic Plane Imaging Survey \citep[MAGPIS,][]{helfandetal06-1} which imaged portions of the first quadrant of the Galaxy at 20\,cm and 90\,cm with the Very Large Array (VLA). We retrieved images from the MAGPIS web site\footnote{http://third.ucllnl.org/gps/}, which also includes 6\,cm VLA, 21\,$\mu$m Midcourse Space Experiment \citep[MSX,][]{eganetal03-1}, and 24\,$\mu$m Spitzer MIPSGAL \citep{careyetal09-1} images. In this way we created composite colour-coded images of 10 arc-minutes around the five targets studied in this paper (see Figs.~\ref{g:figure1}-\ref{g:figure5}). 

\section{Source classification}
\label{sec:SpT}
\begin{table*}
\small\addtolength{\tabcolsep}{-3.5pt}
\begin{center}
\caption{Parameters}\label{t:table4}
\begin{tabular}{ccccccccccccc}
\hline
Name            & SpT         & \Teff       & Rad             & \Av\          &\Nh                           & H     & \Ks   & \Mk\               & d                  & Count rate            &\Lx (0.2-12\,keV)                          &  Nature  \\
3XMM\,J         &             & [K]         & [\Rsun]         &               & [$\times10^{22}$\,cm$^{-2}$]  &       &       &                    & [kpc]              & [\ctss]               & [$10^{32}$\ergs]                           &          \\
\hline
\smallskip
174347.4-292309 &  M2\,I      &  3\,500     & 640             & $\sim$\,26    & 4.6                         & 7.897 & 6.159 & -10.2$^\ddag$              & $>$4.5             & $0.022\,\pm\,0.006$  & $>\,6$          & ?           \\
\smallskip
180920.3-201857 &  WN8        & 40\,000     &  30$\,\pm\,$10  & 23$\,\pm\,$3  & $4.1\,\pm\,0.5$             & 9.225 & 7.719 & -6.79$^\dag$\,--\,$-9.29$  & 1.6-5              & $0.012\,\pm\,0.006$  & 0.4\,--\,4      & WCB         \\
\smallskip
184541.1-025225 &  Ofpe/WN9   & 20\,000     &  79             &  26$\,\pm\,$3 & $4.6\,\pm\,0.5$             & 8.700 & 7.014 & -7.9$\,\pm\,$0.1          & 2.4$\,\pm\,$0.3    & $0.047\,\pm\,0.007$  & 2.3-5.3         & WCB/SFXT     \\
\smallskip
185210.0+001205 &  M2\,I      &  3\,500     & 430             &  $\sim$\,15   & $5.4\,\pm\,0.2$             & 9.487 & 7.852 & -9.06                     & 11                 & $0.009\,\pm\,0.002$  & $<\,15$         & ?            \\
190144.5+045914 &  Be         & --          & --              & 15$\,\pm\,2$  & $2.7\,\pm\,0.2$             & 9.763 & 8.841 & -3.15\,--\,-4.43          & 1.0\,--\,1.9       & $0.033\,\pm\,0.002$  & 0.9\,--\,3.0    & $\gamma$-Cas \\
\hline
\end{tabular}
\end{center}
\begin{justify} 
\textbf{Notes.} We list the 3XMM IAUNAME of the five targets observed at the telescope, the spectral type of the infrared association derived from the spectra taken at the WHT, effective temperature and radius derived from the SED fit, \Av\ estimated from the infrared colour (see text for more details), \Nh\ calculated using the relation from \cite{predehletal95-1}. Observed H and \Ks\ 2MASS magnitudes are followed by the absolute magnitude in the \Ks\ band. We adopted absolute magnitudes from \cite{crowtheretal06-1}$^\dag$ and \cite{coveyetal07-1}$^\ddag$. Distances are calculated from the adopted absolute magnitudes. Finaly we list the count rate and X-ray luminosity in the 0.2--12\,keV band, followed by the likely nature of the X-ray source.
\end{justify}
\end{table*}

In this section we describe the spectral type and stellar parameters for each of the sources observed at the WHT. We estimated the X-ray flux in the 0.2\,--\,12\,keV band assuming a thermal model with kT\,=\,5\,keV (see Section~\ref{sec:cands}) and the Galactic \Nh\ we obtained for each object. All parameters are given in Table~\ref{t:table4} and some important lines with their equivalent width are listed in Table~\ref{t:table5}. 

\subsection{\sourceone\ (\one)}
The infrared spectra of \one\ reveal absorption of many atomic lines typical of late K and early M stars, such as Si\,I, Mg\,I, Fe\,I, K\,I, Al\,I, Na\,I, Ca\,I and the CO band (see Fig.~\ref{g:spectra}). We used the equivalent width (EW) of the CO band-head at 16\,187\,\AA\ and the Fe absorption lines at 22\,263\,\AA\ and 22\,387\,\AA\ as indicators of the temperature and luminosity class of the star. We compared the depth of the CO band head  (EW$_{\mathrm{CO\,16187\,}\AA}\,=\,6\,\pm\,1$\,\AA) and the sum of the EWs of the Fe\,I 22\,263\,\AA\ and the Fe\,I at 22\,387\,\AA\ lines (EW$_{\mathrm{Fe\,22263}\,AA}$+EW$_{\mathrm{Fe\,22387}\AA}\,=\,4\,\pm\,1$\,\AA) with values observed in dwarf, giant and supergiant stars \citep{foersteretal00-1}. These values are consistent with a red giant or a red supergiant with effective temperature close to 3500\,K, corresponding to a spectral type M2\,III-I. The EW of the CO-bandhead at 2.3\,$\mu$m (EW$_{\mathrm{CO\,2.3}\mu\,m}\,=\,41\,\pm\,5$\,\AA) is compatible with a supergiant star \citep{negueruelaetal10-2} although such high values have also been found in some giant stars \citep{messineoetal14-1}.

In order to construct the SED we complemented the 2MASS and GLIMPSE photometric values with information gathered from VizieR. At a distance of 1.6\arcsec\ from \one\ there is a WISE detection \citep{wrightetal10-1}, WISE\,J174347.51-292309.5 \citep{cutrietal12-1}. WISE provides photometry for the whole sky in four bandpasses centered at 3.4, 4.6, 12 and 22\,$\mu$m (W1, W2, W3, and W4 respectively). With magnitudes equal to $5.458\,\pm\,0.038$, $4.118\,\pm\,0.042$, $3.982\,\pm\,0.038$, $2.927\,\pm\,0.075$ in filters W1, W2, W3, and W4 respectively, WISE\,J174347.51-292309.5 W1 and W2 magnitudes are not consistent with GLIMPSE I1 and I2 magnitudes. This discrepancy is likely due to the higher spatial resolution of the GLIMPSE survey compared to WISE, rather than due to photometric variability. Nearby resolved GLIMPSE sources may indeed be blended and contribute to WISE magnitudes. We fitted the SED of \one\ with two models of spectral types M0\,III and M2\,I from \cite{castelli+kurucz04-1}, with (\Teff,$\log(g)$)\,=\,($3800$\,K, $1.34$) and (\Teff,$\log(g)$)\,=\,($3450$\,K, $-0.06$) respectively. The (H-\Ks) colour indicates a large extinction, 22$\,<$\,Av\,$<$\,27. We fixed extinction to \Av\,=\,26 magnitudes, equal to the total interstellar extinction in the area \citep[][]{schultheisetal99-1}. According to the empirical spectral type radius relation from \cite{vanbelleetal99-1}, assuming a giant star, the radius is $40^{+9}_{-7}$\Rsun, for an error in the spectral type determination of $\pm1$ sub-type. From the fitting flux scaling factor we estimated a distance to the star of $310^{+70}_{-50}$\,pc. We are not aware of the existence of such a nearby high extinction region and we thus conclude that the star must be a supergiant. Supergiant stars exhibit a wide range of luminosities \cite[see e.g.][]{meynet+maeder00-1,marigoetal08-1} and as a result a wide range of absolute magnitudes is expected and observed \citep{levesqueetal05-1}. Therefore, no reliable distance may be derived from the photometry. The high extinction observed indicates that the source must be rather distant. According to \cite{marshalletal06-1} the \Ak\ extinction in the line of sight ($l\,II$,$b\,II$)\,=\,($359.5^\circ$,$0^\circ$) reaches a value close to 2.5 at a distance larger than 4.5\,kpc. At a distance of 4.5\,kpc the source absolute magnitude \Mk\ is equal to -10.07, a value consistent with those of red supergiant stars \citep{levesqueetal05-1}. Fixing the distance, from the best SED fit, we obtained a radius of about 640\,\Rsun, a value consistent with known red supergiant stars. A larger distance to the source would imply a brighter absolute magnitude and a larger radius.

\one\ was detected in two different observations with a mean count rate of $0.022\,\pm\,0.006$\,\ctss\ in the 0.2\,--\,12\,keV band. We derived an unabsorbed X-ray luminosity of $\Lx\,>\,6\times\,10^{32}$\,\ergs (\Nh\,=$4.6\times10^{22}$\,cm$^{-2}$). No intrinsic absorption was assumed.

On five occasions, Chandra detected the hard source CXOGC\,174347.4-292309 (RA = 17:43:47.50, DEC = -29:23:09.9, r$_{95\%}$\,=\,0.7\arcsec) \citep{munoetal09-1,mauerhanetal09-1,hongetal09-1} at a distance of 1.7\arcsec\ from the XMM-Newton position. Given the positional errors of \one\ (1.2\arcsec) and of CXOGC\,174347.4-292309, the Chandra and XMM detections are likely to be associated with the same source. Importantly, the improved Chandra position is consistent with the GLIMPSE candidate \citep{munoetal09-1}. Long-term X-ray flux variations were detected between different Chandra observations \citep{munoetal09-1}. The ratio between maximum and minimum observed fluxes is in the range 2.3 to 8.2, and has an associated time span of approximately five years. Maximum flux was detected in September 2006. The two XMM-Newton observations in which the source was detected span less than one year, both of them in 2006 (February and September). No significant flux variation was detected between these two observations nor within the observations. Assuming the X-ray luminosity of \Lx\,=$\,6\,\times\,10^{32}$\,\ergs\ (derived from the XMM-Newton observations) as a maximum value the ratio between the minimum to maximum flux observed by Chandra implies a minimum X-ray luminosity of \Lx$\,\sim\,7.3\,\times\,10^{31}$\,\ergs.

\begin{figure}
\includegraphics[width=\linewidth,angle=0]{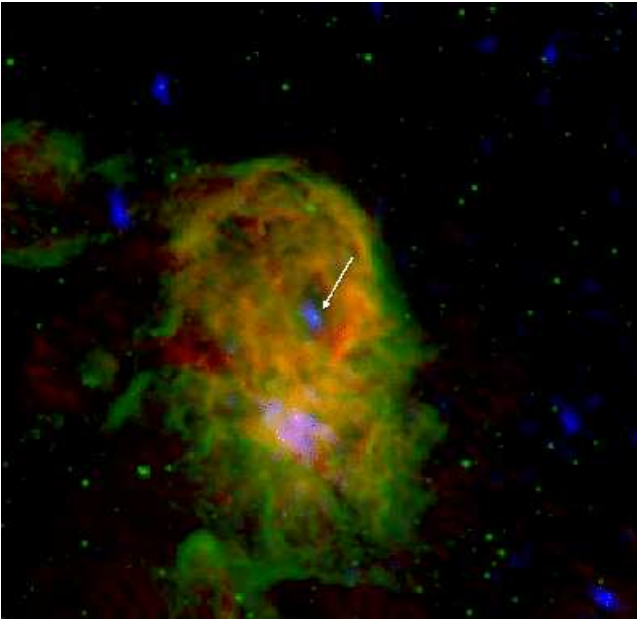}
\caption{Ten arc-minute square RGB image centred around \two\ (red = radio 20-cm, green = I3-GLIMPSE filter, blue = EPIC-pn XMM-Newton). North is up and East is left. }\label{g:figure2}
\end{figure}

\subsection{\sourcetwo\ (\two)}
The WHT J, H and K band spectra of \two\ reveal very broad emission lines, in particular from the complex at 12\,788\,--\,12\,818\,\AA, formed by C\,III, He\,I, He\,II, and H\,I (see Fig.~\ref{g:spectra}), and from the complex at 21\,652\,--\,21\,663\,\AA, formed by He\,II, He\,I and H\,I. The presence of such emission lines is typical of WN stars \citep{sharaetal09-1}. The strong emission  He\,II\,$\lambda$21\,891\,\AA\ line indicates a WN7-8h spectral type. We estimated the ratio of the EW of the 21\,891\,\AA\ He\,II line to the EW of the complex at 21\,663\,\AA\ to be $0.1$, and between the 21\,891\AA\ and the 21\,126\,\AA\ \HeII\ lines to be $0.5$. These values are in agreement with a WN8 spectral type \citep{figeretal97-1,rosslowe+crowther15-1}. 

\begin{table}
\small\addtolength{\tabcolsep}{-4.5pt}
\begin{center}
\caption{Equivalent widths (in \AA) of some important emission lines (in $\mu$m).}\label{t:table5}
\begin{tabular}{cccccccc}
\hline
Name             & He\,II+P$\beta$ & He\,II & He\,I & He\,I & He\,I/NI\,II & He\,II+Br$\gamma$ & He\,II  \\
3XMM\,J          & 1.281           & 1.692  & 1.700 & 2.058 & 2.115        & 2.165             & 2.189    \\
\hline
\smallskip
180920.3-201857  & 71             & 2      & 9    & -     & 19          &  64              & 8      \\
\smallskip
184541.1-025225  & 90             &  -     & 3    & 17   &   -          &  40              & 2      \\
190144.5+045914  & 6              &  -     & 1    &  9   &   -          &  10              & 3      \\
\hline
\end{tabular}
\end{center}
\end{table}

We followed the method in \cite{crowtheretal06-1} to determine the distance to the source. From the 2MASS colours and assuming intrinsic colours of WN7-8h stars from \cite{crowtheretal06-1} we calculated the colour excess E(J\,--\,\Ks) and E(H\,--\,\Ks). We used the extinction relations \Ak\,=$\,0.67^{+0.07}_{-0.06}\times\,$\,E(J\,--\,\Ks) and \Ak\,=$\,1.82^{+0.30}_{-0.23}\times\,$E(H\,--\,\Ks) from \cite{indebetouwetal05-1} to derive a mean value \Ak$\,=\,2.63^{+0.35}_{-0.28}$. Using the extinction relation \Ak$\,=\,0.114\times$\Av\ for \Rv\,$=\,3.1$ from \cite{cardellietal89-1} we estimated \Av$\,=\,23\,\pm\,3$. 
\cite{rosslowe+crowther15-1} revised the infrared absolute magnitudes of the largest sample of WR stars to date. For WN7ha and WN9ha stars they obtained a mean value of \Mk\ of $-7.24$ and $-6.34$ respectively, with a spread of about 0.3 magnitudes at each subtype. From the observed \Ks\ magnitude, the mean \Ak\ and assuming an absolute magnitude \Mk\ equal to the mean value for WN7a and WN9ha stars ($\Mk\,=\,-6.79\,\pm\,0.42$) we estimated a distance to the source of $2.4^{+0.8}_{-0.7}$\,kpc. 

We compared the observed infrared spectra with model spectra from \cite{hammanetal04-1} and found a best match for a WNL with an effective temperature of about 40\,000\,K. Using this \Teff\ and a distance to 2.4\,kpc implies a radius of $30\,\pm\,10$\,\Rsun. Fixing the extinction and distance, no black body can fit the observed SED, even when discarding the GLIMPSE I1 and I2 magnitudes which could be confused with surrounding nebular emission. We therefore left the extinction as a free parameter and obtained a best fit for $\Ak\,\sim\,3.5$ ($\Av\,\sim\,30$). Assuming a mean absolute magnitude for WN7a and WN9ha we obtained a distance $\approx\,1.6\,$kpc, which is an extremely small distance for such a high extinction.

In Figure\,\ref{g:figure2} we show a 10 arc minute composite colour image around \two. The observed infrared nebulosity is likely associated with \two, i.e. the strong wind and ionising photons produced by this hot massive star blow away the surrounding material forming a cavity around the star, and ionising the gas in the nebulae, which is also seen as radio emission (see Figure\,\ref{g:figure2}). Indeed, the X-ray source is in the infrared dark cloud SDC\,G10.156-0.340 \citep[IRDC,][]{peretto+fuller09-1}, embedded in an infrared bubble \citep[N\,1 from ][]{churchwelletal06-1,deharvengetal10-1}. According to \cite{deharvengetal12-1} this bubble is in an H\,II region that forms part of a more complex structure, the star-forming region W\,31 at spectrophotometric distance $3.4\,\pm\,0.3$\,kpc \citep{blumetal01-1}. 
At a distance of about 1.5\arcmin\ from \two\ the source 3XMM\,J180926.9-201930 was detected by the XMM-Newton as an extended X-ray source. This extended source has a very hard spectrum and is likely associated to the OB cluster W31 IR Cluster. 
\cite{steadetal11-1} estimated the cluster to be formed by about 35 candidates with extinction $A_\mathrm{j}\,=\,4.3^{+1.4}_{-0.6}$ ($\Ak\,=\,1.7^{0.6}_{-0.2}$). Although the lower limit of the extinction \Ak\ we obtained for \two\ is consistent with the upper limit of such a value, the distance to the cluster is higher than to \two\ by at least 500 pc. Moreover, in a recent study \cite{sannaetal14-1} find a distance to the cluster of $5.0\,\pm\,0.5$\,kpc based on parallax measurements, a value inconsistent with our distance estimate for \two. It is a bit surprising that \two\ has a higher extinction than W31 while being foreground to W31. Assuming that the source is associated with the cluster, i.e. located at 5\,kpc and is absorbed by  $\Ak\,\sim\,1.7$ implies $\Mk\,\sim\,-8.39$. Under these conditions, no black body can fit the observed SED. Leaving the extinction as a free parameter yields $\Ak\,\sim\,3.5$ and $\Mk\,\sim\,-9.29$, a value higher than observed in typical WRs. 

\two\ was detected in 13 different observations with a mean count rate of $0.012\,\pm\,0.006$\,\ctss\ and with no significant flux variation. Assuming \Ak$\,=\,2.63^{+0.35}_{-0.28}$, corresponding to $\Nh\,\sim\,5.5\,\times10^{22}$\,cm$^{-2}$ \citep{predehletal95-1}, the unabsorbed X-ray luminosity is in the range $4\,\times\,10^{31}\,$--$\,4\,\times\,10^{32}$\,\ergs\ depending on the assumed distance (1.6\,kpc\,--\,5\,kpc) to the source. The bolometric luminosity $\log(\Lbol/\Lsun)$ is in the range of 6.29 to 7.29 (BC$_\mathrm{K}\,=\,-4.2$ \citep{mauerhanetal10-1}) yielding $\log(\Lx/\Lbol)\,=\,-8.2$, a value lower than the mean value for WN stars but within the observed range \citep{mauerhanetal10-1}. 

\begin{figure}
\includegraphics[width=\linewidth,angle=0]{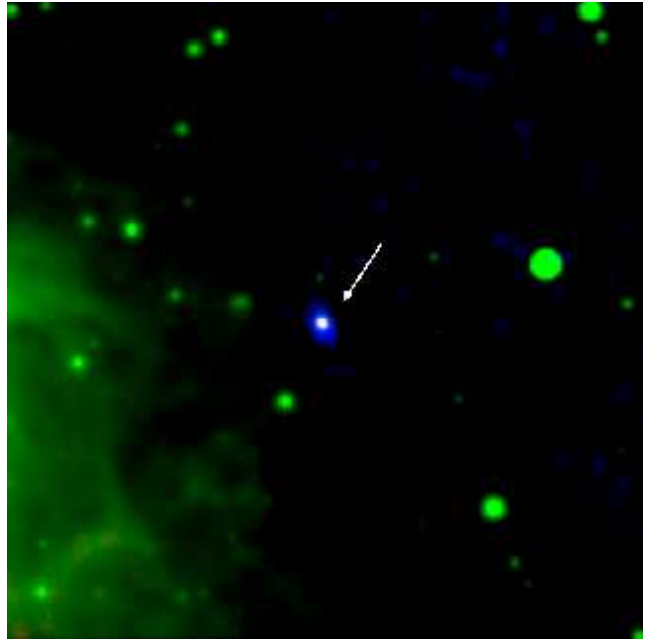}
\caption{Ten arc-minute square RGB image centred around \three\ (red = radio 20\,cm, green = 24\,$\mu$m, blue = EPIC-pn XMM-Newton). North is up and East is left. }\label{g:figure3}
\end{figure}
\subsection{\sourcethree\ (\three)}
The infrared spectra of \three\ reveal strong and narrow emission lines of H\,I, blended with He\,I, and of some ionised species, in particular Fe\,II and Mg\,II. The H band spectra displayed the forbidden line [Fe\,II] at 16\,770\,\AA\ blended with the H\,I (11\,--\,4) line. The absence of the 16\,923\,\AA\ He\,II line implies a spectral type later than O. The spectra also shows Na\,I 22\,060\,\AA\ and 22\,090\,\AA\ in emission. These lines are typical of Ofpe/WN9 and early Be and B[e] stars \citep{morrisetal96-1,hansonetal98-1}. Whether the source is a luminous blue variable star (LBV), i.e. a source displaying spectroscopic variations between both spectral types, is an open question, and more observations are required to elucidate the nature of the source. 

The 90\% error circle of \three\ also contains WISE\,J18454111-025225.5 which is probably associated with the GLIMPSE identification. In Figure~\ref{g:spectra} we present the observed SED. We determined the mean extinction following the same procedure as for \two. The mean \Ak\ value is $3.0^{+0.4}_{-0.3}$. We assumed an effective temperature of 20\,000\,K and a radius of 79\,\Rsun, corresponding to the Ofpe/WN9 star CXOGC\,J174516.1-284909 \citep{mauerhanetal10-1} and fitted a black body curve. From the best fit we obtained a distance to the source of $2.4\,\pm\,0.3$\,kpc, implying an absolute magnitude \Mk$\,=\,-7.9\,\pm\,0.1$. \cite{negueruelaetal11-1} studied in detail several nearby sight-lines. These areas showed moderate extinction up to distances of 3 to 3.5\,kpc and with an abrupt increase between 6 to 7\,kpc ($\Ak\,>\,1.5$) due to dark clouds associated to the Scutum arm. At the calculated \Ak\, the source would be even more distant. Assuming the source is at about 2.4\,kpc it would be in the inter-arm region. Nevertheless, it could be that the source has some local obscuration and or emission, especially since the source has [W1]-[W4]\,=\,2.6, a colour atypical of interstellar absorption.

About 3.6\arcsec\ from \three\ there is the radio emitter GPS5\,029.718-0.031, detected in the in the 5\,GHz VLA survey of the galactic plane \citep{beckeretal94-1}. The radio position and flux of GPS5\,029.718-0.031 is consistent with those of the compact radio source [WBH2005]\,29.719-0.031 \citep{whiteetal05-1}, about 1\arcsec away from the X-ray position but within the combined position errors. The GLIMPSE source is probably associated to the radio source GPS5\,029.718-0.031 \citep{hoareetal12-1}. The radio emission might be ascribed to a massive young stellar object (MYSO) associated to collimated jets and winds, or it could originate in a new born H\,II region. The detected radio emission at 6 and 21\,cm \citep{beckeretal94-1} is consistent with a flat spectral index $\alpha\,\sim\,0$ ($S_\nu\,\propto\,\nu^\alpha$), indicating a non-thermal emission, and excluding a possible MYSO nature of the source, objects with a typical $\alpha$ value $\sim\,+0.6$. X-ray emission associated to high-mass stars has been suggested as an ionising source of UCH\,II regions \citep{hofneretal02-1}. We conclude that the radio emission detected in the combined X-ray plus radio 90\% confidence error circle of \three\ is associated to a radio loud ultra compact H\,II region (UCH\,II).  This ultra-compact H\,II region, detected at radio wavelengths could explain the observed [W1]-[W4] colours. We adopted a distance to the source of 2.4\,kpc.

The three X-ray detections of \three\ have a mean count rate equal to $0.047\,\pm\,0.007$\,\ctss\ in the 0.2\,--\,12\,keV band, implying an unabsorbed X-ray luminosity $\Lx\,=\,3.7^{+1.6}_{-1.4}\,\times\,10^{32}$\,\ergs. We assumed the bolometric correction from \cite{mauerhanetal10-1} derived for the Ofpe/WN9 star CXOGC\,J174516.1-284909 and obtained $\log\Lbol\,=\,6.21$\,\Lsun, and thus $\log(\Lx/\Lbol)\,=\,-7.2$, value consistent with the canonical value of $-7$.

\begin{figure}
\includegraphics[width=\linewidth,angle=0]{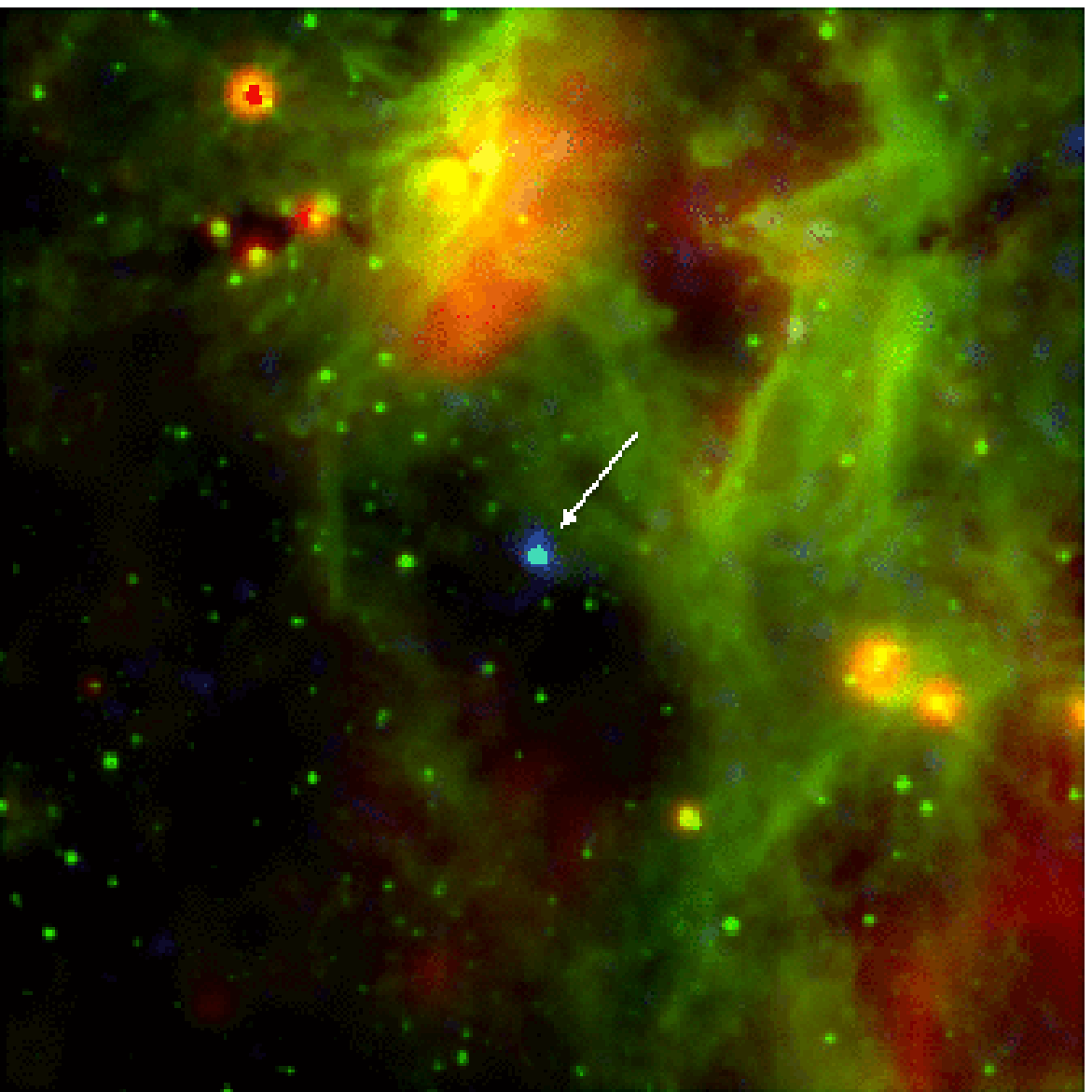}
\caption{Ten arc-minute square RGB image centred around \four\ (red = MIPSGAL 24\,$\mu$m, green = GLIMPSE 8.0\,$\mu$m, blue = EPIC-pn XMM-Newton). North is up and East is left. }\label{g:figure4}
\end{figure}
\subsection{\sourcefour\ (\four)}
As for \one, the infrared spectra of \four\ reveal absorption lines of many atomic species typical of late K and early M stars (Si\,I, Mg\,I, Fe\,I, K\,I, Al\,I, Na\,I, Ca\,I and the CO band, see Fig.~\ref{g:spectra}). Besides the Na\,I 12\,683\,\AA\ emission line in the J band spectrum, no other emission line could be identified in the spectra. Following the same approach as for \one\, we measured the EW of several absorption lines to determine the spectral type and luminosity class of the star. With an EW$_{\mathrm{CO\,23227\,}\AA}\,=\,20\,\pm\,1$\,\AA, an Fe\,I 22\,263\,\AA\ to Fe\,I at 22\,387\,\AA\ EW ratio equal to 5\,$\pm$\,2 and EW$_{\mathrm{CO\,2.3\mu\,m}}\,=\,43\,\pm\,5$\,\AA\, the spectrum of the star is consistent with that of a red supergiant with effective temperature close to 3\,500\,K, corresponding to a spectral type M2\,I \citep{foersteretal00-1,negueruelaetal10-2}. 

\four\ has the optical counterpart USNO\,0902-0385469 in the USNO\,-\,B1.0 catalogue \citep{monetetal03-1}, at a distance of 0.26\arcsec\ and mean magnitudes equal to 20.9, 17.7, and 15.2 in the B, R and I filters respectively. The source was detected in the 2XMMi-DR3 and was already listed as having IR and optical counterparts by \cite{nebotgomezmoranetal13-1}. At a distance of about 2\arcsec\ from \four\ there is the WISE detection WISE\,J185210.06+001207.3 with magnitudes $6.411\,\pm\,0.038$, $5.404\,\pm\,0.025$, $4.675\,\pm\,0.017$, and $3.923\,\pm\,0.040$ in filters W1, W2, W3 and W4 respectively. WISE magnitudes are in agreement with GLIMPSE magnitudes. We constructed the SED and concluded that it cannot be explained by a single atmosphere component when comparing with the \cite{castelli+kurucz04-1} model with \Teff\,=\,3450\,K and $\log(g)\,=-0.06$, corresponding to a spectral type M2\,I. 
Assuming an intrinsic colour (J-\Ks) of 0.87 equal to the values found for M2\,I stars \citep{negueruelaetal10-2} we derived an extinction \Ak\,$\sim\,1.75$ (\Av\,$\sim\,15$). According to \citep{marshalletal06-1} the extinction in the line of sight ($l\,II$,$b\,II$)\,=\,($33.25^\circ$,$0^\circ$) reaches that value at about 11\,kpc, giving us an upper limit on the distance to the source. No single atmosphere model is able to reproduce the observed SED from the B to the W4 bands. To match the blue part of the SED (B to \Ks\ bands) the needed attenuation is \Av\,$\sim\,18$, displaying an excess at longer wavelengths. To match the infrared part of the SED (H to W4 bands) requires \Av\,$\sim\,28$ displaying an excess at shorter wavelengths. The extinction inferred from the infrared colours is closer to the extinction obtained through SED fitting in the blue part of the spectra than to the red part. \four\ is in a dark cloud \citep{peretto+fuller09-1} within a molecular cloud and an H\,II region \citep{andersonetal09-1}, which is consistent with a high extinction. It could be that the GLIMPSE source is confused with this H\,II region at longer wavelengths, explaining the disagreement between the observed SED and that of a red supergiant star. Assuming an upper limit of 11\,kpc to the source and an extinction of \Ak\,$\sim\,1.75$ we obtained an absolute magnitude of M$_\mathrm{J}=-8.2$. From the SED best fit to the blue part of the spectra we estimated that the radius of the source is around 430\,\Rsun. 

\four\ was detected in two XMM-Newton observations, with mean count rate $0.009\,\pm\,0.002$\,\ctss, implying an X-ray luminosity of $1.5\,\times\,10^{33}$\,\ergs\ (at a distance of 11\,kpc. No variability was observed between the two observations. 

\begin{figure}
\includegraphics[width=\linewidth,angle=0]{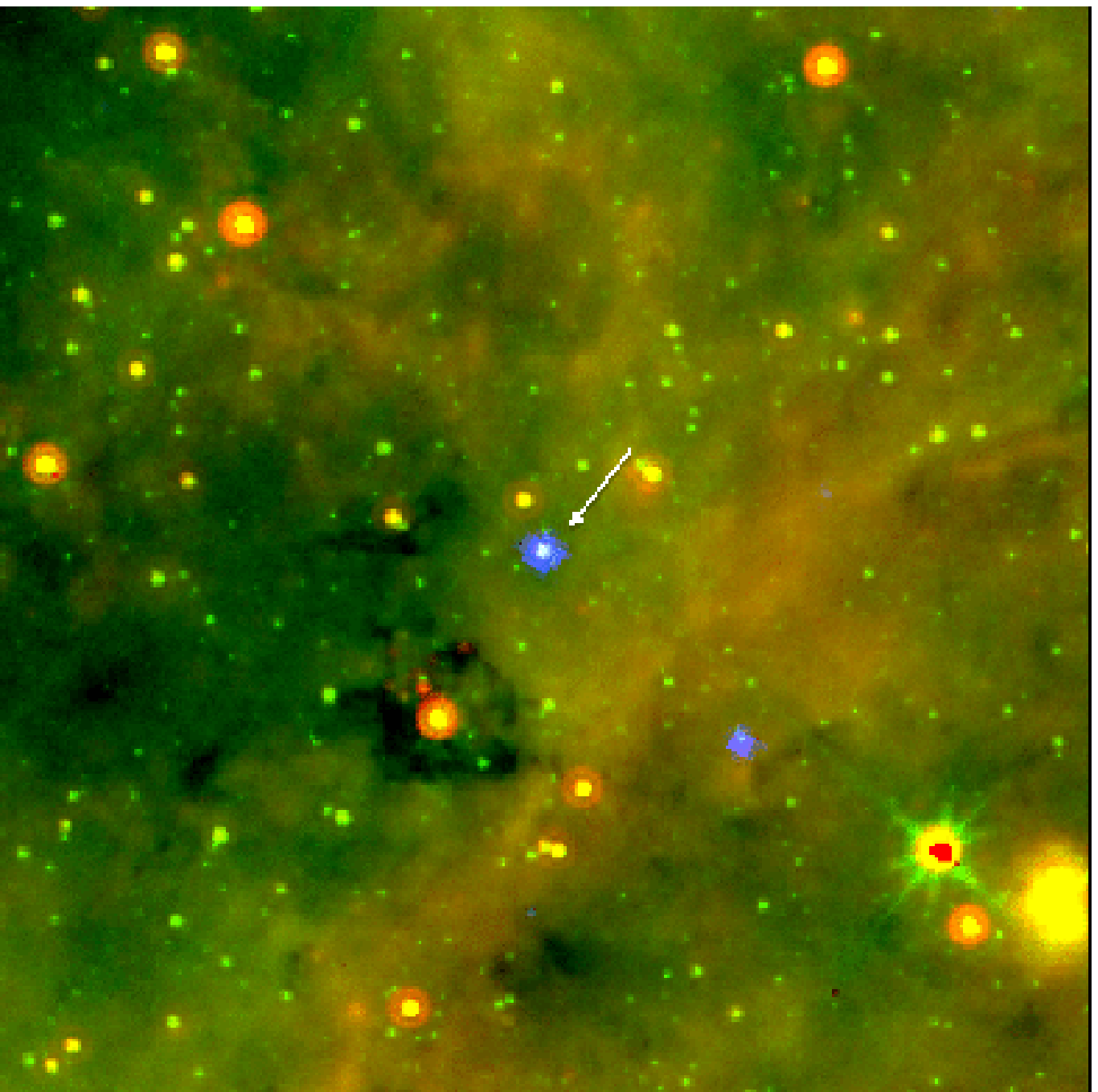}
\caption{Ten arc-minute square RGB image centred around \five\ (red = MIPSGAL 24\,$\mu$m, green = GLIMPSE 8.0\,$\mu$m, blue = EPIC-pn XMM-Newton). North is up and East is left. }\label{g:figure5}
\end{figure}

\subsection{\sourcefive\ (\five)}
The infrared spectra of \five\ show the H\,I Brackett and Pfund series in emission, which together with the He\,I also seen in emission, indicate that the source is probably an early type Be star, with spectral type O9e\,--\,B3e \citep{clark+steele00-1,steele+clark01-1}, where the X-ray and UV photons originating from the hot B star ionise the circumstellar disc giving rise to strong hydrogen recombination emission lines. Photospheric absorption lines are visible in the K band spectrum. Emission lines of He\,II, and Fe\,I are also present in the spectra (see Fig.~\ref{g:spectra}). 

The GLIMPSE and MIPSGAL images at 8\,$\mu$m and 24\,$\mu$m respectively revealed an extended dark cloud in the direction of \five\ (see RGB composite image in Fig.~\ref{g:figure5}). At 8\,$\mu$m, emission is dominated by poly-cyclic aromatic hydrocarbons, and at 24\,$\mu$m emission is dominated by very small dust grains. The emission from this warm dust can explain the high extinction observed for \five, for which we derived \Ak\,$=\,1.69^{+0.21}_{-0.15}$ (\Av$\,=\,15\,\pm\,2$) magnitudes from the (H-\Ks) colour and assuming the intrinsic colours of O9.5 stars from \cite{martins+plez06-1} (note that these colours are independent of the stellar luminosity class). However, the source was also detected in the AllWISE catalogue, AllWISE\,J190144.57+045914.8, with magnitudes that are not consistent with a reddened Rayleigh distribution (see Fig.~\ref{g:spectra}), suggesting some local excess, probably from the circumstellar Be disk. Therefore we cannot use the observed infrared colour as a reliable indicator of the extinction and the \Av\ value should be considered as an upper limit to the extinction. At shorter wavelengths the USNO photographic magnitudes are not accurate enough to constrain the reddening. 

Late-O and early-B supergiants show absorption lines in the infrared spectra, while the brightest blue supergiants show spectra with strong wind emission lines \citep{hansonetal05-1}, both inconsistent with our infrared spectra. Since we don't know the luminosity class of the star we assumed different tabulated magnitudes for O9.5 stars with luminosity classes V and III from \cite{martins+plez06-1} and derived maximum distances of about 1.0 to 1.9\,kpc depending on the luminosity class. The possible link of this system to the class of $\gamma$-Cas analogues will be discussed in Section~\ref{sec:discu-high}

X-ray emission from \five\ was first detected by ASCA, AX\,J190144+0459 \citep{sugizakietal01-1}, as extended diffuse emission, however, the extension was not confirmed by subsequent XMM-Newton observation \citep{yamaguchietal04-1}. A power law model with $\Gamma$\,=\,1.7  \cite{yamaguchietal04-1} derived \Nh\,=\,$3.3\times10^{22}$\,cm$^{-2}$ and a \Nh\ corrected flux of $5.1\times10^{-13}$\,\ergcms\ in the 0.5-10\,keV band. In the range of distances we derived this implies \Lx$\,\sim\,6.4\,\times\,10^{31}-2.1\,\times\,10^{32}$\,\ergs.
\cite{corraletal14-1} fitted the source to different models and present the results in the XMM-Newton spectral-fit catalogue\footnote{\url http://xraygroup.astro.noa.gr/Webpage-prodec/index.html}. In the energy range from 0.5 to 10 keV, the best fit is found for an absorbed thermal plasma with kT\,$=\,9.27^{+3.69}_{-2.23}$\,keV, \Nh$\,=\,2.99^{+0.59}_{-0.28}\times\,10^{22}$\, atoms\,cm$^{-2}$, with a corresponding absorption corrected flux of $3.14\,\pm\,0.34\times10^{-13}$\,\ergcms, implying $\Lx\,$ in the range of $\sim\,4.0\,\times\,10^{31}$ to $1.3\,\times\,10^{32}$\,\ergs.

\section{Discussion}
\label{sec:dis}
In this section we discuss the possible origin of the hard X-ray emission observed in these five X-ray sources, where infrared counterparts pointed to three high-mass stars and two descendants of high-mass stars. We then discuss the nature of all the hard X-ray sources found in this study. 
\subsection{High-mass stars}
\label{sec:discu-high}
The origin of hard X-ray emission in massive stars is still in debate. There are mainly three competing scenarios: intrinsic X-ray emission, accretion on a compact object and colliding winds in a binary system. These three scenarios have been discussed in great detail in \cite{mauerhanetal10-1}. We here briefly summarise the main properties of systems in the different scenarios. 

The intrinsic X-ray emission of early type stars is typically thermal, soft (kT$\,\sim\,0.5$\,keV) and due to radiative shocks occurring within the high velocity wind \citep{lucy+white80-1}. This mechanism is thought to be at work in a wide range of massive stars, including Wolf-Rayet stars \citep{skinneretal10-1}. Unusually hard X-ray emission, although rare in single objects, has been detected in some massive stars (e.g. $\theta^{1}$-Orionis~C). As a possible explanation to the hard X-ray emission (kT$\,\sim\,3$\,keV) it has been suggested that the high magnetic fields observed ($\ge\,1$\,kG) could confine and channel the stellar wind from the poles into the equator where a shock could heat the plasma up to $10^7$\,K, thus producing hard X-ray emission \citep{gagneetal05-1}. \cite{mauerhanetal10-1} calculated that the strength of the magnetic field should be $\gtrsim\,5$\,kG in order to confine the fast winds observed in the WR and O stars of their sample, which is similar to ours. Since so far there is no observational evidence of such high magnetic fields in WR stars, we disregard this scenario for our WR candidates.

In the second scenario, a compact object accretes matter from a massive star, i.e. high-mass X-ray binaries (HMXB). There are three types of HMXBs: those accreting material from the circumstellar decretion disc of a Be star (Be-HMXB), those accreting material from the wind of a supergiant star (SGXB), and those accreting from a Roche lobe filling supergiant star \citep{chaty11-1}. Since Be-HMXBs display \Lx$\,>\,10^{33}$\,\ergs, the low observed X-ray luminosity of the Be star counterpart of \five\ indicates that the source is most probably not a classical Be-HMXB in quiescence. 

Supergiant high-mass X-ray binaries (SGXBs), i.e. a neutron star or a black hole accreting from the strong radiative wind emitted by an OB supergiant star or by a WR star, have persistent X-ray luminosities \Lx$\,\sim\,10^{35}-10^{36}$\,\ergs, show rapid variations and are rather constant on long time scales \citep[see][and references therein]{sgueraetal06-1}. The X-ray luminosities observed in our sample of massive stars are too low to be compatible with SGXBs.

However, a sub-class of accreting sources associated with supergiant stars, the Supergiant fast X-ray transients (SFXTs) show quiescent luminosities in the range \Lx$\,\sim\,10^{32}-10^{33}$\,\ergs, consistent with those of our five stars, while exhibiting flares lasting less than one day and reaching up to \Lx$\,\sim\,10^{36}$\,\ergs\ \citep[][]{negueruelaetal06-1}. The physical origin of the energetic flares is still not clear, it can be either due to the accretion of clumps from the supergiant wind or to some gating mechanism acting close to the compact accreting object. The X-ray light curves of the only source possibly associated with an early supergiant counterpart (\three) does not show any flaring activity characteristic of SFXTs. However, the wide range of inactivity duty cycles displayed by SFXTs \citep{romano2014} does not allow us to exclude a SFXT nature for this particular source.

Likewise, none of the five sources reported in this paper may qualify as an HMXBs undergoing Roche lobe overflow which usually exhibit relatively steady X-ray luminosities of the order of \Lx$\,\sim\,10^{36-38}$\,\ergs. 

The third possible explanation for hard X-ray emission in high-mass stars is the colliding-wind binary interpretation. In this scenario opposing winds generated in two massive stars collide producing a hot shock that is the source of hard thermal X-rays. There are many cases reported in the literature of such systems \citep{mauerhanetal10-1}, where WR\,+\,O stars are the most frequent association. X-ray luminosities are in the $10^{32}-10^{34}$\,\ergs\ range, i.e. compatible with the X-ray luminosities observed in \two\ and \three. Although in our spectra we did not detect any additional spectral lines hinting at the presence of a O companion star, the lower optical luminosity of these objects would not allow us to detect the OB dwarf spectrum. \two\ has very broad emission lines, associated to a very fast wind. It could also be that these emission lines are blended with a second WR star, which would not be resolved given the available resolution. 

\begin{figure}
\begin{center}
\includegraphics[width=\linewidth]{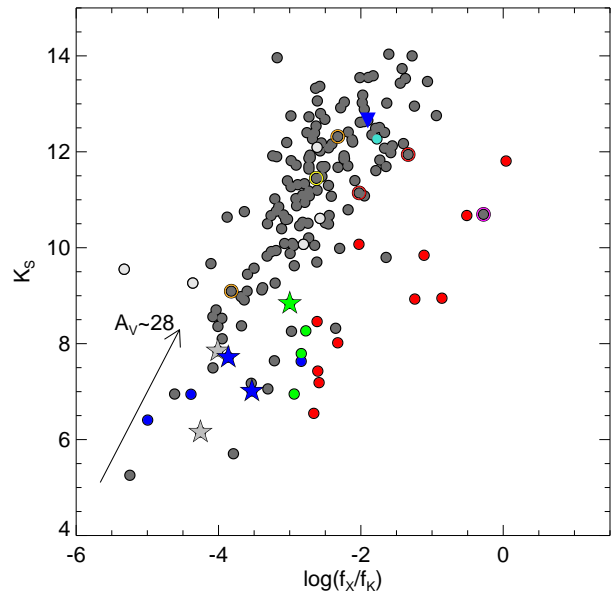}
\caption{Infrared \Ks\ magnitude versus X-ray-to-infrared flux ratio (in the 2\,--\,12 keV band) for hard X-ray sources. Plotted with filled circles are confirmed stars (2 B and 3 G stars) (light grey), HMXBs (red), Be stars (green), and Wolf-Rayet stars (blue). The only AGN from the sample is shown as a blue filled upside down triangle and a Nova as a turquoise filled circle. Unidentified sources are shown in dark grey, surrounded by a coloured ring when a candidate classification of the source is available: candidate HMXBs (red), candidate YSO (yellow), candidate AGB stars (orange), candidate LMXB (magenta). Targets observed at the telescope are highlighted with a star symbol colour coded according to their classification. The effect of extinction assuming a population of hard X-ray massive stars (model 3 in Section~\ref{sec:cands}) is shown with an arrow.} \label{g:logfxfk}
\end{center}
\end{figure}

Finally, $\gamma$-Cas analogues constitute a distinct class of hard X-ray emitting Be stars \citep{smith2015a}. Confined into a narrow range of spectral types ranging from B0e to B1.5e with dwarf or giant luminosity classes, their X-ray emission appears essentially thermal with temperatures larger than 10\,keV and luminosities of a few $\,10^{32}$\,\ergs. Both X-ray temperatures and luminosities are at variance with those displayed by normal Be stars. The origin of the X-ray emission in these objects is debated, it can either be associated with accretion onto a white dwarf companion or due to magnetic interaction between the star and the inner part of the decretion disc \citep{smith+robinson99-1}. However, a recent analysis of the long term optical and X-ray light curves of $\gamma$-Cas strongly favour magnetic activity as the source of the hard X-ray emission \citep{motch2015}. The X-ray luminosity and hard thermal spectrum of \five\ indicate that the star is a likely new $\gamma$-Cas analogue.

We conclude that the X-ray emission of sources \two\ and \three\ is consistent with that expected from radiative shocks in the massive winds of single WR or Ofpe stars or from a wind-colliding binary. The supergiant \three\ may also qualify as a SFXT in quiescence and the giant Be \five\ is consistent with the class of $\gamma$-Cas analogues.

\one\ and \four\ were classified as M2I stars based on the IR spectra taken at the WHT of 2MASSJ\,17434750-2923098 and 2MASSJ\,18521006+0012073 respectively. Red supergiant stars are descendents of OB stars that went through episodes of strong mass loss during their evolution on the giant branch. Although so far no RSG has been confirmed in X-rays, Sct\,X-1 has been suggested as a possible HMXBs with a red supergiant companion \citep{kaplanetal07-1}. This system has been classified as a likely symbiotic binary. Symbiotic binaries are wide binaries in which a white dwarf or a neutron star accretes matter from the wind of a companion red giant star (note that Sct\,X-1 may have either a late giant or a supergiant luminosity class). They have soft X-ray spectra, and show optical emission line spectra. Recent studies, based on Swift/XRT observations, have nevertheless shown that this class of objects can also have hard X-ray spectra \citep{lunaetal13-1}. Although we did not detect any emission line in the infrared spectra of \one\ and \four\, which would rule out a possible symbiotic nature of the source, there are a few symbiotic binaries lacking emission lines \citep{munarietal02-1,hyenesetal13-1}.  The nature of the X-ray emission of \one\ and \four\ remains unclear. Although they could belong to the class of HMXB with a late supergiant donor stars, it is more likely that 2MASSJ\,17434750-2923098 and 2MASS\,18521006+0012073 are spurious infrared identifications of \one\ and \four.

\subsection{The nature of the hard X-ray sources}
We found 690 3XMM-2MASS-GLIMPSE unique associations with reliable X-ray parameters which allowed us to discern between soft and hard X-ray sources. Making use of SIMBAD and VizieR we compiled the list of previously classified objects (see Section~\ref{sec:cands}). 
The fraction of identified sources in the soft and hard X-ray bands are similar (18\% versus 16\%). In Table~\ref{t:table6} we list the number of classified X-ray sources per class of object. Soft X-ray emitters are dominated by stars with spectral types O--M, representing about 90\% of identified soft X-ray sources. 
Hard X-ray sources are dominated by HMXBs, $41\%$ of the classified hard X-ray sources. Nevertheless the population of Wolf-Rayet and O-B stars being discovered in the hard X-ray band is constantly increasing. In this study we add three more sources to this new population (see Fig.~\ref{g:fraction_SIMBAD}). About $81\%$ of the identified hard X-ray sources are high-mass stars either single or in binary systems (WR+Be+HMXB+O/B from Table~\ref{t:table6}).

However, the nature of six hard sources identified from the literature remains unclear. Source 3XMM\,J182833.7-103702 was classified as a LMXB due to the occurence of type I X-ray bursts \citep{cornelisseetal02-1}. The X-ray burst showed low brightness, and furthermore it did not show a clear exponential cooling decay as can be seen from their figure 1. Since so far the spectral type of the secondary star is not confirmed we still consider this source as a LMXB candidate (note that this is not the soft LMXB presented in Table~\ref{t:table6}). Sources 3XMM\,J155033.7-540916 and 3XMM\,J183255.1-100741 have been classified as AGB candidates and 3XMM\,J184841.0-023929 as a YSO candidate by \cite{robitailleetal08-1}. Sources 3XMM\,J182746.3-120605 and 3XMM\,J182814.3-103728 have been classified as candidate HMXBs by \cite{motchetal10-1} (XGPS\,10 and XGPS\,15 respectively). The hard sample source 3XMM\,J182746.3-120605 is the same as XGPS-10 in \cite{motchetal10-1}. We note that the position used for XGPS-10 in the latter work (RA = 18:27:45.49 , Dec = -12:06:06.7) was derived from an early XMM-Newton reduction pipeline in 2001 and is very significantly offset by 12.2 arcsec with respect to that now provided by the 3XMM catalogue. Accordingly, the faint possible candidate deemed as uncertain by \cite{motchetal10-1} cannot be the actual counterpart. Our work shows that XGPS-10 is in fact identified with the nearby bright K=11.94 infrared source. To conclude, since the nature of these six sources is still not confirmed we prefer to leave them as unidentified (i.~e., among the 146 Hard Unid in Table~\ref{t:table6}), meaning that a confirmation of their nature is needed. 

Finally, three stars classified as active coronae have hard X-ray spectra. These three stars have all G spectral type. Although active stellar coronae have typically soft X-ray spectra, many BY~Draconis and RS\,CVn binaries, i.~e. main sequence and evolved binaries, have been detected in hard X-rays \citep[e.~g. II\,Peg][]{ostenetal07-1}. The three G stars in this sample could belong to these types of objects. 
\begin{table}
\begin{center}
\caption{Statistics of source type.}\label{t:table6}
\setlength{\tabcolsep}{0.3em}
\begin{tabular}{ccccccccccc}
\hline
     & Total & WR   & Be   & HMXB   & LMXB & Nova & O/B & A-M  & AGN & Unid \\
     &       &      &      &        &      &      &     &      &     &      \\
\hline                                              
Hard &   173 &  5   &  4     &   11   &      0 &      1 &   2  &     3   &      1 &  146  \\
     &       &  3\% &  2\%   &    6\% &        &    1\% & 1\%  &   2\%   &    1\% &  84\% \\ 
Soft &   517 &  7   &  0     &    0   &      1 &      0 &  30  &    55   &      0 &  424  \\
     &       &  1\% &  0\%   &  0\%   &      0 &    0\% & 6\%  &  11\%   &   0\%  &  82\% \\
All  &   690 & 12   &  4     &   11   &      1 &      1 &  32  &    58   &      1 &  570  \\  
     &       &  2\% & $<$1\% &    2\% & $<$1\% & $<$1\% & 5\%  &    8\%  & $<$1\% &  83\% \\
\hline 
\end{tabular}
\end{center}
\begin{justify} 
\textbf{Notes.} The table shows the fraction of each type of source among a sample of 640 X-ray sources with infrared counterparts. The cross-match procedure and the selection criteria applied to this sources are presented in Sections~\ref{sec:cross-match} and ~\ref{sec:cands}.
\end{justify} 
\end{table}

\begin{table*}
\caption{HMXBs and Be stars identified from the literature.}
\begin{tabular}{llclll}
\hline
Source Name & Sp Type & $\fxfk$ & HR3  & Ks & Note \\
\hline
HD 119682       & B0.5Ve     & -2.56   & -0.21 & 6.95  & $\gamma$-Cas analogue\\
SS 397          &  B0.5Ve    & -2.60   & +0.08 & 8.27  & $\gamma$-Cas analogue\\
NGC 6649 WL 9   & B1-1.5IIIe & -2.73   & +0.15 & 7.79  & $\gamma$-Cas analogue\\
\hline
IGR J16318-4848 & SgB[e]     & -2.60   & +0.89 & 7.19  & supergiant X-ray binary \\
EXO 1722-363    & B0-1Ia     & -0.53   & +0.96 & 10.67 & wind fed Sg accreting neutron star\\
Sct X-1         & late Sg    & -2.67   & +0.92 & 6.55  & supergiant X-ray binary \\
AX J1910.7+0917 & B          & +0.04   & +0.82 & 11.81 & early B-type supergiant star \\
AX J1749.2-2725$^\dag$ & B3  & -2.04   & 0.976 & 10.07 & wind-fed accreting neutron stars\\
\hline
IGR J16465-4507 & O9.5Ia    & -1.08   & +0.51 &  9.84  & supergiant fast X-ray transient \\
XTE J1739-302   & O8.5Iab   & -2.55   & +0.62 &  7.43  & supergiant fast X-ray transient \\
IGR J17544-2619 &  O9Ib     & -2.26   & +0.50 &  8.02  & supergiant fast X-ray transient \\
IGR J18410-0535 & B1Ib      & -1.26   & +0.93 &  8.93  & supergiant fast X-ray transient \\
IGR J18450-0435 & O9Ia      & -0.82   & +0.59 &  8.95  & supergiant fast X-ray transient \\
IGR J18483-0311 & B0.5Ia    & -2.59   & +0.82 &  8.46  & supergiant fast X-ray transient \\
\hline
\end{tabular}
\begin{justify} 
\textbf{Notes.} $^\dag$ The infrared counterpart of AX\,J1749.2-2725 from which a spectral type B was determined has infrared magnitudes fainter than those we determine in this paper, implying the infrared counterpart is different. \cite{kauretal10-1} derived the X-ray position from XMM and Chandra observations and infrared imaging and spectroscopy was carried out at the NTT, e.g. with a better resolution than 2MASS. We note that our identification of the true counterpart of the source might be wrong.
\label{highmasstable}
\end{justify} 
\end{table*}

We show the \Ks\ magnitude versus X-ray-to-infrared flux ratio \fxfk\ for hard sources in Fig.~\ref{g:logfxfk}, highlighting the position of the different types of objects found (HMXBs, WR, Be stars and late-type stars). In our hard X-ray source sample, Wolf-Rayet stars dominate at X-ray-to-infrared flux ratios $\fxfk\,<\,-2.6$, Be stars cover a very narrow range of values (although there are only a few of these systems), with mean $\fxfk\,\sim\,-2.6$, and HMXB dominate at higher values, $\fxfk\,>\,-2.7$. It should be stressed that the sources we selected for identification at the telescope in our pilot study are in no way representative of the entire set of pre-selected hard X-ray sources. First, we selected the brightest infrared sources for reasons of efficiency. Since our X-ray sample is flux limited this implied selecting sources at the low end of the X-ray-to-infra-red flux ratio distribution. In addition, as shown in Fig. \ref{g:hr3hr4}, our pilot sample lies among the most reddened or intrinsically red stars. This is due in part to the presence of two possible M2I interlopers and to the two Wolf-Rayet stars. Finally, due to the location of the William Herschel Telescope, we could only survey the Northern part of the Galactic plane covered by GLIMPSE.

More information on the likely source content of the hard sample can be derived from identifications made through archival catalogues and from the literature. Table \ref{highmasstable} lists HMXBs and  $\gamma$-Cas analogues present in our sample.

\begin{figure*}
\begin{center}
\includegraphics[width=0.63\linewidth]{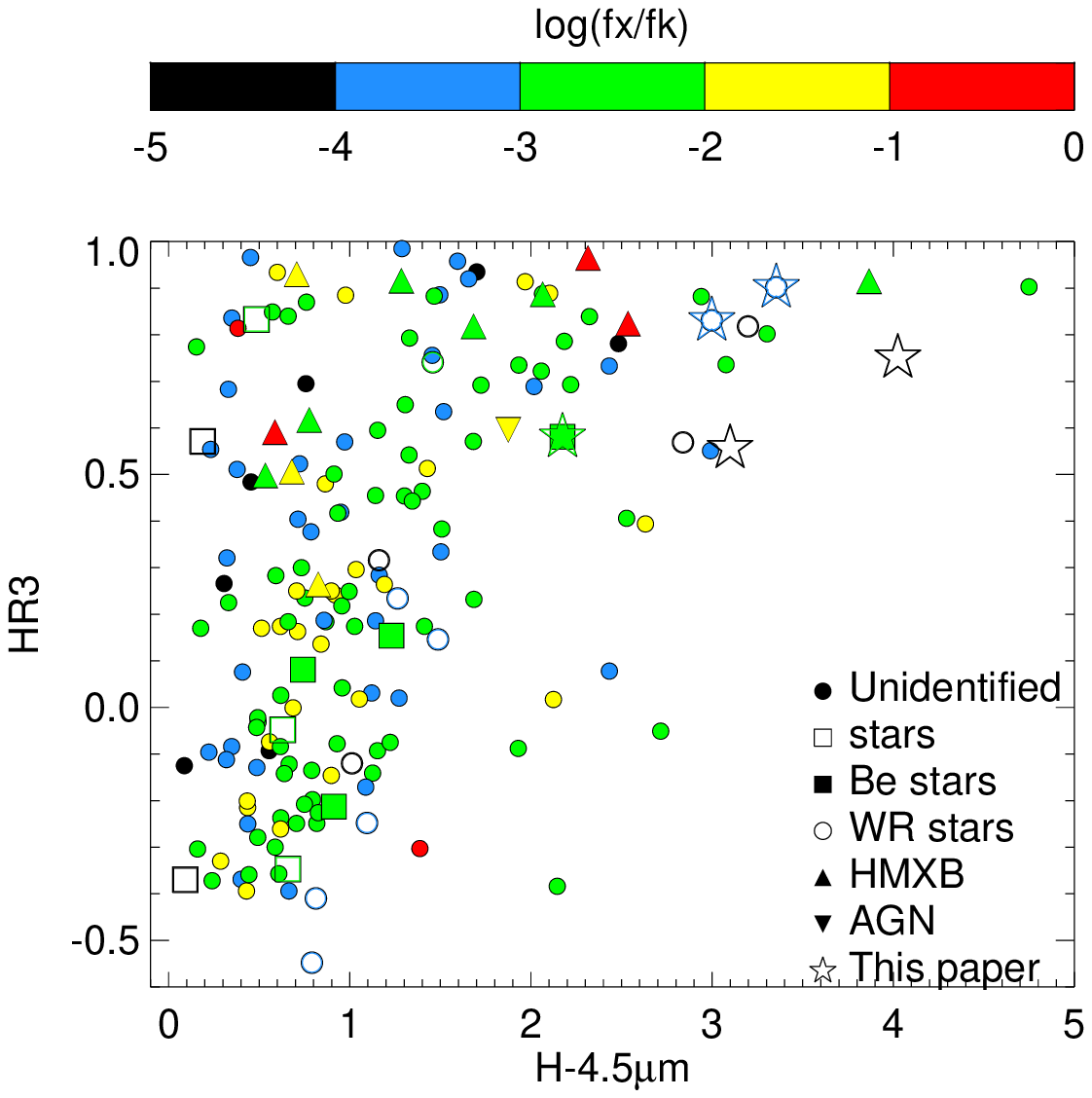}\\
\includegraphics[width=0.63\linewidth]{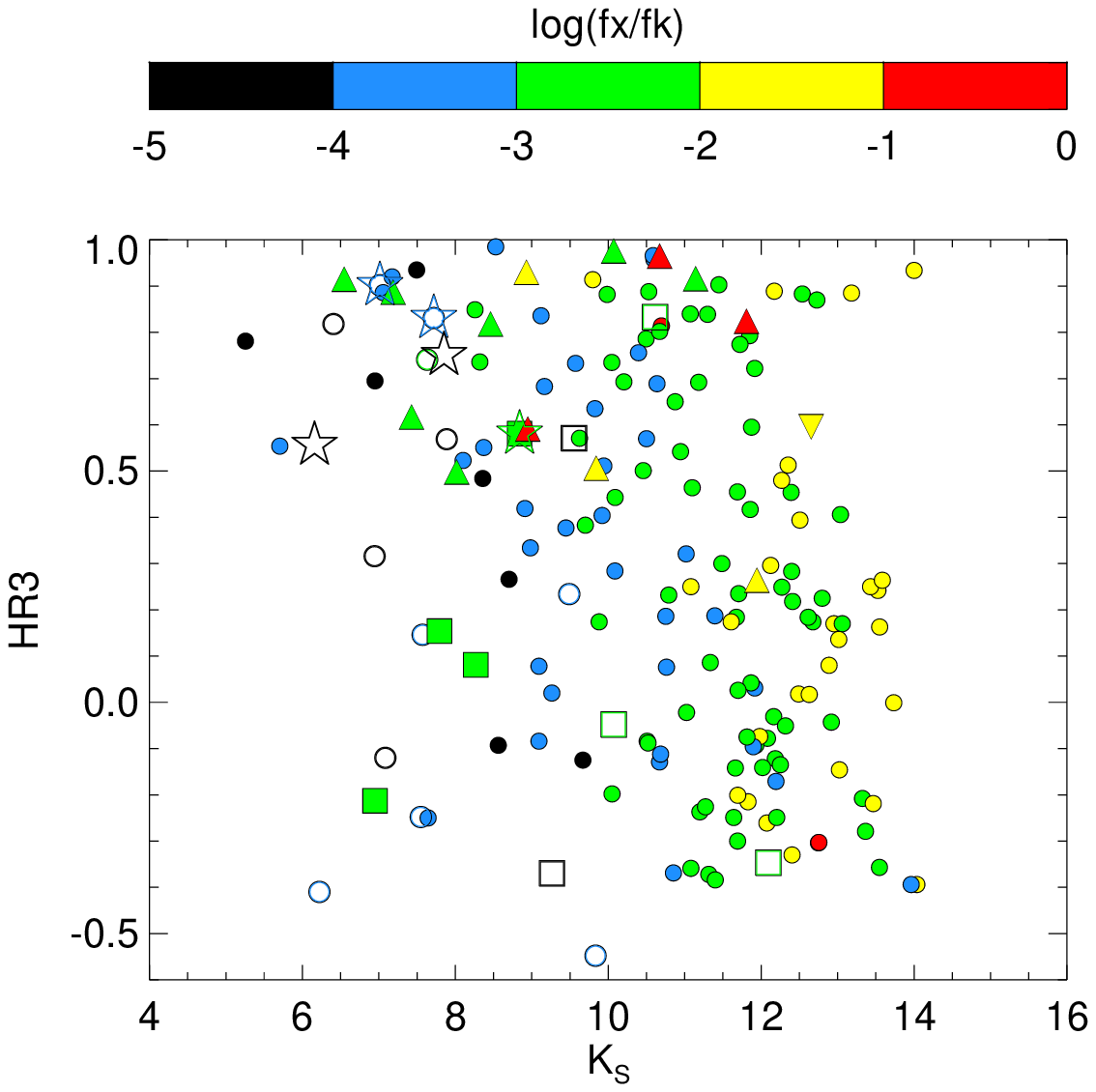}
\caption{Colours of hard X-ray sources. In both figures symbols are colour coded according to the ratio of X-ray-to-infrared fluxes. Unclassified sources are shown as filled circles, WRs as open circles (we also included soft WRs in these two plots), HMXBs as triangles, Be stars as filled squares and other stars with open squares. The position of the only AGN is shown with an upside down triangle. Targets observed in this study are highlighted with a five pointed star symbol.}\label{g:kshr3}
\end{center}
\end{figure*}

\begin{figure}
\begin{center}
\includegraphics[width=\linewidth]{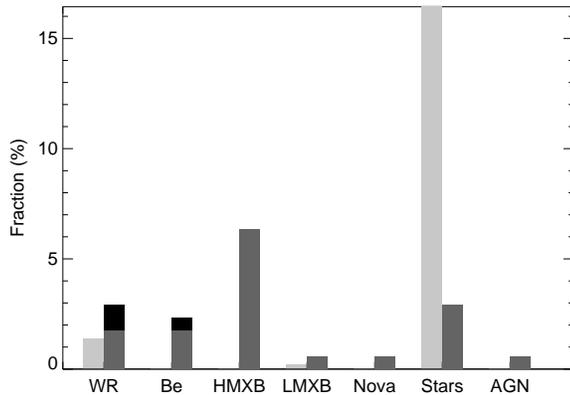}
\caption{Fraction of soft sources (light grey) and hard sources (grey) per object type. In black we show the position of the three X-ray sources for which we give a classification in this study. \label{g:fraction_SIMBAD}}
\end{center}
\end{figure}

We found that the various classes of sources were best separated in a 3-D space whose axis are an infrared colour (giving information on both intrinsic colours and amount of reddening) HR3 (which is the best defined proxy of the X-ray spectral hardness) and the X-ray-to-infrared flux ratio (which provides information on the emission mechanism). The top panel in Fig.~\ref{g:kshr3} shows the distribution of the hard sample in this 3-D space, while introducing an IR magnitude allow us to sort out identified sources from the rest of the unidentified population (see the bottom panel in Fig.~\ref{g:kshr3}).

The group of four $\gamma$-Cas analogues (three previously identified) share similar X-ray and infrared colours and exhibit very comparable average X-ray-to-infra-red flux ratios. As illustrated in the two panels in Fig.~\ref{g:kshr3}, many hard sources have properties similar to those of $\gamma$-Cas analogues, albeit with fainter magnitudes and expected larger distances and may indeed be good candidates for belonging to this particular class.

Very few active coronae have hard enough X-ray spectra to qualify as a hard X-ray source. However, several of the hard sources with HR3 $\leq$ 0.6, \fxfk\ in the range of -2 to -3 and no infra-red excess could still be identified with extreme active coronae.

Compared to  $\gamma$-Cas analogues, Wolf-Rayet stars display on average softer X-ray spectra. Consequently, several of the WR stars identified in the literature actually failed to qualify as hard sources due to their too soft HR4, although their HR3 were fulfilling our conditions. It is thus unlikely that many WR stars appear in our selection of hard sources with HR3 $\leq$ 0.2. However, four WR stars, two identified in the literature and two from our work, are found among our sample of hard sources with HR3 $\geq$ 0.2. Accordingly, we expect to discover new X-ray selected WR stars in the region comprising the hardest X-ray sources with low to medium X-ray-to-infrared flux ratios. We note that the WRs in our sample fall in \cite{mauerhanetal10-1} best selection areas in the (\Ks-8.0\,$\mu$\,m) colour (see their Figure~8).

The top panel in Fig.~\ref{g:kshr3} shows the presence of several hard sources in the region bounded by HR3 $\geq$ 0.4,  \fxfk\ $\geq$ -3 and low to high infrared excess. This area is populated by supergiant X-ray binaries and prospects are therefore relatively good that we find new massive X-ray binaries in this sample. Their relatively modest X-ray-to-infrared flux ratios would rather suggest SFXTs or classical Be/X-ray transients in the close to quiescence state. We note that among our relatively small sample of high-mass X-ray binaries SFXTs seem to display bluer IR colours.

The nature of the population of faint (Ks $\geq$ 11) sources with \fxfk\ in the range of -2 to -1 and HR3 $\leq$ 0.5 (yellow spots in Fig.~\ref{g:kshr3}) remains essentially unconstrained since only one identified source shares these properties, an AGN. Our identified AGN has \fxfk\,=\,-1.90, HR3 = 0.60, Ks = 12.65 and H\,-\,4.5$\,\mu$ = 1.88. We determined the total galactic extinction in the line of sight using the Schlegel maps \citep{schlegeletal98-1}, and found no relation between the measured HR2 or HR3 and the total \Nh\ for hard sources with $\Ks\,\geq\,11$ and $\fxfk\,\geq\,-2.5$. The nature of this group of faint sources is thus not dominated by a population of background AGN. Likewise, the significantly higher X-ray-to-infrared flux ratio displayed by these faint sources compared to brighter ones suggests that  they do not represent a reddened and remote population of sources identified at shorter distances.
 
\section{Conclusions}
\label{sec:summ}
We presented a study which aims at finding hard X-ray high mass stars with low- to intermediate- X-ray -luminosity. For that purpose we cross-correlated the 3XMM-DR4 catalogue with the 2MASS and GLIMPSE catalogues. We pre-selected 173 hard X-ray sources with bright infrared counterparts and with harder X-ray spectra (hardness ratio) than expected for a normal stellar corona. Among these 173 hard X-ray sources, 25 were previously classified: 3 WRs, 3 Be stars, 11 HMXBs, 1 LMXB, 1 Nova, 5 stars and 1 AGN. In this pilot study we carried out infrared spectroscopy of five new high-mass star candidates. We derived spectral types, and stellar parameters for these systems. We confirmed three sources as being high mass hard X-ray stars with spectral types WN7-8h, Ofpe/WN9 and Be and with X-ray luminosities in the range $\sim\,10^{32}-10^{33}$\,\ergs\ (0.2\,--\,12\,keV). One source is compatible with a colliding-wind binary, another source is compatible with both a colliding-wind binary and a supergiant fast X-ray transient and one source is a $\gamma$-Cas analogue candidate. We classified two 2MASS sources as supergiant stars, both with a M2\,I spectral type, and concluded that they are likely not the true counterpart of the X-ray sources, whose nature is still unkown, although they could also belong to the class of HMXBs with a late supergiant star. Finally, the distribution of hard X-ray sources in the parameter space made of X-ray hardness ratio, infrared colours and X-ray-to-infrared flux ratio suggests that many of the unidentified sources are new $\gamma$-Cas analogues, WRs and low \Lx\ HMXBs. However, the nature of the X-ray population appearing at \Ks $\geq$ 11 and exhibiting average X-ray-to-infrared flux ratios remains unconstrained.

\section*{Acknowledgements}
We would like to thank our referee I.~Negueruela for his useful comments which significantly helped improving the quality of this work. 
We also thank P.~A.~Crowther and C.~K.~Rosslowe for sharing with us their new calibrated infrared Wolf-Rayet absolute magnitudes.
We acknowledge financial support from the ARCHES project (7th Framework of the European Union, nº 313146). FJC acknowledges financial support from the Spanish Ministerio de Econo\'ia y Competitividad under project AYA2012-31447. 
This publication makes use of data products from the Two Micron All Sky Survey, which is a joint project of the University of Massachusetts and the Infrared Processing and Analysis Center/California Institute of Technology, funded by the National Aeronautics and Space Administration and the National Science Foundation. 
This work is based in part on observations made with the Spitzer Space Telescope, which is operated by the Jet Propulsion Laboratory, California Institute of Technology under a contract with NASA. 
This research has made use of NASA's Astrophysics Data System. 
This research has made use of the SIMBAD and VizieR databases, operated at CDS, Strasbourg, France.
Based on spectral-fitting results from the XMM-Newton spectral-fit database, an ESA PRODEX funded project, based in turn on observations obtained with XMM-Newton, an ESA science mission with instruments and contributions directly funded by ESA Member States and NASA.

\bibliographystyle{mn2e} 
\bibliography{references}

\begin{appendix}
\section{Notes on B and Be hard X-ray stars}
We briefly discuss the nature of the hard B and Be stars found in this study via cross-correlations with SIMBAD and our spectroscopic observations. \\
\noindent
\textbf{3XMM\,J134632.5-625524} The optical counterpart of 3XMM\,J134632.5-625524, HD\,119682, was found to be a B0.5Ve star at a distance of 1.3\,kpc and with X-ray luminosity \Lx$\,\sim\,10^{32}$\,\ergs, belonging to the class of $\gamma$-Cas analogues \citep[see][and references therein]{torrejonetal13-1}. It is a blue straggler belonging to the open cluster NGC\,5281 at a distance of 1.3\,kpc.\\
\noindent
\textbf{3XMM\,J181046.7-194421} This X-ray source is likely associated to HD\,312792 which was classified as B5 by \cite{nesterovetal95-1}. Visual inspection of the optical spectra used for stellar classification was not possible. No estimates for the stellar luminosity class and distance to the star were found in the literature. \\
\noindent
\textbf{3XMM\,J182746.3-120605} aka XGPS-I\,J182745-120606 is identified with XGPS-10 \citep{motchetal10-1}. We note that the position used for XGPS-10 in the latter work (RA = 18:27:45.49 , Dec = -12:06:06.7) was derived from an early XMM-Newton reduction pipeline in 2001 and is very significantly offset by 12.2 arcsec with respect to that now provided by the 3XMM catalogue. Accordingly, the faint possible candidate deemed as uncertain by \cite{motchetal10-1} cannot be the actual counterpart. Our work shows that XGPS-10 is identified with the nearby bright K=11.94 infrared source. \\
\noindent
\textbf{3XMM\,J182814.3-103728} Based on the optical spectrum, which revealed H$\alpha$ emission and Paschen absorption lines, \cite{motchetal10-1} classified the source as a possible B star (source XGPS-15). Although the authors noted the possibility of the optical source being an interloper. At a distance of about 10\,kpc, consistent with the strong extinction observed for this source, its X-ray luminosity is $\sim\,2.4\,\times\,10^{34}$\,\ergs. The source was classified by \cite{motchetal10-1} as a HMXB candidate, with a giant or supergiant star as secondary star. As for the case above, if the optical counterpart is a blue supergiant the source would not belong to the class of Be stars. \\
\noindent
\textbf{3XMM\,J182828.0-102401} A spectrum taken for the optical counterpart USNO-A2.0\,0750-13261865 revealed a B3 spectral type \citep{motchetal10-1} lacking emission lines. No luminosity class and distance values are available in the literature. \\
\noindent
\textbf{3XMM\,J183327.7-103524} Optical identification of the X-ray source 3XMM\,J183327.7-103524 was firstly presented by \cite{motchetal03-1}. The optical spectrum of SS\,397 revealed a B0\,V star. \cite{nebotgomezmoranetal13-1} estimated a distance to the source of 1.5\,kpc and an X-ray luminosity of $\Lx\,=\,4.4\times\,10^{32}$\,\ergs, and suggested the source to be a likely $\gamma$-Cas analogue. \\
\noindent
\textbf{3XMM\,J183328.3-102408} Based on spectroscopic observations of the optical counterpart USNO0750-13549725\,C, \cite{nebotgomezmoranetal13-1} classified the source as a B1-1.5\,IIIe star. They estimated a distance to the star of 1.7 kpc, a value consistent with that to the cluster NGC6649. USNO0750-13549725\,C is the brightest star in the cluster and it is a blue straggler. It's X-ray luminosity is \Lx$\,=\,6\,\times\,10^{32}$\,\ergs. The source belongs to the $\gamma$-Cas analogue class.\\
\noindent
\textbf{3XMM\,J190144.5+045914} We classified the source in this study as a Be star, at a distance in the range 1--1.8\,kpc and a luminosity of \Lx\,=$0.9--3.0\,\times\,10^{32}$\,\ergs. This source belongs to the class of $\gamma$-Cas analogues. \\
Out of the eight hard B stars presented in this section, two are Be/X-ray high mass binary candidates with X-ray luminosities $\Lx\,\sim\,10^{34}$\,\ergs\ (both with a giant or supergiant secondary star), four are $\gamma$-Cas analogues with $\Lx\,\sim\,10^{32}$\,\ergs, and two are B stars with no emission lines detected so far. 

\end{appendix}

\end{document}